\documentclass{article}

\usepackage{arxiv}
\usepackage[utf8]{inputenc} 
\usepackage[T1]{fontenc}    
\usepackage{microtype}      
\usepackage{booktabs}       
\usepackage{amsmath}
\usepackage{amsfonts}
\usepackage{amssymb}
\usepackage{nicefrac}       
\usepackage{url}            
\usepackage{hyperref}       
\usepackage{cleveref}       
\usepackage{graphicx}
\usepackage{doi}
\usepackage{mhchem}
\usepackage{stmaryrd}
\usepackage{adjustbox}
\usepackage{esint}
\usepackage{bbold}
\usepackage{subcaption}
\usepackage{float}
\usepackage{mathrsfs}
\usepackage{soul}
\usepackage{xcolor}
\usepackage{tcolorbox}
\usepackage{tikz}
\usepackage{placeins}

\newcommand{\filledrhombus}{
    \tikz \fill[gray] (0,0) -- (0.12cm,0.12cm) -- (0.24cm,0) -- (0.12cm,-0.12cm) -- cycle;
}
\newcommand{\filledblackcircle}{%
    \tikz \fill[black] (0,0) circle (0.12cm);%
}
\newcommand{\outlinedwhitecircle}{%
    \tikz \draw[black, thick, fill=white] (0,0) circle (0.11cm);%
}

\usepackage{geometry}
\graphicspath{{./figures/}}

\title{Adaptive mesh refinement algorithm for CESE schemes on quadrilateral meshes}

\newif\ifuniqueAffiliation

\uniqueAffiliationtrue

\ifuniqueAffiliation 
\author{Lisong Shi\thanks{alternative email: ls.shi@connect.polyu.hk} \\
    Department of Aeronautical and Aviation Engineering\\
    The Hong Kong Polytechnic University\\
    \texttt{ls.mark.shi@gmail.com} \\
    \And
    Chaoxiong Zhang \\
    Department of Mechanical Engineering\\
    The Hong Kong Polytechnic University\\
    \texttt{21099246d@connect.polyu.hk} \\
    \And
    Chih-Yung Wen \thanks{corresponding author} \\
    Department of Aeronautical and Aviation Engineering\\
    The Hong Kong Polytechnic University\\
    \texttt{chihyung.wen@polyu.edu.hk}
}

\begin{document}
\maketitle

\begin{abstract}
This study presents constructions of the space-time Conservation Element and Solution Element (CESE) methods to accommodate adaptive unstructured quadrilateral meshes. Subsequently, a novel algorithm is devised to effectively manage the mesh adaptation process for staggered schemes, leveraging a unique cell-tree-vertex data structure that expedites the construction of conservation elements and simplifies the interconnection among computational cells. The integration of second-order $a$-$\alpha$, Courant number-insensitive, and upwind CESE schemes with this adaptation algorithm is demonstrated. Numerical simulations focusing on compressible inviscid flows are carried out to validate the effectiveness of the extended schemes and the adaptation algorithm. 
\end{abstract}


\section{Introduction}
\label{sec1}

In the field of computational fluid dynamics, the resolution of the mesh plays a crucial role in determining the accuracy of the obtained results. This is particularly significant when dealing with problems involving shock or combustion waves. For instance, in scenarios related to deflagration to detonation transition, inadequate mesh resolution, leading to increased numerical diffusion, can contaminate the results. It has been observed that a rapid transition to detonation occurs with coarse resolution, whereas refining the resolution reveals a deflagration wave~\cite{fan2022numerical}. Additionally, in simulations utilizing fixed meshes, a significant number of computational cells may be wasted on smooth regions, a situation exacerbated in transient problems. To address this issue and minimize computational resources without compromising the fidelity of the physics, adaptive mesh refinement (AMR) serves as an effective approach. Finite Volume Methods (FVM)~\cite{cant2022unstructured} and the Discontinuous Galerkin (DG) method~\cite{papoutsakis2018efficient} have been extensively integrated with AMR methods. AMR enables the concentration of computational load in areas of interest, such as shocks, contact surfaces, and vortices. In recent years, AMR methods have found application in various physical problems, including shock waves~\cite{sun1999conservative,wang2022prediction}, two-phase flows~\cite{schmidmayer2019adaptive,schmidmayer2020ecogen}, detonation waves~\cite{deiterding2003parallel,gallier2017detonation}, cosmology~\cite{o2005introducing}, shock-flame interaction~\cite{khokhlov1999interaction}, and reactive shock-bubble interaction~\cite{fan2022numerical}.

In mesh adaptation, two prominent approaches are the block-structured method and the cell-based method. The block-structured approach involves overlaying coarse meshes with patches of finer meshes. A notable solver based on this approach is AMROC developed by Deiterding~\cite{deiterding2003parallel}, which has been widely utilized in addressing detonation problems~\cite{deiterding2009parallel,liang2014effects,cai2016adaptive}. Recently, AMROC has been extended to support curvilinear meshes, as evidenced by its adaptation by Peng~\cite{peng2023three}, allowing for more flexible mesh structures beyond Cartesian grids. Another example of block-structured AMR solver is PeleC, developed by Henry de Frahan et al.~\cite{henry2023pelec}, which utilizes the AMReX library for mesh infrastructure~\cite{katz2020preparing,zhang2019amrex,zhang2021amrex}.

On the other hand, the cell-based approach involves operating on individual cells independently, offering greater flexibility in cell adaptation. Examples of cell-based AMR frameworks include PARAMESH~\cite{macneice2000paramesh} and Athena\texttt{++}~\cite{stone2020athena++}. Efficient algorithms for managing root and leaf cells are crucial in the cell-based approach. For instance, in the fully threaded tree (FTT) structure~\cite{khokhlov1998fully}, dual/quad/oct-tree configurations are designed for one-/two-/three-dimensional simulations, enabling high flexibility in individual cell management. Locating neighbors of a specific leaf cell is a critical aspect of cell-based structured AMR. While a straightforward yet inefficient strategy involves traversing the cell's tree to the root cell and then searching for neighbors from neighboring root cells to leaf cells, more optimized approaches exist. Specialized data structures like cell-edge data structures~\cite{sun1999conservative} or innovative cell-based dual-tree AMR algorithms~\cite{schmidmayer2019adaptive} can streamline the neighbor searching process. During mesh refinement, both cells and faces can be split, allowing direct connections among cells through faces, facilitating rapid neighbor searches at the expense of additional memory. This methodology has found success in applications such as two-phase flows and multi-component reactive flows~\cite{fan2022numerical,schmidmayer2020ecogen}.

The majority of the aforementioned AMR frameworks are primarily tailored for Cartesian meshes, or more generally, structured meshes. This inherent characteristic facilitates the implementation of dynamic load balancing (DLB) and parallel computation with relative ease. In contrast, unstructured AMR offers greater flexibility in mesh topology, albeit at the cost of increased memory consumption per cell. Dune~\cite{bastian2021dune} and ParFUM~\cite{lawlor2006parfum} stand as notable examples of unstructured AMR frameworks. However, it is essential for researchers to meticulously select the most suitable strategies based on the specific requirements of a given problem.

Despite the various advantages and limitations reviewed above, the prevailing focus in current AMR methodologies remains on non-staggered numerical schemes. However, as a special type of finite volume method, the space-time conservation element and solution element (CESE) schemes~\cite{chang1995method,wang2010improved,shen2015robust,shi2023numerical}, feature a unique approach where physical variables are resolved and retained at both the primal and staggered control volumes in an alternating fashion. In this sense, the conservative variables are continuous at the interface of the adjacent control volumes. This unique characteristic eliminates the need for a Riemann solver to update conservative variables explicitly, although in the upwind CESE scheme, a Riemann solver is utilized for computing spatial derivatives without directly impacting the computation of conservative variables. Three primary types of second-order CESE schemes are recognized in the literature: $a$-$\alpha$~\cite{chang2000application}, Courant number insensitive (CNI)~\cite{chang2002courant}, and upwind~\cite{shen2015characteristic,shen2016characteristic} CESE schemes. These schemes have demonstrated favorable numerical characteristics and computational efficiency~\cite{jiang2020space}. Existing applications utilizing CESE schemes predominantly rely on structured Cartesian meshes~\cite{fedorov2004evolution,shen2017maximum,guan2018numerical,fan2019numerical,zhang2020effects}, coordinate-transformed meshes~\cite{shi2023numerical,yang2018upwind}, unstructured tetrahedrons/hexahedral grids~\cite{wang19993,zhang2002space} or hybrid meshes~\cite{wen2018extension,shen2018positivity}.

The challenges associated with dynamically changing mesh topology are particularly pronounced in CESE schemes due to their staggered marching strategy. Significant disparities emerge when contemplating the design of cell-based AMR strategies for either CESE schemes or non-staggered FVM schemes. In FVM, numerical fluxes are added through cell boundaries, and temporal integration is typically accomplished using high-order algorithms. The FVM scheme itself tends to be less susceptible to significant impacts from the AMR process. Conversely, the staggered approach inherent in CESE schemes can lead to complex topologies when attempting to implement AMR without compromising conservation. The following sections highlight that the AMR procedure not only influences mesh redefinition but also has a profound effect on the fundamental definitions of basic elements within CESE schemes. This complexity underscores the need for careful consideration and specialized adaptations when applying AMR techniques to CESE schemes to ensure both accuracy and conservation properties are preserved effectively.

In Jiang et al.~\cite{jiang2012solving}, the CNI CESE scheme on two-dimensional (2D) Cartesian meshes was extended with the AMR framework PARAMESH~\cite{macneice2000paramesh} for solving magnetohydrodynamic (MHD) problems~\cite{liu2022numerical}. However, a loss of conservation was identified due to the mismatch of neighboring conservation elements. Subsequently, Fu et al.~\cite{fu2013simulation} proposed a new definition of conservation elements to ensure that conservation is well-preserved on adaptive Cartesian meshes. Refinement or merging was achieved by inserting or deleting vertices on grid edges, and different approaches to inserting and assigning derivatives on the newly refined grids were explored. On the other hand, unstructured meshes offer the convenience of conforming to complex geometries. To extend the capabilities of the CESE scheme to a broader range of physical problems, there is a need for a generalized AMR strategy tailored to staggered schemes on more adaptable meshes. However, implementing a suitable conservation element dynamically in unstructured meshes can be highly intricate. Additionally, the staggered nature presents challenges for the numerical implementation of the scheme with AMR. As a result, the schemes and the corresponding AMR algorithm for staggered schemes on unstructured meshes remain unavailable.

This paper introduces a novel method for developing an AMR algorithm for CESE schemes and integrating recent advancements in CESE within this AMR framework. The main contributions of this study include:

(1) Devising a novel data structure and an AMR strategy customized for staggered schemes.

(2) Formulating an algorithm for splitting cells and constructing conservation elements to ensure full conservation on general quadrilateral meshes.

(3) Extending three CESE schemes ( $a$-$\alpha, \mathrm{CNI}$, and upwind CESE schemes) in conjunction with this AMR approach.

The subsequent sections of this paper are structured as follows: First, Sec.~\ref{secrev} provides a concise overview of one-dimensional (1D) CESE schemes and identifies the challenges involved in developing an adaptation algorithm for these schemes. Section~\ref{sec2} outlines the construction and formulations of CESE schemes on split quadrilateral meshes. In Sec.~\ref{sec3}, the adaptive algorithm for staggered numerical schemes and its detailed implementations are introduced. Section~\ref{sec4} demonstrates the numerical tests conducted for the proposed algorithm. Section~\ref{secperf} presents the computational efficacy of the current AMR algorithm. Finally, Sec.~\ref{secconc} provides a summary and suggests potential future enhancements.

\section{Brief review of the 1D CESE scheme}
\label{secrev}

\subsection{1D $a$-$\alpha$ CESE scheme}
\label{sec1d}

\begin{figure}[t]
\centering
\includegraphics[width=11cm, trim=3cm 1cm 3cm 4cm, clip]{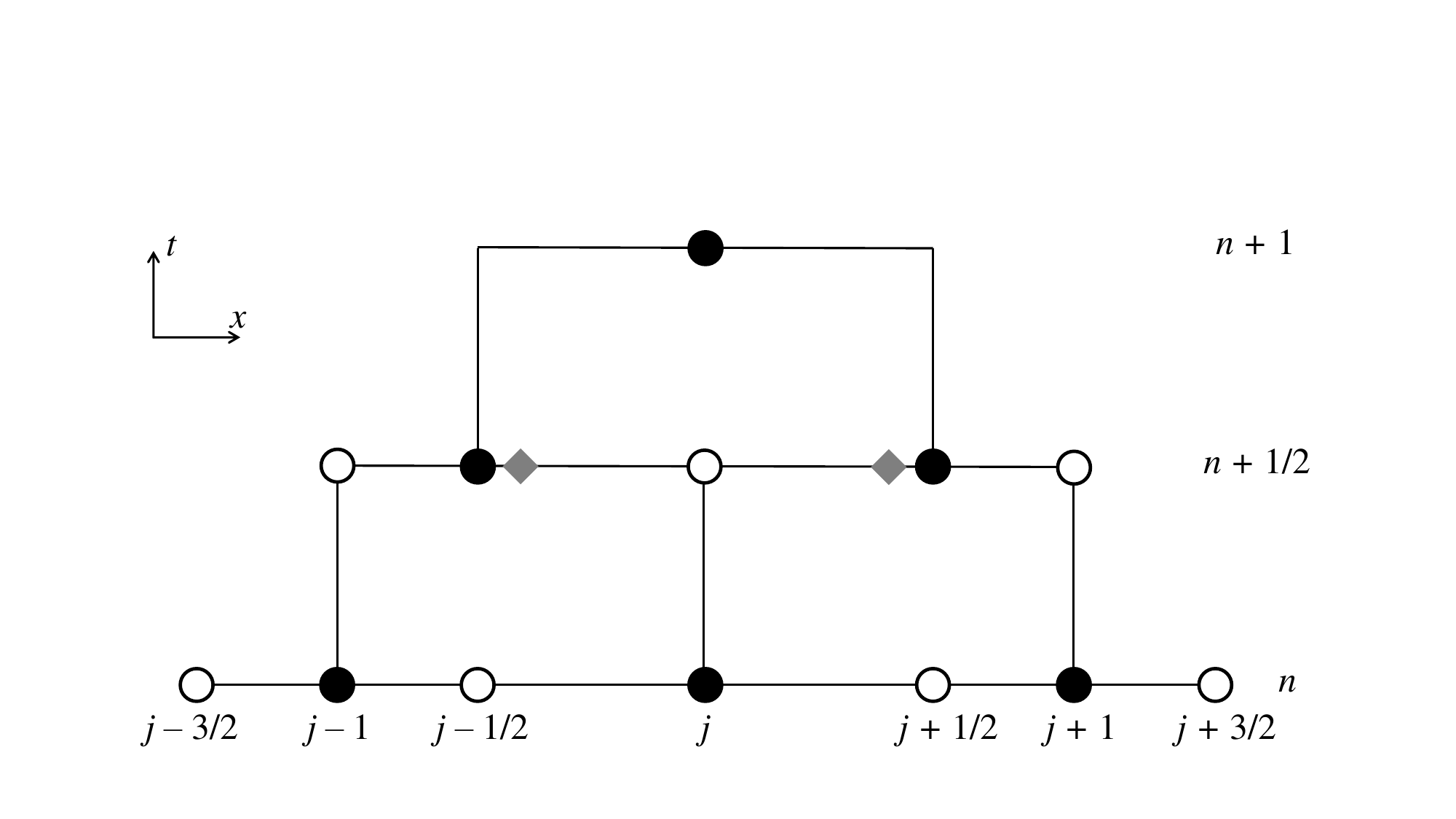}
\caption{Schematic of 1D CESE on non-uniform meshes. $\protect\filledblackcircle$ and $\protect\outlinedwhitecircle$ represent two sets of grid points, and $\protect\filledblackcircle$ represents the point where variables are stored (solution point), $\protect\filledrhombus$ represents the center of the line segment.}\label{1dcese}
\end{figure}

Here, we employ the $a$-$\alpha$ CESE scheme~\cite{chang2000application} to present a succinct overview of its fundamental framework. Consider the 1D scalar conservation law expressed as \begin{equation}
\frac{\partial u}{\partial t}+\frac{\partial f(u)}{\partial x}=0. \label{scalareq}
\end{equation}
The spatial discretization is achieved using non-uniform meshes, as depicted in Fig.~\ref{1dcese}. Denote the point at $(j,n)$ as $p_j^n$. These computational cells are delineated by cell vertices such as $p_{j-1/2}$ and $p_{j+1/2}$. The CESE scheme updates the physical variables in an alternating manner. For each computational step $n \rightarrow n+1$, the computation is split into two half-steps: $n \rightarrow n+1/2$ and $n+1/2 \rightarrow n+1$. In either half-step, the physical values ($u$ and its spatial derivative $u_x$) are stored alternatively in space. In the current study, these values are stored at the cell centers $p_{j-1}$, $p_{j}$, and $p_{j+1}$ for integer time steps such as $t^n$ and $t^{n+1}$, while values are stored at the cell vertices $p_{j-1/2}$ and $p_{j+1/2}$ for half time steps such as $t^{n+1/2}$. During the first half-step, the values at the cell vertices $p_{j-1/2}^{n+1/2}$ and $p_{j+1/2}^{n+1/2}$ are computed. Subsequently, in the second half-step, the values at the cell centers $p_{j}^{n+1}$ are computed.

The CESE methods unify the treatment of space and time, where the integration of fluxes is computed in a similar way. For the 1D scheme, assume $u$ and $u_x$ at each solution point at $t^n$ are known. Define a closed rectangular space-time region named as a conservation element (CE) for each solution point at $t^{n+1/2}$ where the physical values are to be computed. For example, the conservation element corresponding to $p_{j-1/2}^{n+1/2}$ is formed by points $p_{j}^{n}$, $p_{j}^{n+1/2}$, $p_{j-1}^{n+1/2}$, and $p_{j-1}^{n}$.

By defining $\mathbf{h}=(f, u)$, Eq.~(\ref{scalareq}) can be rewritten using Gauss's divergence theorem as 
\begin{equation}
\oint_{S} \mathbf{h} \cdot \mathrm{d} \mathbf{s} = \iint_{\mathrm{CE}} \nabla \cdot \mathbf{h} \mathrm{d} v = 0, \label{1dgauss}
\end{equation}
where $S$ represents the surface of the closed space-time region, and $\mathrm{d} \mathbf{s} = \mathrm{d} \delta \cdot \mathbf{n}$ with $\mathrm{d} \delta$ being an infinitesimal length and $\mathbf{n}$ the corresponding unit outward normal vector. To complete the integration in Eq.~(\ref{1dgauss}), the solution element (SE) for each solution point is defined. For example, solution element for the solution point $p_{j}^{n}$ is defined as two cross lines 
$\overline{p_{j}^{n-1/2} p_{j}^{n+1/2}} \cup \overline{p_{j-1/2}^{n\vphantom{1/2}} p_{j+1/2}^{n\vphantom{1/2}}}$.
In each solution element, the variable $u$ and its flux $f$ are assumed linear and can be approximated by a first-order Taylor expansion,
\begin{equation}
\begin{aligned}
& u(x, t) = u_j^n + (u_x)_j^n (x - x_j) + (u_t)_j^n (t - t^n), \quad (x, t) \in (\mathrm{SE})_j^n \\
& f(x, t) = f_j^n + (f_x)_j^n (x - x_j) + (f_t)_j^n (t - t^n). \quad (x, t) \in (\mathrm{SE})_j^n
\end{aligned}
\end{equation}

The subscripts $x, t$ of ${u}$ or ${f}$ indicate the corresponding spatial or temporal derivatives.
By applying the chain rule, the derivatives of $f$ are described as $f_x=\frac{\partial f}{\partial u}u_x$ and $f_t=\frac{\partial f}{\partial u}u_t$, where  $u_t=-f_x$ as derived from Eq.~\ref{scalareq}.
Then from Eq.~\ref{1dgauss} we can compute the $\tilde u$ at the center ($\filledrhombus$ in Fig.~\ref{1dcese}) of $\overline{p_{j-1}^{n+1/2} p_{j}^{n+1/2}}$ as
\begin{equation}
\tilde u=\frac{1}{2}\left(U_\mathrm{L}+U_\mathrm{R}\right)+\frac{\Delta t}{\Delta x}\left(F_\mathrm{L}-F_\mathrm{R}\right),
\end{equation}
with $\Delta t=t^{n+1/2}-t^{n}$, and
\begin{equation}
\begin{aligned}
& U_\mathrm{L}=u_{j-1}^n+\frac{x_{j-1/2}-x_{j-1}}{2} \left(u_x\right)_{j-1}^n, \\
& U_\mathrm{R}=u_j^n-\frac{x_j-x_{j-1/2}}{2} \left(u_x\right)_j^n, \\
& F_\mathrm{L}=f_{j-1}^n+\frac{\Delta t}{2} \left(f_t\right)_{j-1}^n, \\
& F_\mathrm{R}=f_j^n+\frac{\Delta t}{2} \left(f_t\right)_j^n.
\end{aligned}
\end{equation}

Furthermore, we can get the one-sided derivatives as
\begin{equation}
\begin{aligned}
\left(u_x^{-}\right)_j^n & =\frac{\tilde u-\left[u_{j-1}^{n}+\Delta t \left(u_t\right)_{j-1}^{n}\right]}{({x_j-x_{j-1}}) / 2}, \\
\left(u_x^{+}\right)_j^n & =\frac{\left[u_{j}^{n}+\Delta t\left(u_t\right)_{j}^{n}\right]-\tilde u}{({x_j-x_{j-1}}) / 2}
\end{aligned}
\end{equation}
To suppress oscillations, a weighted average function is used
\begin{equation}
\left(u_x\right)_j^n=W\left(\left(u_x^{-}\right)_j^n,\left(u_x^{+}\right)_j^n, \alpha\right),
\end{equation}

\begin{equation}
W\left(x^{-}, x^{+}, \alpha\right)=\frac{\left|x^{+}\right|^\alpha x^{-}+\left|x^{-}\right|^\alpha x^{+}}{\left|x^{+}\right|^\alpha+\left|x^{-}\right|^\alpha+\epsilon}.
\end{equation}
Here, the adjustable parameter $\alpha$ can take values of 0, 1, or 2, and $\epsilon$ is a small value to avoid division by zero. After computing the derivatives, one needs to interpolate the values from the center of $\overline{p_{j-1}^{n+1/2} p_{j}^{n+1/2}}$ to $p_{j-1/2}^{n+1/2}$.
Using the similar technique, the values at solutions points at $t^{n+1}$ can be computed with the information at $t^{n+1/2}$, completing a full time-step integration.
This $a$-$\alpha$ CESE scheme has been shown to be robust. However, it is sensitive to minimal Courant number. The CNI scheme~\cite{chang2002courant} was proposed to mitigate this drawback by approaching the non-dissipative $a$ scheme when decreasing the Courant number. Moreover, a class of characteristic CESE schemes~\cite{shen2015characteristic,shen2016characteristic} was proposed to be both Courant number insensitive and able to accurately capture material interfaces in multiphase flows. These three schemes share the same staggered approach, with major difference in formulating the strategies in computing spatial derivatives.

\subsection{The basic idea behind AMR of 1D CESE and FVM}

\begin{figure}[t]
\begin{subfigure}{\textwidth}
  \centering
  \includegraphics[width=11cm,  trim=3cm 0.5cm 3cm 13cm, clip]{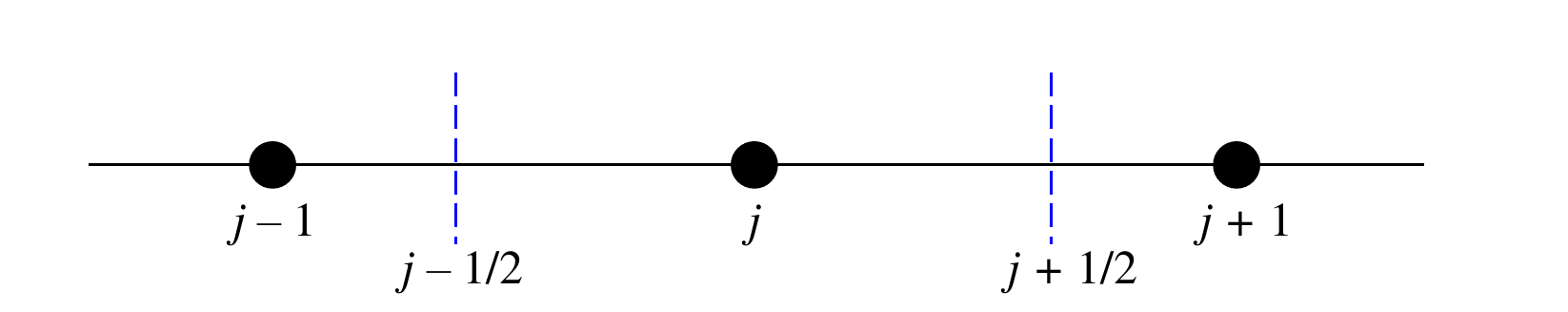}
  \caption{1D FVM on non-uniform meshes.}
\end{subfigure}
\begin{subfigure}{\textwidth}
  \centering
  \includegraphics[width=11cm,  trim=3cm 0.5cm 3cm 13cm, clip]{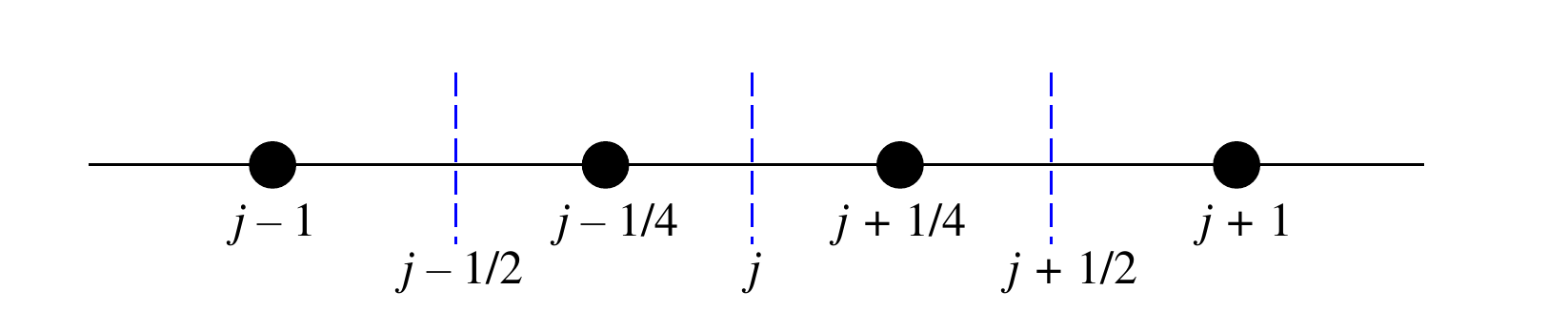}
  \caption{1D FVM after split.}
\end{subfigure}
\caption{Schematics of 1D FVM on non-uniform meshes with/without mesh adaption. The blue dash lines indicate the interfaces between cells.}\label{fvm}
\end{figure}

If we extend the above staggered approach in the CESE schemes to adapted meshes, it will be significantly different from the FVM methods. Figure~\ref{fvm} illustrates the basic idea when applying AMR to the FVM method. The update of $u_j^n$ to $u_j^{n+1}$ (Fig.~\ref{fvm}a) for the FVM method can be expressed as
\begin{equation}
\frac{\mathrm{d} u_j}{\mathrm{d} t}=\frac{1}{x_{j+1/2}-x_{j-1/2}}\left(f_{j-1/2}-f_{j+1/2}\right), \label{fvmeq}
\end{equation}
where $f_{j\pm1/2}$ represents the numerical flux and introduces a Riemann problem:
\begin{equation}
f=\mathscr R(u_\mathrm{L},u_\mathrm{R}),
\end{equation}
Here, $u_\mathrm{L}$ and $u_\mathrm{R}$ are the variables at the left and right sides of the interface. High-order FVM schemes can be achieved by using for example Monotonic Upstream-centered Scheme for Conservation Laws (MUSCL)~\cite{van1979towards} reconstructions and high-order temporal integration~\cite{shu1988efficient,gottlieb2001strong}.

If we split the 1D meshes, for example, the cell $j$ is split into two smaller cells $j-1/4$ and $j+1/4$ (Fig.~\ref{fvm}b). Apparently, there is no fundamental difference in flow integration from the unsplit situations. The major effort lies in implementing a proper data structure managing the adaptation process.

\begin{figure}[t]
\centering
\includegraphics[width=11cm, trim=3cm 1cm 3cm 4cm, clip]{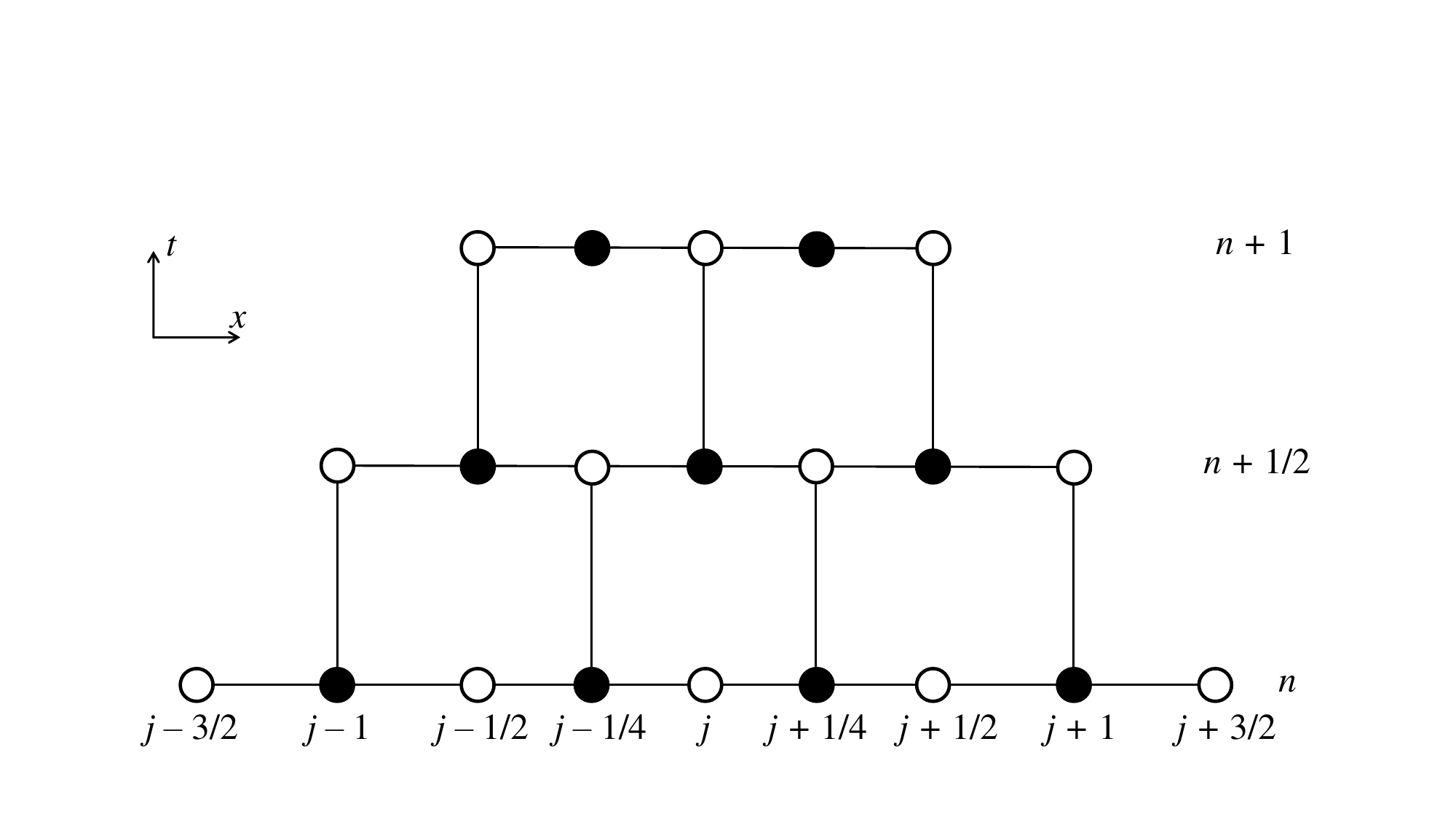}
\caption{Schematic of 1D CESE after splitting the cell $j$ in Fig.~\ref{1dcese}.}\label{1dcesesplit}
\end{figure}

However, in CESE schemes, if we split the cell $j$ as shown in Fig.~\ref{1dcesesplit}, due to the staggered marching nature, it is necessary to not only create the child cells but also add vertices and manage the linkage among the cell centers and vertices appropriately.  Even though the 1D scenario is relatively simple, most available AMR libraries seem unable to manage this kind of topology, not even to say 2D unstructured root meshes. If traditional AMR strategies are forcibly implemented in CESE schemes, mismatches between conservation elements will occur.
When extended to multi-dimensional Cartesian situations, the challenges become more complex. For FVM, the AMR in 2D meshes doesn't exhibit significant differences, with all basic elements remaining in rectangular shapes. But for AMR in 2D staggered meshes, in order to maintain conservation, maintaining conservation requires careful design of arbitrary polygonal conservation elements~\cite{fu2013simulation}, as will be further elaborated in subsequent sections. Thus, a proper data structure, adaptation strategy, and properly extended CESE schemes are desired for multi-dimensional root meshes.

\section{CESE on adaptive quadrilateral meshes}
\label{sec2}

\subsection{The Euler equations}
\label{subsec2-1}

Here, we focus on 2D compressible inviscid flows, governed by the Euler equations:
\begin{equation}
\frac{\partial \mathrm{U}}{\partial t}+\frac{\partial \mathrm{F}(\mathrm{U})}{\partial x}+\frac{\partial \mathrm{G}(\mathrm{U})}{\partial y}=0,\label{eulerEq}
\end{equation}
The vector of conservative variables and the corresponding fluxes are
\begin{equation}
\mathrm{U}=\left[\begin{array}{c}
\rho  \\
\rho u \\
\rho v \\
E
\end{array}\right], \quad \mathrm{F}=\left[\begin{array}{c}
\rho u \\
\rho u^{2}+p \\
\rho u v \\
(E+p) u
\end{array}\right], \quad \mathrm{G}=\left[\begin{array}{c}
\rho v \\
\rho u v \\
\rho v^{2}+p \\
(E+p) v
\end{array}\right],
\end{equation}
where $\rho, u, v, p, E$ are density, velocities, pressure, and total energy, respectively. The energy per unit volume is defined as
\begin{equation}
E=\frac{p}{(\gamma-1)}+\frac{1}{2} \rho\left(u^{2}+v^{2}\right),
\end{equation}
and $\gamma$ is the specific heat ratio. In this study, the main focus is on developing robust AMR strategies and extending CESE schemes. The algorithms outlined in this paper can be generally applied to various flow scenarios such as viscous flows~\cite{guo2004extension}, reactive flows~\cite{shi2017assessment}, MHD~\cite{feng2007novel}, etc., with minimal modifications to the proposed AMR framework, which is beyond the current scope of this work.

\subsection{Mesh topology}
\label{sectopo}
\begin{figure}[t]
\centering
\includegraphics[width=15cm, trim=5cm 0cm 6cm 4cm, clip]{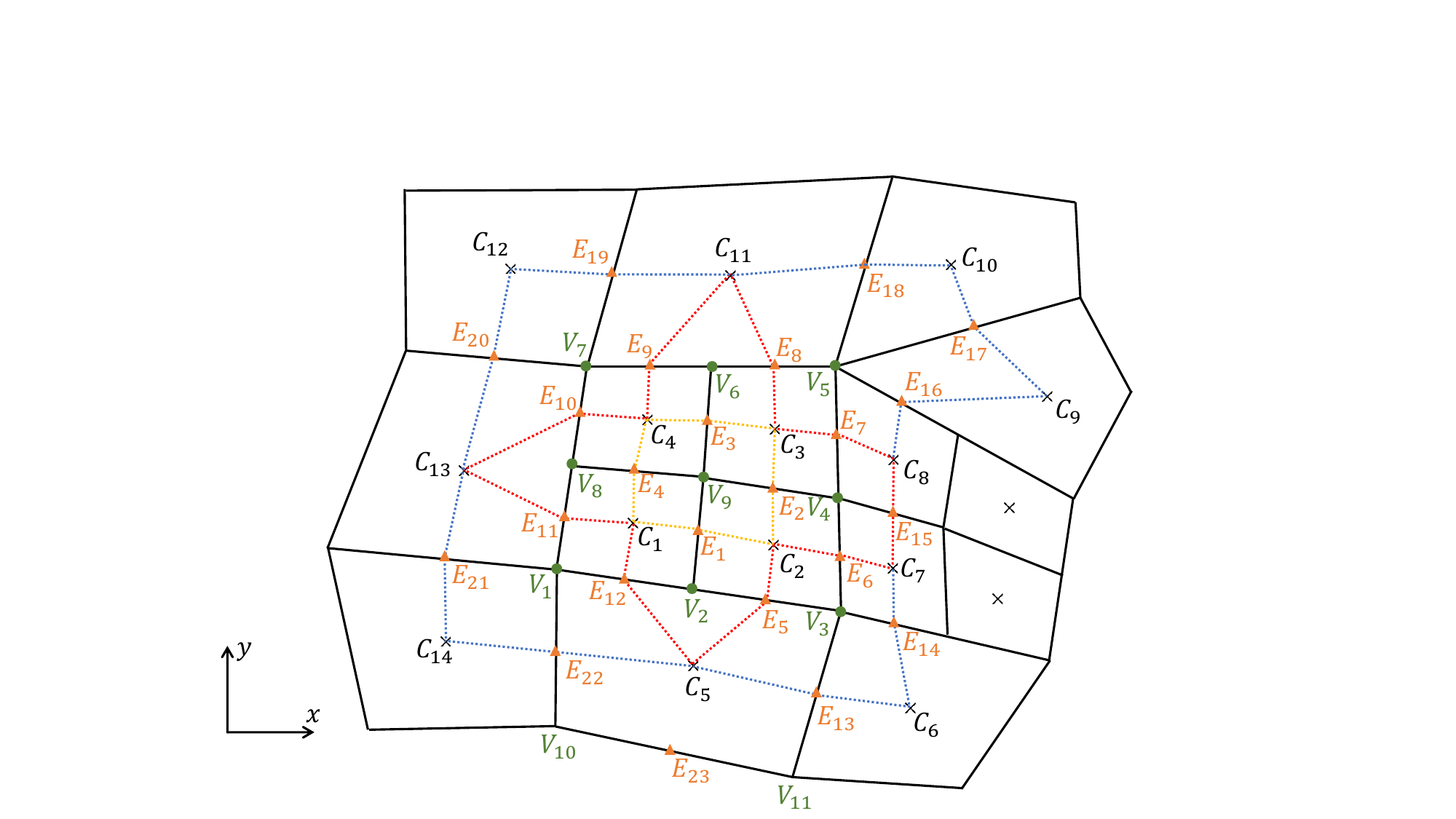}
\caption{Meshes of split quadrilateral meshes and topologies for cells and vertices. $C$ represents the centroid of a cell, $V$ denotes a vertex, and $E$ indicates the center of an edge. Solid lines depict the edges of the unsplit/split cells, while dashed lines show the projections of the sub-CEs associated with $V_{1 \sim 9}$ on the $x$-$y$ plane. }\label{splitQuadMesh}
\end{figure}

As a common feature of the CESE method, each time-step is divided into two half-timesteps: the first and the second half-timesteps. Initially, information regarding the conservative variables and their spatial derivatives is assumed to be stored at cell centers. During the first half-timestep, the focus is on updating the values at vertices based on the information at cell centers. Subsequently, during the second half-timestep, the aim is to update the values at cell centers based on the information at surrounding vertices. The detailed procedure for this splitting process will be further elaborated in Sec.~\ref{sec3}. Here, we assume that some cells in the unstructured quadrilateral meshes have already been split, as illustrated in Fig.~\ref{splitQuadMesh}. To facilitate easy reference throughout the present study, several important symbols are defined as follows:

(1) $C$: represents either the centroid of a quadrilateral cell or the quadrilateral cell itself.

(2) $V$: denotes a vertex, with the additional note that a vertex possesses a level variable upon its creation.

(3) $E$: indicates the center of a line segment.

(4) $\ell$: signifies the level of the cells/vertices, with the level of the root cell defined as 0.

(5) $\ell_{\max}$: denotes the maximum level allowed for refinement.

(6) $\xi, \xi_{\text {split}}$, and $\xi_{\text {join}}$: represent the refinement indicator and critical values.

In Fig.~\ref{splitQuadMesh}, only essential cells, vertices, or line centers are designated for clarity. The cells are organized within cell-trees, where each primary quadrilateral cell (parent) can be subdivided into four child cells. For instance, cell $C_{5}$, defined by the corners $V_{10,11,3,1}$, exists at level $\ell$, while cell $C_{1}$, with corners $V_{1,2,9,8}$, is at level $\ell+1$. The cells $C_{1 \sim 4}$ at level $\ell+1$ are the four child cells derived from the parent cell $C_{0}$ at level $\ell$, specified by corners $V_{1,3,5,7}$. There are a total of nine vertices associated with these four child cells, denoted as $V_{1 \sim 9}$.

While there is no hierarchical tree structure for vertices, a level number is consistently assigned to them upon creation, and this level remains unchanged. The level of a vertex corresponds to the level of the cell to which the vertex belongs as a corner. For example, vertices $V_{1,3,5,7,10,11}$ share the same level as cells $C_{0,5}$, i.e., level $\ell$. Conversely, vertices $V_{2,4,6,8,9}$ are at level $\ell+1$. Notably, $V_{1}$ serves as corners for cells $C_{13}, C_{14}$ and $C_{5}$ at level $\ell$ and $C_{1}$ at level $\ell+1$, maintaining the level of $V_{1}$ at $\ell$ since it was created concurrently with cells $C_{13}, C_{14}, C_{5}$, and $C_{0}$. Even when $V_{1}$ becomes a corner of $C_{1}$ following the split from $C_{0}$, its level remains unaltered. On the other hand, vertices $V_{4}$ and $V_{9}$ are introduced during the construction of cells $C_{1 \sim 4}$, thus being at level $\ell+1$.

The number of cells connected to a vertex is determined by the local mesh topology. Based on the specified conditions and by imposing a constraint ensuring that the difference in levels between neighboring cells does not exceed one, cells connected to a vertex are either at the same level or one level higher. For instance, $V_{2}$ is linked to three cells ($C_{5}, C_{2}$, and $C_{1}$), similar to $V_{6}$ and $V_{8} ; V_{3}$ is connected to four cells ($C_{5}, C_{6}, C_{7}$, and $C_{2}$), as are $V_{1}, V_{4}, V_{7}$, and $V_{9}$; and $V_{5}$ is connected to five cells ($C_{3}, C_{8}, C_{9}, C_{10}$, and $\left.C_{11}\right)$.

\begin{figure}[t]
\centering
\begin{subfigure}{0.31\textwidth}
  \centering
  \includegraphics[width=\linewidth, trim=12.2cm 0cm 13cm 7cm, clip]{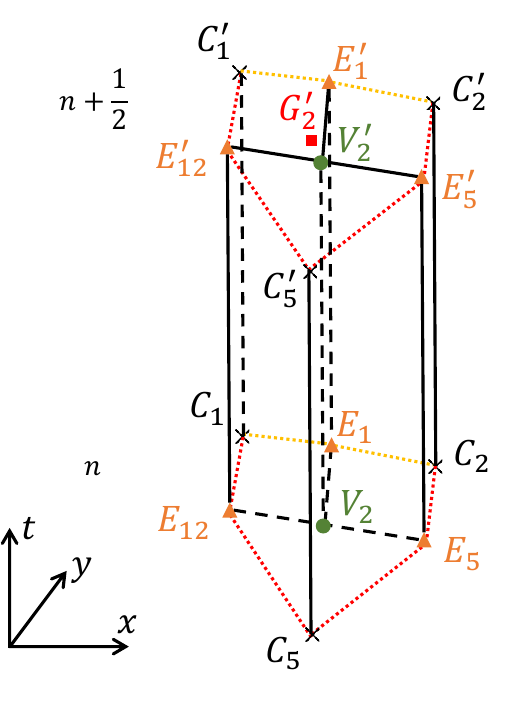}
  \caption{}
\end{subfigure}
\begin{subfigure}{0.31\textwidth}
  \centering
  \includegraphics[width=\linewidth, trim=14cm 0cm 15.5cm 14cm, clip]{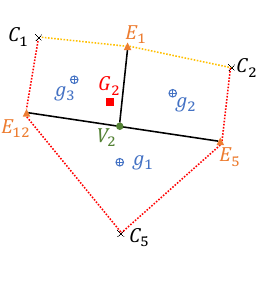}
  \caption{}
\end{subfigure}
\begin{subfigure}{0.28\textwidth}
  \centering
  \includegraphics[width=\linewidth, trim=12cm 0cm 14.5cm 7cm, clip]{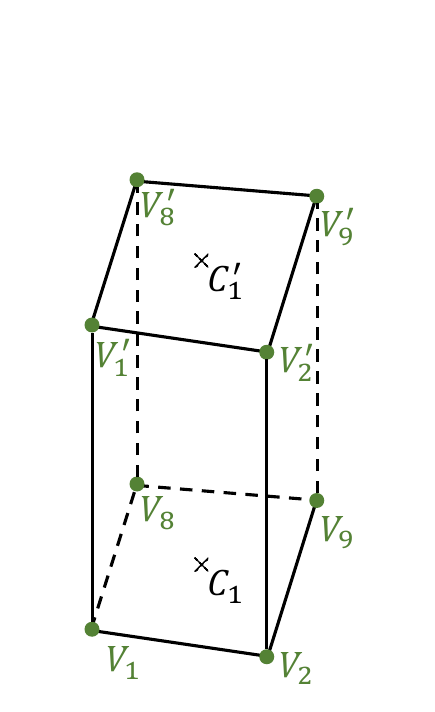}
  \caption{}
\end{subfigure}
\caption{(a) Definition of CE$\left(V_{2}^{\prime}\right)$ and its three sub-CEs, (b) Projection of sub-CEs on the $x$-$y$ plane, (c) Definition of SE$\left(C_{1}\right)$.}\label{defineCEnSE}
\end{figure}

To complete the space-time integration, without loss of generality, a representative conservation element associated with $V_{2}^{\prime}$, denoted as CE$\left(V_{2}^{\prime}\right)$, is defined in Fig.~\ref{defineCEnSE}a. The conservation element serves as the building block of CESE schemes that the computations are based on the integration over it.
For 2D scenarios, it is defined as a temporal excursion of a closed polygon in space. Throughout the following sections, the notation for points (such as cell centroids, edge centers, vertices, etc.) at step $n+1 / 2$ is indicated by a prime superscript, those at step $n+1$ a double prime superscript, and those at step $n$ without any superscript. The time interval $\Delta t$ is defined as either $t^{n+{1}/{2}}-t^{n}$ for the first half-step or $t^{n+1}-t^{n+{1}/{2}}$ for the second half-step.

Here, the cylinder $C_{5} E_{5} C_{2} E_{1} C_{1} E_{12}$-$C_{5}^{\prime} E_{5}^{\prime} C_{2}^{\prime} E_{1}^{\prime} C_{1}^{\prime} E_{12}^{\prime}$ is defined as CE$\left(V_{2}^{\prime}\right)$. This conservation element can be further subdivided into three subordinate CEs (sub-CE), $E_{12} C_{5} E_{5} V_{2}$-$E_{12}^{\prime} C_{5}^{\prime} E_{5}^{\prime} V_{2}^{\prime}($sub-CE$_{1})$, $E_{5} C_{2} E_{1} V_{2}$-$E_{5}^{\prime} C_{2}^{\prime} E_{1}^{\prime} V_{2}^{\prime}($sub-CE$_{2})$, and $E_{12} V_{2} E_{1} C_{1}$-$E_{12}^{\prime} V_{2}^{\prime} E_{1}^{\prime} C_{1}^{\prime}($sub-CE$_{3})$. It is worth noting that the lines $E_{12} V_{2}$ and $V_{2} E_{5}$ are colinear (while lines $C_{1} E_{1}$ and $E_{1} C_{2}$ are generally not colinear), hence sub-CE$_{1}$ is effectively a triangular prism, whereas the other two sub-CEs are quadrilateral cylinders. The surface of a sub-CE is designated as an outer surface if it is shared with the CE; otherwise it is classified as an inner surface of the CE.

The centroid of the polygon $C_{5} E_{5} C_{2} E_{1} C_{1} E_{12}$ (the projection on the $x$-$y$ plane of CE$\left(V_{2}^{\prime}\right)$) is denoted as $G_{2}$. Typically, $G_{2}$ does not coincide with the vertex $V_{2}$. The centroids of the projections on the $x$-$y$ plane of the three sub-CEs are denoted as $g_{1}, g_{2}$, and $g_{3}$, respectively (Fig.~\ref{defineCEnSE}b). Furthermore, a solution element is defined as a region where variables are considered continuous. For example, the solution element corresponding to $C_{1}$, denoted as SE$\left(C_{1}\right)$, is defined as $V_{1} V_{2} V_{9} V_{8}$-$V_{1}^{\prime} V_{2}^{\prime} V_{9}^{\prime} V_{8}^{\prime}$. Notably, as physical variables are stored at the cell center $C_{i}$ at arbitrary $t=t^{n}$, the focus in the first half-step is on computing the physical variables at cell vertices $V_{i}$. Hence, the conservation elements are defined for vertices at $t=t^{n+1/2}$, while the solutions elements are designated for centers at $t=t^{n}$.

In the second half-step, the calculation of variables at cell centers is undertaken. Consequently, the conservation elements specified in this half-step are linked to cell centers at $t=t^{n+1}$, while solution elements are associated with cell vertices at $t=t^{n+1/2}$. The number of vertices connected to a cell is intricately tied to the mesh's topology, and these connected vertices formulate the definition of conservation element associated to the cell center.

For instance, the conservation element corresponding to point $C_{1}^{\prime \prime}$, CE$\left(C_{1}^{\prime \prime}\right)$, is defined as $V_{1}^{\prime} V_{2}^{\prime} V_{9}^{\prime} V_{8}^{\prime}$-$V_{1}^{\prime \prime} V_{2}^{\prime \prime} V_{9}^{\prime \prime} V_{8}^{\prime \prime}$, and its four sub-CEs are $V_{1}^{\prime} E_{12}^{\prime} C_{1}^{\prime} E_{11}^{\prime}$-$V_{1}^{\prime \prime} E_{12}^{\prime \prime} C_{1}^{\prime \prime} E_{11}^{\prime \prime},
 V_{2}^{\prime} E_{1}^{\prime} C_{1}^{\prime} E_{12}^{\prime}$-$V_{2}^{\prime \prime} E_{1}^{\prime \prime} C_{1}^{\prime \prime} E_{12}^{\prime \prime}, V_{9}^{\prime} E_{4}^{\prime} C_{1}^{\prime} E_{1}^{\prime}$-$V_{9}^{\prime \prime} E_{4}^{\prime \prime} C_{1}^{\prime \prime} E_{1}^{\prime \prime}$, and $V_{8}^{\prime} E_{11}^{\prime} C_{1}^{\prime} E_{4}^{\prime}$-$V_{8}^{\prime \prime} E_{11}^{\prime \prime} C_{1}^{\prime \prime} E_{4}^{\prime \prime}$. A more complex conservation element, such as CE$\left(C_{5}^{\prime \prime}\right)$ comprises $V_{10}^{\prime} V_{11}^{\prime} V_{3}^{\prime} V_{2}^{\prime} V_{1}^{\prime}$-$V_{10}^{\prime \prime} V_{11}^{\prime \prime} V_{3}^{\prime \prime} V_{2}^{\prime \prime} V_{1}^{\prime \prime}$, and its five sub-CEs detailed as $V_{1}^{\prime} E_{22}^{\prime} C_{5}^{\prime} E_{12}^{\prime}$-$V_{1}^{\prime \prime} E_{22}^{\prime \prime} C_{5}^{\prime \prime} E_{12}^{\prime \prime}, V_{10}^{\prime} E_{23}^{\prime} C_{5}^{\prime} E_{22}^{\prime}$-$V_{10}^{\prime \prime} E_{23}^{\prime \prime} C_{5}^{\prime \prime} E_{22}^{\prime \prime}, V_{11}^{\prime} E_{13}^{\prime} C_{5}^{\prime} E_{23}^{\prime}$-$V_{11}^{\prime \prime} E_{13}^{\prime \prime} C_{5}^{\prime \prime} E_{23}^{\prime \prime}, V_{3}^{\prime} E_{5}^{\prime} C_{5}^{\prime} E_{13}^{\prime}$-$V_{3}^{\prime \prime} E_{5}^{\prime \prime} C_{5}^{\prime \prime} E_{13}^{\prime \prime}$, and $V_{2}^{\prime} E_{12}^{\prime} C_{5}^{\prime} E_{5}^{\prime}$-$V_{2}^{\prime \prime} E_{12}^{\prime \prime} C_{5}^{\prime \prime} E_{5}^{\prime \prime}$. Notably, the centroid of the projection of the conservation element corresponded to a cell center consistently coincide with the cell' centroid $C$, contrasting the situation with vertices~(Fig.~\ref{defineCEnSE}b). Furthermore, SE$\left(V_{1}^{\prime}\right)$ is defined as $C_{14}^{\prime} E_{22}^{\prime} C_{5}^{\prime} E_{12}^{\prime} C_{1}^{\prime} E_{11}^{\prime} C_{13}^{\prime} E_{21}^{\prime}$-$C_{14}^{\prime \prime} E_{22}^{\prime \prime} C_{5}^{\prime \prime} E_{12}^{\prime \prime} C_{1}^{\prime \prime} E_{11}^{\prime \prime} C_{13}^{\prime \prime} E_{21}^{\prime \prime}$. Similar definitions can be extrapolated for other vertices.

In brief, the above structure comprises cell-trees, where the apexes of these trees are the root unstructured quadrilateral meshes. Cells that do not have any child are referred to as leaf cells. Furthermore, this structure includes a list that encompasses all the vertices. Simply put, each leaf cell is dynamically linked with the associated vertices, in addition to a data package that contains all the physical variables (essential conservative variables U and the spatial derivatives $\mathrm{U}_x$ and $\mathrm{U}_y$). Similarly, each vertex is dynamically linked with the connected cells. All the adaptation procedures and element redefinition will heavily be reliant on this connectivity. We call this special data structure for mesh as \emph{cell-tree-vertex} structure. It is important to note that despite the cells having connections to parent and child cells, they do not actively participate in the neighbor-searching routine. This is a unique advantage over the conventional FTT approach. These parent-child relationships are only utilized during a refine/merge operation. The construction of conservation elements and solution elements is largely dependent on the cell-vertex connectivity. The primary objective of constructing such cell-trees and linkages among cells and vertices is to accommodate the staggered numerical scheme, which alternates between cell centroid and vertex. Surprisingly, this data structure also provides an efficient and direct searching method for unstructured meshes without traversing the entire tree. The process of finding neighbor cells of any cell only requires accessing its connected vertices' information to determine the neighborhood, significantly reducing the searching operation overhead. This will be elaborated further in Sec.~\ref{sec3}. All active cells are directly and independently involved in the construction of the entire domain. In other words, once a cell is split, it is removed from the computational domain, and its child cells assume the connectivity with the surrounding cells and vertices.

\subsection{CESE schemes for split meshes}
In the present AMR framework, we extend both the second-order central schemes~\cite{chang2000application,chang2002courant} and the recently proposed upwind scheme by Shen et al.~\cite{shen2016characteristic} to accommodate the general split unstructured quadrilateral meshes illustrated in Fig.~\ref{splitQuadMesh}. In traditional CESE schemes for unstructured meshes, vertices are constrained to cell corners. However, in CESE schemes designed for split meshes, a unique scenario arises where a vertex may be positioned at the corners of certain cells while simultaneously being located at the midpoint of an edge in another cell. Though the rationales deriving these schemes may vary, it becomes apparent upon closer examination that all three schemes share a common core: the equations governing the update of conserved variables remain consistent. The primary divergence among them lies in the methodologies employed for computing spatial derivatives. For a more comprehensive review of the fundamental CESE schemes, readers can refer to works such as those by Jiang et al.~\cite{jiang2020space} and Wen et al.~\cite{wen2023space}. In this section, we will provide the formulation for the first half-step concerning an arbitrary vertex $V^{\prime}$. The formulation for the second half-step pertaining to an arbitrary cell center closely resembles the first halfstep and is therefore omitted here for brevity. As already mentioned in Sec.~\ref{sec1d} and Sec.~\ref{sectopo}, updating the conservation variables U and its spatial derivatives $\mathrm{U}_{x}$ and $\mathrm{U}_{y}$ is crucial at each solution point in the computational process.

\subsubsection{$a$-$\alpha$ CESE scheme}
By imposing the space-time conservation of Eq. \eqref{eulerEq} on the closed ensemble CE$\left(V^{\prime}\right)$,
\begin{equation}
\oiint_{S} \mathbb{H} \cdot \mathrm{d} \mathbf{s}=\iiint_{\mathrm{CE}} \nabla \cdot \mathbb{H} \mathrm{d} v=0,
\end{equation}
where $\mathbb{H}=(\mathrm{F}, \mathrm{G}, \mathrm{U}), S$ represents the surface of the closed space-time region, and $\mathrm{d} \mathbf{s}=\mathrm{d} \delta \cdot \mathbf{n}$ with $\mathrm{d} \delta$ being an infinitesimal area and $\mathbf{n}$ the corresponding unit outward normal vector. The integration over the surfaces can be approximated by summing up the fluxes across each individual surface, calculated as the product of the surface area and the average flux on that surface. In the context of a second-order scheme, the average flux is determined at the centroid of the surface and is calculated using first-order Taylor expansion.

Let $m \in[1, M]$, where $M$ is the number of cells connected to a vertex. For convenience, we assume that the arrangement from $C_{m} \rightarrow C_{m+1}$ follows a counterclockwise order. Expanding the aforementioned conservation law for CE$\left(V^{\prime}\right)$, it yields:
\begin{equation}
\iint_{\Omega} \mathrm{U}^{\prime} \cdot \mathrm{d} \mathbf{s}+\sum_{m}\left(\iint_{\Omega_{m, \mathrm{D}}} \mathrm{U}\cdot \mathrm{d} \mathbf{s}+\iint_{\Omega_{m, \mathrm{L}}} \mathcal{F} \cdot \mathrm{d} \mathbf{s}+\iint_{\Omega_{m, \mathrm{R}}} \mathcal{F} \cdot \mathrm{d} \mathbf{s}\right)=0.
\end{equation}
Here, $\mathcal{F}=\mathcal{F}(\mathrm{F}, \mathrm{G}, \mathbf{n})$ represents the flux normal to the corresponding surface, $\Omega$ and $\Omega_{m, \mathrm{D}}$ are the surfaces of the projection of CE and sub-CE$_{m}$ onto the $x$-$y$ plane, with $\operatorname{area}(\Omega)=$ $\sum \operatorname{area}\left(\Omega_{m, \mathrm{D}}\right)$. Additionally, $\Omega_{m, \mathrm{L}}$ and $\Omega_{m, \mathrm{R}}$ denote the two outer surfaces of sub-CE$_{m}$. For instance, considering CE$\left(V_{2}^{\prime}\right)$ in Fig.~\ref{defineCEnSE}a, $\Omega_1$ corresponds to $C_{5}^{\prime} E_{5}^{\prime} C_{2}^{\prime} E_{1}^{\prime} C_{1}^{\prime} E_{12}^{\prime}$, while $\Omega_{1, \mathrm{D}}, \Omega_{1, \mathrm{L}}$, and $\Omega_{1, \mathrm{R}}$ respectively represent $E_{12} C_{5} E_{5} V_{2}, E_{12} C_{5} C_{5}^{\prime} E_{12}^{\prime}$, and $C_{5} E_{5} E_{5}^{\prime} C_{5}^{\prime}$. The expressions for $\Omega_{2}$ and $\Omega_{3}$ are similar which are omitted here.

Then, the average value at the centroid $G^{\prime}$ of the polygon at the new half-step can be determined by rearranging the preceding equation:
\begin{equation}
\begin{split}
\mathrm{U}\left(G^{\prime}\right) &=\frac{\iint_{\Omega} \mathrm{U}^{\prime} \mathrm{d} \mathbf{s}}{\operatorname{area}(\Omega)}\\
&=\frac{\sum_{m}\left(\iint_{\Omega_{m, \mathrm{D}}} \mathrm{Ud} \mathbf{s}+\iint_{\Omega_{m, \mathrm{L}}} \mathcal{F} \cdot \mathrm{d} \mathbf{s}+\iint_{\Omega_{m, \mathrm{R}}} \mathcal{F} \cdot \mathrm{d} \mathbf{s}\right)}{\operatorname{area}(\Omega)} \\
&=\frac{\sum_{m}\left(\overline{\mathrm{U}}_{m} \cdot \operatorname{area}\left(\Omega_{m, \mathrm{D}}\right)+\overline{\mathcal{F}}_{m, \mathrm{L}} \cdot \operatorname{area}\left(\Omega_{m, \mathrm{L}}\right)+\overline{\mathcal{F}}_{m, \mathrm{R}} \cdot \operatorname{area}\left(\Omega_{m, \mathrm{R}}\right)\right)}{\operatorname{area}(\Omega)},
\end{split} \label{updatecv1}
\end{equation}
where symbols with an overhead denote the average value on the corresponding surfaces. These averaged values are calculated using the first-order Taylor expansion,
\begin{equation}
\overline{\mathrm{U}}_{m}=\mathrm{U}\left(C_{m}\right)+\mathrm{U}_{x}\left(C_{m}\right) \Delta x_{\mathrm{D}, m}+\mathrm{U}_{y}\left(C_{m}\right) \Delta y_{\mathrm{D}, m},\label{updatecv2}
\end{equation}
\begin{equation}
\overline{\mathcal{F}}_{\eta, m}=\mathcal{F}\left(C_{m}\right)+\mathcal{F}_{x}\left(C_{m}\right) \Delta x_{\eta, m}+\mathcal{F}_{y}\left(C_{m}\right) \Delta y_{\eta, m}+\mathcal{F}_{t}\left(C_{m}\right) \frac{\Delta t}{2}.\label{updatecv3}
\end{equation}
Here, $\eta=\mathrm{L}$ or $\mathrm{R}$, and
\begin{align*}
\Delta x_{\mathrm{D}, m} & =x\left(g_{m}\right)-x\left(C_{m}\right), \\
\Delta y_{\mathrm{D}, m} & =y\left(g_{m}\right)-y\left(C_{m}\right), \\
\Delta x_{\eta, m} & =\frac{x\left(E_{\eta, m}\right)-x\left(C_{m}\right)}{2},  \\
\Delta y_{\eta, m} & =\frac{y\left(E_{\eta, m}\right)-y\left(C_{m}\right)}{2}.
\end{align*}
Here, $E_{\eta, m}$ refers to the centers of line segments directly connected to $C_{m}$ on either left or right side. The derivatives of fluxes can be computed as $\mathrm{F}_{\xi}=\frac{\partial \mathrm{F}}{\partial \mathrm{U}} \mathrm{U}_{\xi}$ and $\mathrm{G}_{\xi}=\frac{\partial \mathrm{G}}{\partial \mathrm{U}} \mathrm{U}_{\xi}$ with $\xi=x, y, t$. The Cauchy-Kowalevski procedure is utilized to obtain $\mathrm{U}_{t}=-\mathrm{F}_{x}-\mathrm{G}_{y}$. These equations complete the updates of conservative variables U. For the updates of spatial derivatives, we employ interpolation within arbitrary SE$\left(C_{m}\right)$ :
\begin{equation}
\mathrm{U}\left(C_{m}^{\prime}\right)=\mathrm{U}\left(C_{m}\right)+\Delta t \cdot \mathrm{U}\left(C_{m, t}\right).
\end{equation}

Establishing the relation based on the information of the variables at the new half-step:
\begin{equation}
\mathrm{U}\left(C_{m}^{\prime}\right)=\mathrm{U}\left(G^{\prime}\right)+\delta x \cdot \mathrm{U}_{x}\left(G^{\prime}\right)+\delta y \cdot \mathrm{U}_{y}\left(G^{\prime}\right),
\end{equation}
where $\delta x=x\left(C_{m}^{\prime}\right)-x\left(G^{\prime}\right), \delta y=y\left(C_{m}^{\prime}\right)-y\left(G^{\prime}\right)$. Here, we define $\delta \mathrm{U}=\mathrm{U}\left(C_{m}\right)+\Delta t \cdot \mathrm{U}\left(C_{t, m}\right)-\mathrm{U}\left(G^{\prime}\right)$. The spatial derivatives, two unknowns, can then be computed by solving the above equations based on information from two neighboring $m$ and $m+1$, such that
\begin{equation}
\mathrm{U}_{x, m}\left(G^{\prime}\right)=\frac{\Delta_{x, m}}{\Delta_{m}}, \mathrm{U}_{y, \mathrm{m}}\left(G^{\prime}\right)=\frac{\Delta_{y, m}}{\Delta_{m}},\label{computederiv}
\end{equation}
\begin{gather*}
\Delta_{m}=\left|\begin{array}{cc}
\delta x_{m} & \delta y_{m} \\
\delta x_{m+1} & \delta y_{m+1}
\end{array}\right|, \Delta_{x, m}=\left|\begin{array}{cc}
\delta \mathrm{U}_{m} & \delta y_{m} \\
\delta \mathrm{U}_{m+1} & \delta y_{m+1}
\end{array}\right|, \Delta_{y, m}=\left|\begin{array}{cc}
\delta x_{m} & \delta \mathrm{U}_{m} \\
\delta x_{m+1} & \delta \mathrm{U}_{m+1}
\end{array}\right|.
\end{gather*}
A weighted average function is used to compute the spatial derivatives \cite{shen2015robust,zhang2002space}:
\begin{equation}
\begin{aligned}
& \mathrm{U}_x\left(G_i^{\prime}\right)=\sum_{m=1}^M W_m \mathrm{U}_{x, m}\left(G_i^{\prime}\right) /\left(\sum_{m=1}^M W_m+\epsilon\right), \\
& \mathrm{U}_y\left(G_i^{\prime}\right)=\sum_{m=1}^M W_m \mathrm{U}_{y, m}\left(G_i^{\prime}\right) /\left(\sum_{m=1}^M W_m+\epsilon\right),\label{averagederiv}
\end{aligned}
\end{equation}
where $\epsilon$ represents a small value to prevent zero denominators, and
$$
W_{m}=\left(\prod_{i=1, i \neq m}^{M} \theta_{m}\right)^{\chi}
$$
with
$$
\theta_{m}=\sqrt{\left[\mathrm{U}_{x, m}\left(G_{i}^{\prime}\right)\right]^{2}+\left[\mathrm{U}_{y, m}\left(G_{i}^{\prime}\right)\right]^{2}}.
$$

\subsubsection{CNI CESE scheme}
The derivation for updating the conservative variables follows the approach of the $a$-$\alpha$ scheme, as expressed in Eqs.~\eqref{updatecv1}-\eqref{updatecv3}. To remedy the excessive dissipative nature of the $a$-$\alpha$ scheme when subjected to minimal Courant numbers, following the work by Chang~\cite{chang2002courant} and Shen \& Parsani~\cite{shen2018positivity}, a new point in Fig.~\ref{defineCEnSE}b, $q_{m}$, is defined for which the coordinates are calculated by
\begin{equation}
x_{i}\left(q_{m}\right)=\frac{v}{v_{0}} x_{i}\left(C_{m}\right)+\left(1-\frac{v}{v_{0}}\right) x_{i}\left(g_{m}\right), \label{defineqm}
\end{equation}
where $v$ and $v_{0}$ represent the local and global Courant numbers, respectively. Utilizing Taylor expansions within SE$\left(G^{\prime}\right)$, we have
\begin{equation*}
\mathrm{U}\left(q_{m}^{\prime}\right)=\mathrm{U}\left(G^{\prime}\right)+\mathrm{U}_{x}\left(G^{\prime}\right) \delta x_{m}+\mathrm{U}_{y}\left(G^{\prime}\right) \delta y_{m},
\end{equation*}
where $\delta x_{m}=x\left(q_{m}\right)-x(G), \delta y_{m}=y\left(q_{m}\right)-y(G)$, and $\mathrm{U}\left(q_{m}^{\prime}\right)$ can be explicitly calculated utilizing the Taylor expansion within SE($C_{m}$). A similar relationship can be established for the $(m+1)^{\text {th }}$ points. Subsequently, the two spatial derivatives can be solved in a manner akin to Eq.~\eqref{computederiv}, culminating with the application of the average function as delineated in Eq.~\eqref{averagederiv}. This approach enables the CNI scheme to transition towards the non-dissipative core scheme as $v \rightarrow 0$, and towards the $a$-$\alpha$ scheme as $v \rightarrow 1$~\cite{chang2003multi}.

\subsubsection{Upwind CESE scheme}
\label{upwindScheme}
The introduction of characteristic-based fluxes into the CESE frameworks commences by imposing the conservation law on each sub-CE, one obtains
\begin{equation}
\oiint_{S_{m}} \mathbb{H} \cdot \mathrm{d} s=\iiint_{\mathrm{CE}_{\mathrm{m}}} \nabla \cdot \mathbb{H} \mathrm{d} v=0 \label{upwindSubEq},
\end{equation}
here $S_{m}$ denotes the surface of the $m^{\mathrm{th}}$ sub-CE. Expands Eq. \eqref{upwindSubEq} to
\begin{equation}
\iint_{\Omega_{m}} \mathrm{U}^{\prime} \cdot \mathrm{d} \mathbf{s}+\iint_{\Omega_{m, \mathrm{D}}} \mathrm{U} \cdot \mathrm{~d} \mathbf{s}+\iint_{\Omega_{m, \mathrm{L}}} \mathcal{F} \cdot \mathrm{d} \mathbf{s}+\iint_{\Omega_{m, \mathrm{R}}} \mathcal{F} \cdot \mathrm{d} \mathbf{s}+\iint_{\Omega_{m, i \mathrm{L}}} \mathcal{F} \cdot \mathrm{d} \mathbf{s}+\iint_{\Omega_{m, i \mathrm{R}}} \mathcal{F} \cdot \mathrm{d} \mathbf{s}=0 \label{upwindExpEq}.
\end{equation}

The first four terms on the left-hand side of Eq. \eqref{upwindExpEq} denote the fluxes across the outer surfaces of the sub-CE. The computation of these fluxes follows the methodologies illustrated in previous $a$-$\alpha$ and CNI schemes, which are derived through Taylor expansion. The last two terms correspond to the fluxes across the inner boundaries, where $\Omega_{m, i \mathrm{L}}$ and $\Omega_{m, i \mathrm{R}}$ signify the surfaces of these inner boundaries. Since these inner boundaries act as interfaces between two solution elements, they are typically treated as discontinuities, resulting in a Riemann problem of the form $\mathcal{F}=\mathscr{R}\left(\mathrm{U}_{\mathrm{L}}, \mathrm{U}_{\mathrm{R}}\right)$. Here, U$_{\mathrm{L}}$ and U$_{\mathrm{R}}$ represent the conservative variables reconstructed at the centroids of the inner surface. The WBAP (Weighted Biased Averaging Procedure limiter) as presented by Li et al.~\cite{li2011multi} is applied for reconstructing the derivatives:
\begin{align}
& \widetilde{\mathrm{U}}_{x, \mathrm{L}}=\mathrm{U}_{x, \mathrm{L}} \operatorname{WBAP}\left(1, \theta_{\mathrm{L}}\right), \label{wbap1}\\
& \widetilde{\mathrm{U}}_{x, \mathrm{R}}=\mathrm{U}_{x, \mathrm{R}} \operatorname{WBAP}\left(1, \theta_{\mathrm{R}}\right), \label{wbap2}
\end{align}
where $\theta_{\mathrm{L}}=\mathrm{U}_{x, \mathrm{R}} / \mathrm{U}_{x, \mathrm{L}}$ and $\theta_{\mathrm{R}}=\mathrm{U}_{x, \mathrm{L}} / \mathrm{U}_{x, \mathrm{R}}$ with
$$
\operatorname{WBAP}(1, \theta)=\left\{\begin{array}{l}
\frac{n+1 / \theta}{n+1 / \theta^{2}}, \text { if } \theta>0 \\
0, \text { else }
\end{array}\right.
$$
The linear weight $n$ is set to 5. Similarly, the derivatives in the $y$-direction can also be reconstructed. Consequently, the inner fluxes can be solved using approximate Riemann solvers~\cite{shen2016characteristic,toro2013riemann} or more simply, the local Lax-Friedrichs (LLF) flux~\cite{shi2023numerical,shen2018positivity}. In this investigation, the rotated Harten-Lax-van Leer contact (HLLC) Riemann solver~\cite{shen2016characteristic,toro1994restoration,levy1993use,nishikawa2008very,ren2003robust} is adopted for enhanced accuracy.

Furthermore, the inner fluxes across neighboring sub-CEs exhibit equal magnitudes but opposite directions. Summing Eq.~\eqref{upwindExpEq} for all sub-CEs results in a balance of the inner flux terms, leading to Eq.~\eqref{updatecv1} and ensuring consistency in the calculations of conservative variables. Practically, the computation simplifies to:
\begin{equation}
\mathrm{U}\left(G^{\prime}\right)=\frac{\sum_{m} \mathrm{U}\left(g_{m}^{\prime}\right) \cdot \operatorname{area}\left(\Omega_{m}\right)}{\operatorname{area}(\Omega)}.\label{avgUEq}
\end{equation}

Upon evaluating the fluxes across outer surfaces and upwind fluxes across inner surfaces, Eq. (17) yields a unique solution for $\mathrm{U}\left(g_{m}^{\prime}\right)$ and
\begin{equation}
\mathrm{U}\left(g_{m}^{\prime}\right)=\mathrm{U}\left(G^{\prime}\right)+\delta x \cdot \mathrm{U}_{x}\left(G^{\prime}\right)+\delta y \cdot \mathrm{U}_{y}\left(G^{\prime}\right),
\end{equation}
where $\delta x=x\left(g_{m}^{\prime}\right)-x\left(G^{\prime}\right), \delta y=y\left(g_{m}^{\prime}\right)-y\left(G^{\prime}\right)$. The methodology for computing spatial derivatives, as outlined in Eqs.~\eqref{computederiv}\text{\&}\eqref{averagederiv}, remains consistent. 

Notably, for all above schemes, after computing $\mathrm{U}\left(G^{\prime}\right)$ and spatial derivatives $\mathrm{U}_{x}\left(G^{\prime}\right)$ and $\mathrm{U}_{y}\left(G^{\prime}\right)$, interpolation is employed to determine values at the vertex $V^{\prime}$ rather than at the centroid of the polygon $G^{\prime}$.

\section{Adaptive algorithm}
\label{sec3}
Section \ref{sec2} outlined in detail of the strategies employed for updating the physical variables using the CESE approach on split meshes. In order to enhance the accessibility of the code algorithm, this section offers a comprehensive presentation of the AMR algorithm and data structure specifically designed for staggered schemes. The algorithms have been implemented using an Object-Oriented-Programming (OOP) style, which enables efficient management of data and mesh structures, thus providing significant flexibility. Emphasis has been placed on ensuring proper connectivity between cells and vertices, addressing the treatment of vertices during the cell refinement process, and establishing conservation elements following the AMR procedure at each time step.

\subsection{Basic restrictions and constructions}
When integrating with the staggered CESE scheme, the time-stepping approach outlined in Sec.~\ref{sec2} differs from that of the conventional FVM schemes. Consequently, a conventional cell-based AMR method cannot be directly applied to the CESE scheme without compromising essential characteristics.

To clarify, we recall the root mesh level as ${\ell_0 \stackrel{\text{def}}{=} 0}$ and the maximum refinement level as $\ell_{\max }$. The refinement ratio is constrained to a factor of 2, such that during a split step, one edge is divided into two smaller edges, resulting in the subdivision of a quadrilateral cell into four child quadrilateral cells. Additionally, the maximum allowable level difference for any Moore neighborhood (defined as any two cells sharing at least one common vertex) of a cell is limited to 1. Here, we designate two cells sharing two vertices as adjacent neighbors and those sharing only one vertex as connected neighbors. The introduction of the vertex class allows us to enforce the aforementioned restriction on level differences, ensuring that the level difference of all cells linked to a specific vertex does not exceed 1. Moreover, this restriction facilitates the automatic construction of a buffer layer near discontinuities. The thickness of this buffer layer can be controlled through a straightforward operation, as elaborated in subsequent sections.

The proposed methodology endeavors to optimize the independence between mesh algorithms and physical algorithms. Essential physical state information, such as $\mathrm{U}, \mathrm{U}_{\eta}, \mathrm{F}, \mathrm{F}_{\eta}, \mathrm{G}, \mathrm{G}_{\eta}$ with $\eta=$ $x, y, t$, is encapsulated within cells and vertices as objects. Recall that conservative variables and spatial derivatives are iteratively updated between the cell centers and vertices during each half-step. To accomplish this, proper definitions for conservation elements and thus sub-CEs are indispensable. Consequently, the cell-tree-vertex structure comprises two fundamental classes: namely the Cell class and the Vertex class. The construction of conservation elements (and sub-CEs) necessitates access to information regarding their connected entities. Specifically, establishing a conservation element associated with a cell center requires knowledge of all the vertices situated on the edges of that cell, whereas creating a conservation element associated with a vertex requires information about all the cells linked to that vertex.

The Cell and Vertex objects are interlinked through addresses to facilitate rapid direct access. Furthermore, it is noteworthy that each cell does not store the information of its neighboring cells directly; rather, it accesses this information through connected vertices. Notably, the adjacent cell for two consecutive vertices on a cell can be identified as the common connected cell of these two vertices $V_{i}$ and $V_{j}$, i.e., $C_{\text {neighbor }} \equiv\left\{C\right.$: cells connected to $\left.V_{i}\right\} \cap\left\{C\right.$: cells connected to $\left.V_{j}\right\}$. For instance, in Fig.~\ref{splitQuadMesh}, to determine the adjacent neighbor of $C_{1}$ on the side over edge $V_{1} V_{2}$, the cells connected to $V_{1}$ include $C_{13}, C_{14}, C_{5}$, and $C_{1}$, while the cells connected to $V_{2}$ include $C_{5}, C_{2}$, and $C_{1}$. The common cell shared by these two vertices, aside from $C_{1}$, is $C_{5}$. This approach expedites easy access to neighbors without traversing through the trees, and this operation is only required once after the cell is impacted by the AMR operation.

Practically, upon cell splitting, the child cells and neighbor cells are labeled as “affected”, and similarly, when four child cells are merged, their parent and neighbors are marked. Only these labeled cells necessitate an update for neighboring information. Throughout the remainder of this paper, the term “neighbor” denotes a “Moore neighbor”. It is important to note that any active cell may link to more than four vertices (e.g., $C_{5}$). The sub-CEs corresponding to a vertex are formulated by linking the vertex, the center of a connected cell, and two segment centers, followed by temporal extrusion. For instance, to establish sub-CEs corresponding to $V_{1}^{\prime}$, the following steps are followed:

(1) Update the connectivity to determine if any connected cells have been impacted by the AMR process. Store the addresses of these connected cells in a specific order, such as the connected cells of $V_{1}: C_{5}$-$C_{1}$-$C_{13}$-$C_{14}$.

(2) Examine the first connected cell, $C_{5}$, which contains information about all connected vertices including $V_{10}$-$V_{11}$-$V_{3}$-$V_{2}$-$V_{1}$. This allows for the identification of the preceding and succeeding vertices to $V_{1}$, namely $V_{2}$ and $V_{10}$, respectively.

(3) Calculate the positions of the centers of the line segments $V_{2} V_{1}$ and $V_{1} V_{10}$, denoted as $E_{12}$ and $E_{22}$, respectively.

(4) Sequence the points $V_{1}, E_{22}, C_{5}$, and $E_{12}$ accordingly, then extrude $V_{1} E_{22} C_{5} E_{12}$ over time, resulting in the formation of a quadrilateral prism represented by $V_{1} E_{22} C_{5} E_{12}$-$V_{1}^{\prime} E_{22}^{\prime} C_{5}^{\prime} E_{12}^{\prime}$. This quadrilateral prism serves as a sub-CE corresponding to $V_{1}^{\prime}$ situated on the side of $C_{5}$.

(5) Repeat steps (2)-(4) for the sides associated with cells $C_{1}, C_{13}$, and $C_{14}$ to construct additional sub-CEs corresponding to $V_{1}^{\prime}$. These four sub-CEs collectively constitute the CE$\left(V_{1}^{\prime}\right)$. It is noteworthy that a vertex may be affiliated with an arbitrary number of sub-CEs, exemplified by the presence of 3 and 5 sub-CEs associated with vertices $V_{2}$ and $V_{5}$, respectively.

In a similar fashion, we can establish the sub-CEs corresponding to a specific cell center, as exemplified by the construction of sub-CEs corresponded to $C_{5}^{\prime \prime}$ :

(1) Update the connectivity and organize the addresses of these connected vertices in a specified sequence, for instance: $V_{3}$-$V_{2}$-$V_{1}$-$V_{10}$-$V_{11}$.

(2) For the vertex $V_{3}$, identify the preceding and succeeding vertices to $V_{3}$ within the list of connected vertices, which are $V_{11}$ and $V_{2}$, respectively.

(3) Determine the positions of the centers of the line segments $V_{11} V_{3}$ and $V_{3} V_{2}$, denoted as $E_{13}$ and $E_{5}$, respectively.

(4) Arrange the points $C_{5}, E_{13}, V_{3}$, and $E_{5}$ sequentially, then extrude $C_{5} E_{13} V_{3} E_{5}$ over time, resulting in the formation of a quadrilateral prism represented by $C_{5}^{\prime} E_{13}^{\prime} V_{3}^{\prime} E_{5}^{\prime}$-$C_{5}^{\prime \prime} E_{13}^{\prime \prime} V_{3}^{\prime \prime} E_{5}^{\prime \prime}$. This quadrilateral prism serves as a sub-CE corresponding to $C_{5}^{\prime \prime}$ on the side of $V_{3}$.

(5) Replicate steps (2)-(4) for the sides associated with $V_{2}, V_{1}, V_{10}$, and $V_{11}$ to construct additional sub-CEs corresponding to $C_{5}^{\prime \prime}$. These five sub-CEs collectively constitute CE$\left(C_{5}^{\prime \prime}\right)$. It is noteworthy that the sub-CE on the $V_{2}$ side forms a triangle prism, as the lines $E_{5} V_{2}$ and $V_{2} E_{12}$ are collinear. Nonetheless, it is regarded as a specialized type of quadrilateral prism.

\subsection{Refinement indicator and smoothing}
The refinement indicators $\xi$ play a crucial role in identifying the appropriate domain of interest for applying mesh adaptation techniques. The evaluation of these indicators, often based on the gradients of physical properties, has been a common practice in computational fluid dynamics research~\cite{khokhlov1998fully,colella1984piecewise}. In cases where a more comprehensive assessment is required, a combination of multiple indicators can be employed. In this study, we focus on a refinement indicator that is contingent upon significant gradients~\cite{schmidmayer2019adaptive}. Within the context of CESE schemes, each computational point retains spatial derivatives, allowing for the calculation of $\xi$ to be self-contained within the cell without relying on finite differencing involving neighboring cells. The formulation of $\xi$ is defined as:
\begin{equation}
\xi=\left\{\begin{array}{cc}
1 & \text { if } \left|\frac{(\Delta X)_{\max }}{X}\right|>\epsilon \\
0 & \text { else } \label{xiEq} 
\end{array}\right. ,
\end{equation}
where $X$ represents various physical quantities such as density, pressure, or velocity magnitude. In this investigation, we opt for a density-based indicator to facilitate the detection of shocks and contact surfaces. The term $(\Delta X)_{\max }$ denotes the maximum variation within the computational cell, while $\epsilon$ serves as an empirical threshold value.

\begin{figure}[t]
\centering
\includegraphics[width=11cm, trim=11cm 0cm 13cm 12cm, clip]{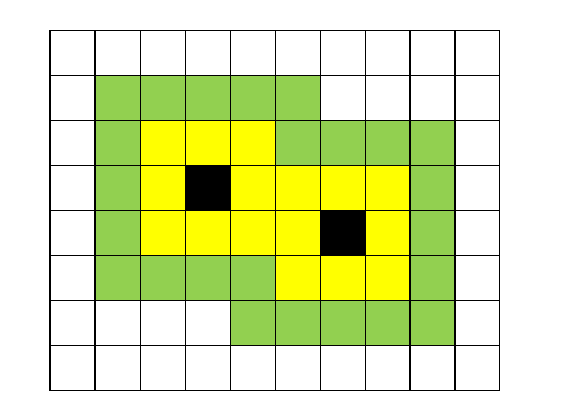}
\caption{Diagram illustrating the buffering process. The black, yellow, and green cells represent the original, first layer, and second layer, respectively, with $\xi$ set to 1. }\label{buffering}
\end{figure}

When the local variation exceeds this threshold, the indicator for the cell is set to 1. When considering an isolated shock, based on Eq. \eqref{xiEq}, the cells containing this shock exhibit significant variations and are prone to refinement. However, cells ahead of the shock may not be adequately refined, and refinement is postponed until the shock is encountered. This delay can result in the smearing of shock structures upon their arrival, potentially compromising the effectiveness of adaptive refinement strategies. To enhance the pre-refinement of cells proximal to discontinuities, a smoothing process for $\xi$ becomes imperative. An arithmetic averaging scheme is employed to update $\xi$ iteratively based on neighboring cells:
\begin{equation}
\xi^{j+1}=\frac{1}{n} \sum_{k=\mathrm{neighbor}} \xi_{k}^{j}\label{bufferEq}
\end{equation}
Here, $n$ represents the number of Moore neighbors linked to a cell, and $j$ denotes the iteration count for smoothing. Numerical experiments suggest that three smoothing iterations, or those equal to $\ell_{\text {max}}$, yield satisfactory results. Alternatively, Schmidmayer~\cite{schmidmayer2019adaptive} proposed the use of a diffusion equation to spread $\xi$ numerically. Additionally, prior to the application of Eq. \eqref{bufferEq} for smoothing, a modification is implemented to generate buffer layers (Fig.~\ref{buffering}) based on the maximum $\xi$ from neighboring cells:
\begin{equation}
\xi^{j+1}=\max \left(\xi^{j} \text { of neighbors }\right)
\end{equation}

This process can also be iterated multiple times, a practice that holds particular significance in ensuring thorough mesh refinement near the discontinuities, particularly in scenarios featuring intricate wave structures. As the number of iterations increases, the resulting fine mesh layer becomes progressively denser. While this approach enhances the ability to capture discontinuities more effectively, it does incur a modest increase in computational cell count. Our tests indicate that employing 1 to 2 layers yields satisfactory coverage of the regions of interest. The aforementioned buffering and smoothing procedures collectively serve to ensure comprehensive refinement near shocks and contact discontinuities.

\subsection{Procedures for refining and merging}
An intricate aspect of the AMR algorithm in staggered schemes on unstructured root meshes involves managing cell-vertex relationships during refine/merge operations. This complexity arises from the fact that any modification to a cell impacts not only itself and its connected vertices but also neighboring cells. The precomputation and smoothing of $\xi$ within each cell establish the fundamental criteria for refine/merge operations. The critical values, denoted as $\xi_{\text {split}}$ and $\xi_{\text {join}}$, serve as thresholds for determining these operations, and as observed in numerous AMR investigations, their optimal values may vary depending on the specific problem.

The process of splitting a quadrilateral cell into four child cells entails creating a vertex at the centroid of the cell and generating vertices at the centers of its edges if these vertices are not already present. Subsequently, the vertex at the cell centroid is connected to the centers of each edge of the cell to form the child cells. It is crucial to note that when a new edge center is established, it must be appropriately linked to the neighboring cell to ensure mutual referencing between vertices and cells, thereby defining the sub-CEs. This connectivity is facilitated by the cell-tree-vertex data structure, enabling the seamless insertion of a new vertex's address into the vertex list of the neighboring cell. Meanwhile, the conservative variables of the child cells are inherited from their parent cell.

When merging four child cells into their parent cell, the deletion of the four child cells and the centroid vertex is a standard operation. Additionally, any hanging vertices (i.e., vertices that do not belong to any of the cells' corners) on the edges of the parent cell are removed if present. The conserved variables of the parent cell are then determined by averaging over the children in space, as indicated in Eq. \eqref{avgUEq}: $\mathrm{U}=\sum_{k=\text {child}} \mathrm{U}_{k} \cdot$area$_{k} /$area. Subsequently, the derivatives of the parent cell are computed using the upwind scheme equations with the limited average procedure, as outlined in Sec.~\ref{upwindScheme}.

Furthermore, all cells subject to split/merge operations, along with their neighboring cells, are designated as affected. This marking ensures that unnecessary operations are avoided for cells not involved in the ongoing processes. It is imperative to revise the definitions of conservation elements and solution elements following mesh operations to accurately reflect the altered cell configurations.

\begin{figure}[t]
\centering
\includegraphics[width=15cm, trim=1cm 0cm 3cm 5cm, clip]{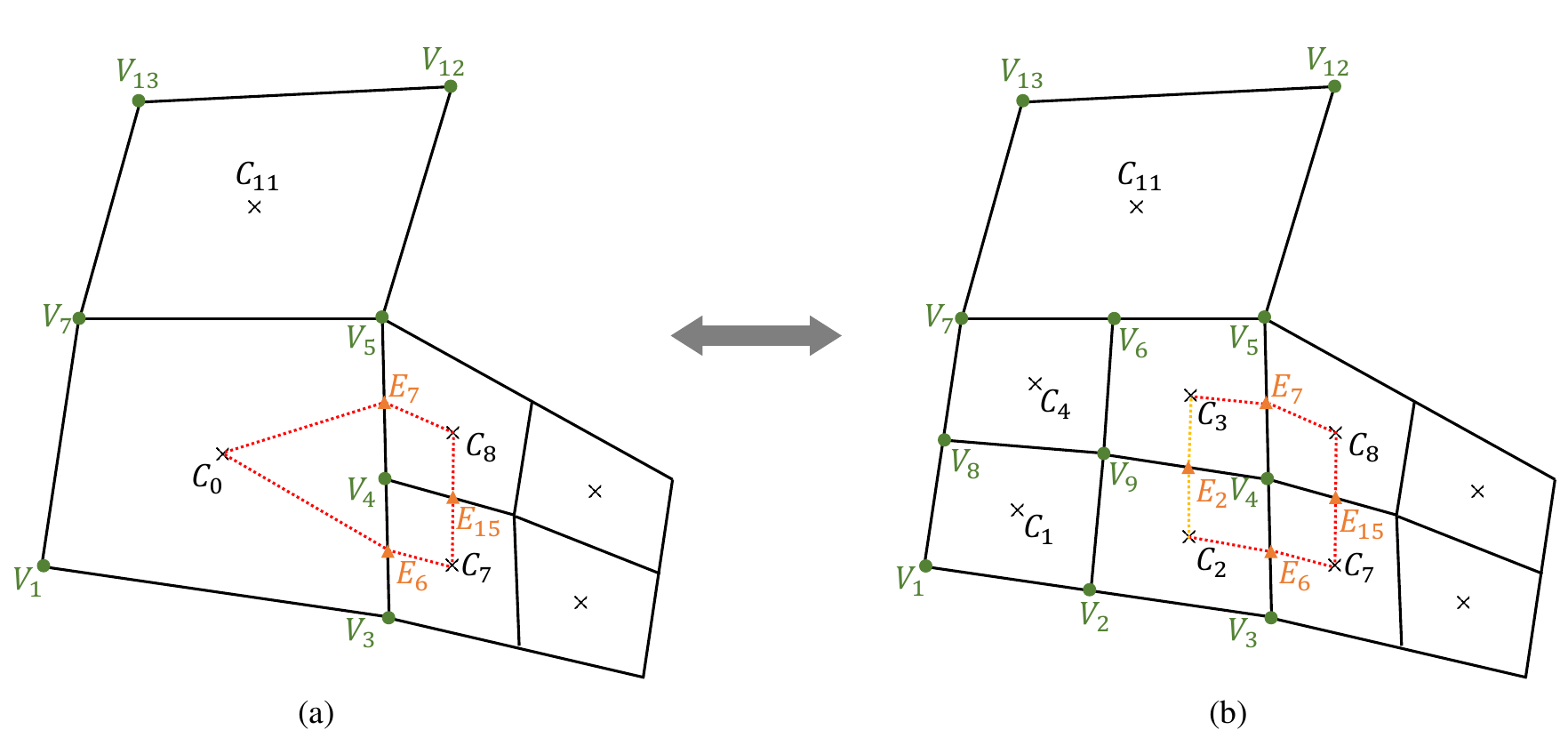}
\caption{Illustration of cell splitting and merging. The projection of CE$\left(V_{4}\right)$ on the $x$-$y$ plane changes before and after mesh refinement, as indicated by the dashed lines.}\label{splitMerge}
\end{figure}

As depicted in Fig.~\ref{splitMerge}, during the split procedure, vertices $V_{9}, V_{2}, V_{6}$, and $V_{8}$ are created and added to the vertex list. Since $V_{6}$ is newly created, it is connected to $C_{11}$, and similar connections are established for $V_{2}$ and $V_{8}$. Cells $C_{1}, C_{2}, C_{3}$, and $C_{4}$ are created and added to the cell list, and they are linked to the vertices at their parent cell's centroid, corners, and edge centers. Physical variables are assigned to these child cells based on those of the parent cell. Subsequently, the linkage between the parent cell and all the vertices is disconnected. Conservation elements and sub-CEs of affected points are redefined to accommodate the changes resulting from the splitting operation, such as the projection of CE$\left(V_{4}\right)$ on $x$-$y$ plane change from $E_{6} C_{7} E_{15} C_{8} E_{7} C_{0}$ to $E_{6} C_{7} E_{15} C_{8} E_{7} C_{3} E_{2} C_{2}$. During the merging procedure, the averaged conservative variables and spatial derivatives in the parent cell are computed. The linkage between child cells $C_{1}, C_{2}, C_{3}, C_{4}$ and the vertices is disconnected, and the hanging nodes (vertices $V_{9}, V_{2}, V_{6}$, and $V_{8}$ ) are deleted. Finally, the parent cell is reconnected with the corresponding vertices, and conservation elements and sub-CEs are redefined to reflect the updated configuration.

\begin{figure}[t]
\centering
\includegraphics[width=16cm, trim=4.1cm 0cm 3cm 5cm, clip]{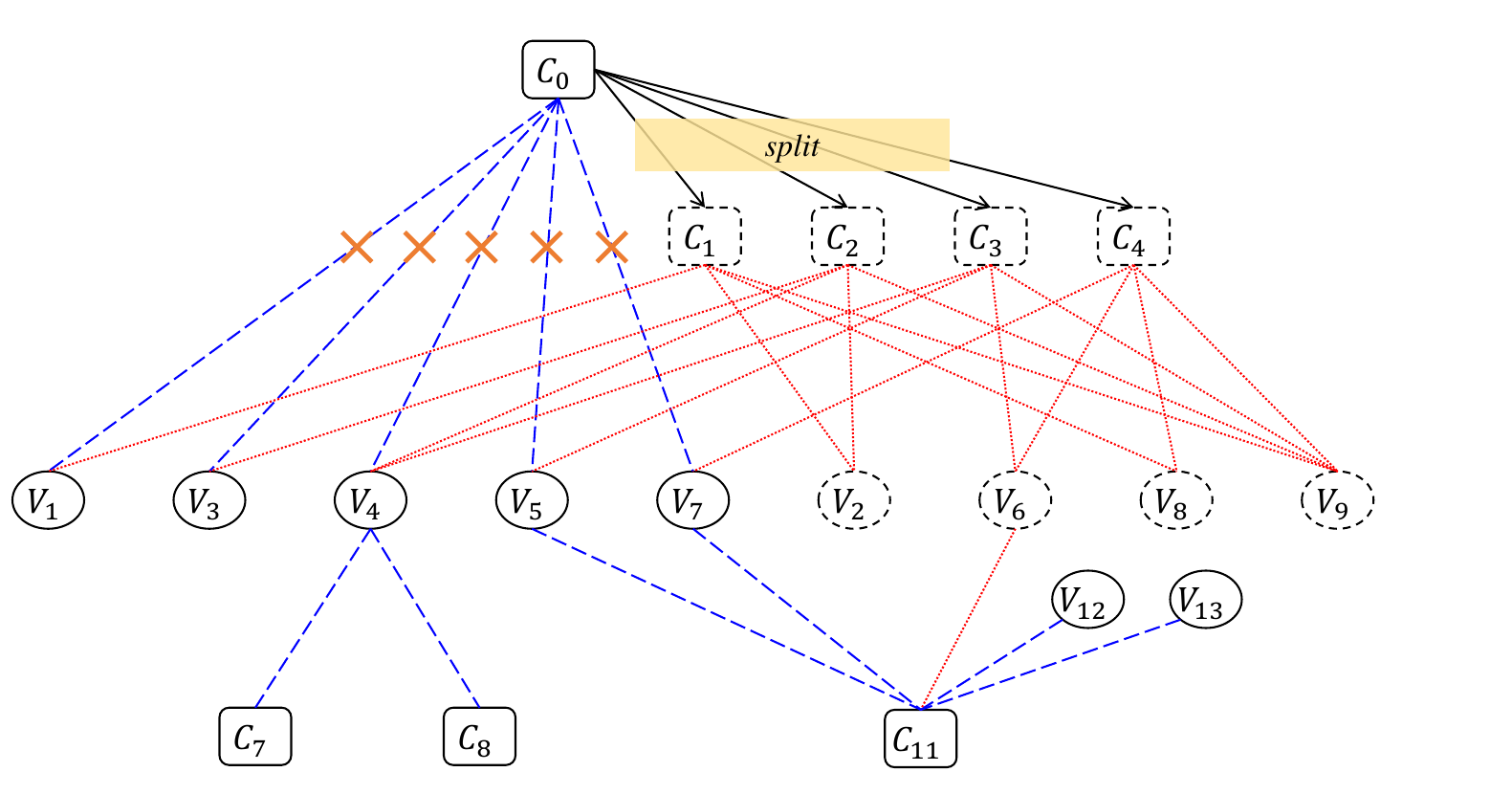}
\caption{Simplified cell-tree-vertex data structure for a cell-vertex staggered scheme. Blue dashed lines represent cell-vertex linkages pre-refinement, while red dotted lines indicate new linkages postrefinement. Black arrows show a parent cell splitting into four child cells, and orange crosses signify broken linkages after refinement. }\label{dataStructure}
\end{figure}

Without additional data for physical variables, miscellaneous, derived linkages among cells and vertices, etc., we will now analyze the simplified connectivity diagram (Fig.~\ref{dataStructure}) in the context of mesh refinement (Fig.~\ref{splitMerge}b). Upon the division of cell $C_{0}$, four new vertices $V_{2,6,8,9}$ are generated, leading to the creation of four distinct child cells $C_{1,2,3,4}$ by establishing connections with the corresponding vertices. Subsequently, the physical attributes of the child cells are interpolated. Notably, it is imperative to remove the linkage between $C_{0}$ and the vertices to avoid potential errors during subsequent operations such as conservation element construction and neighbor identification, as these vertices may still be associated with $C_{0}$. Furthermore, in the event that a vertex, such as $V_{6}$, is generated along an edge, it should also be linked to $C_{11}$. The process of merging is essentially the reverse operation, and the detailed elaboration of this procedure is omitted.

To facilitate implementation, the overall AMR procedure is divided into refining and merging components. Both segments encompass multiple iterations for manipulation at all levels. It is crucial to enforce a constraint where no more than one level difference is permitted between any two neighboring cells during any operation. If a refinement or merging operation result in a larger level discrepancy, this operation must be abandoned. The strategy for AMR can be succinctly summarized as follows:

(1) Compute the refinement indicator $\xi$ for each active cell based on the stored derivatives within the cell.

(2) Perform buffering and smoothing operation on $\xi$ for each cell.

(3) Execute the refinement loops from the root level to the maximum level. During each iteration, assess if $\xi>\xi_{\text {split}}$ and if $\left|\ell_{\text {refined}}-\ell_{\forall \text {neighbor}}\right| \leq 1$. If these conditions are met, the cell may undergo splitting. Simultaneously, incorporate all new vertices into the vertex list. Notably, when a new vertex is generated on a cell edge, it should also be linked to the neighboring cell.

(4) Conduct the merging loops from level $\ell_{\max}-1$ to the root level. For each iteration, verify if all four child cells satisfy the criteria $\xi<\xi_{\text {join}}$ and ascertain if the estimated $\xi$ for the parent cell also meets the merging criteria, along with $\left|\ell_{\text {merged}}-\ell_{\forall \text {neighbor}}\right|\leq1$. If these conditions are fulfilled, the child cells may be merged.

(5) Remove any hanging vertices and associated information from the relevant cells.

(6) Update conservation element and sub-CE information for all affected cells and vertices.

\subsection{Flowchart for CESE with AMR}
\begin{figure}[t]
\centering
\includegraphics[width=9cm, trim=10cm 0cm 10cm 0cm, clip]{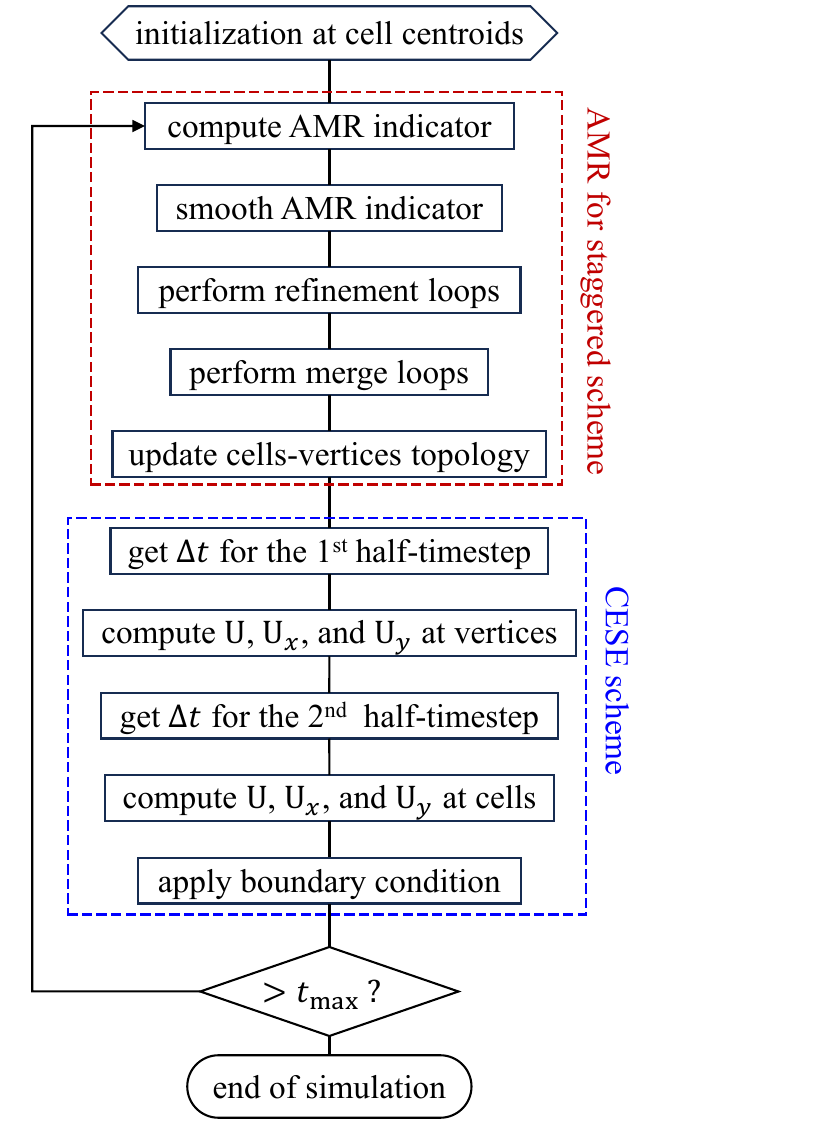}
\caption{Flowchart outlining CESE schemes integrated with AMR for staggered schemes.}\label{flowchart}
\end{figure}

In the preceding sections, a detailed explanation has been provided regarding the constructions of CESE schemes to accommodate split quadrilateral meshes. Additionally, a novel AMR strategy tailored for staggered schemes, relying on cell-tree-vertex structures, has been introduced. The integration of these components into the current comprehensive algorithm is visually depicted in Fig.~\ref{flowchart}.
It is important to highlight that all AMR operations are carried out exclusively on cells before the commencement of flow integration. As described in Sec.~\ref{sec2}, the core computations are organized based on conservation elements. At each vertex, where physical variables are updated in the first half-step, and at each cell center during the second half-step, sub-CEs are constructed accordingly, facilitating the evaluation of fluxes.
An additional notable feature of the current algorithm is its capacity to handle cells of varying levels consistently when advancing the variables. Once the AMR component is completed, the data structure exerts minimal influence on the complexity of the update in the CESE scheme. This refined AMR algorithm and associated data structure effectively support the seamless implementation of all three CESE schemes.

\section{Numerical examples}
\label{sec4}
In this section, simulations are conducted to analyze both steady and unsteady flows using either Cartesian or unstructured root meshes. Notably, for problems initially defined on a Cartesian mesh, a consistent unstructured solver is employed for computation. The Cartesian meshes are generated by the in-house code, while the generation of unstructured root meshes is facilitated using Gmsh~\cite{geuzaine2009gmsh}. The entire code implementation is realized in \texttt{C++}. The density gradient is selected as the sole parameter for computing the refinement indicator, denoted as $X$ in Eq. \eqref{xiEq}. Mesh refinement parameters $\left(\epsilon, \xi_{\text {split}}, \xi_{\text {join}}\right)$ are problem-specific and are adjusted to ensure the discontinuities are resolved by the finest cells. The mesh adaptation algorithm is executed in each time-step. In these simulations, an ideal gas with a specific heat ratio of $\gamma=1.4$ and a Courant number of 0.8 are considered. The rotated HLLC method is chosen for computing the flux across inner surfaces in the upwind CESE scheme.

Subsequently, the Sod shock problem is utilized to evaluate the response at different refinement levels. The regular shock reflection problem is selected as an example for steady flows. Furthermore, the 2D Riemann problem is utilized to compare the performance of various CESE schemes. Additionally, the shock over wedge problem and double Mach reflection are investigated to evaluate performance for transient flows on adapted unstructured meshes. Finally, a supersonic flow over cylinder problem is examined to demonstrate the capability of the present algorithms in addressing complex domain structures.

\subsection{Sod shock problem}
\begin{figure}[t]
\centering
\includegraphics[width=10cm, trim=1cm 0cm 1cm 1cm, clip]{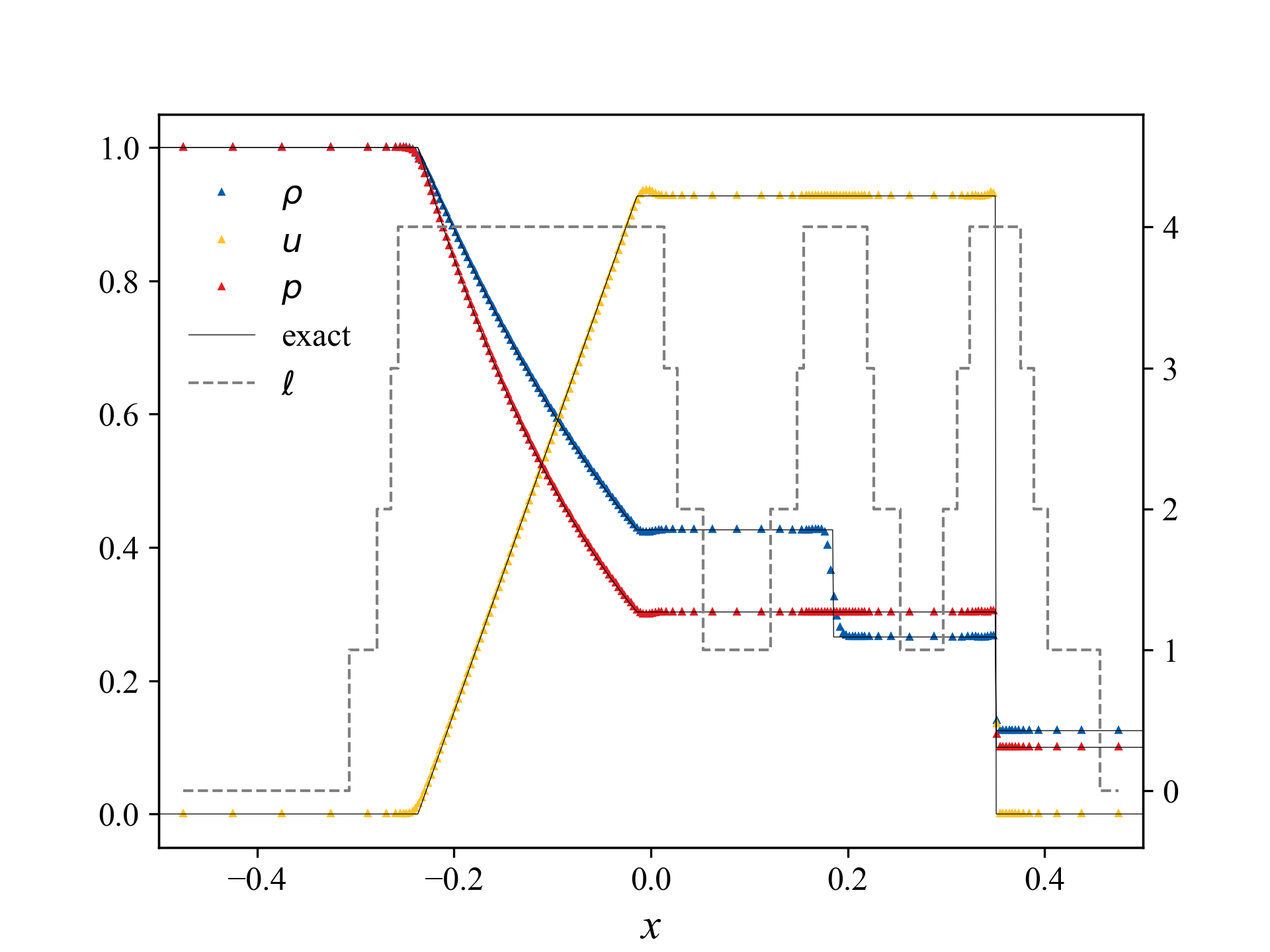}
\caption{Sod shock problem solved utilizing the CNI scheme with $\Delta x=\frac{1}{20}$ and $\ell_{\max }=4$ at $t=0.2$. The exact solutions are depicted by solid black lines, simulation results are shown as dots at the centers of computational cells, and dashed lines indicate cell refinement levels.}\label{sodAllVar}
\end{figure}

\begin{figure}[H]
\centering
\includegraphics[width=10cm, trim=1cm 0cm 1cm 1cm, clip]{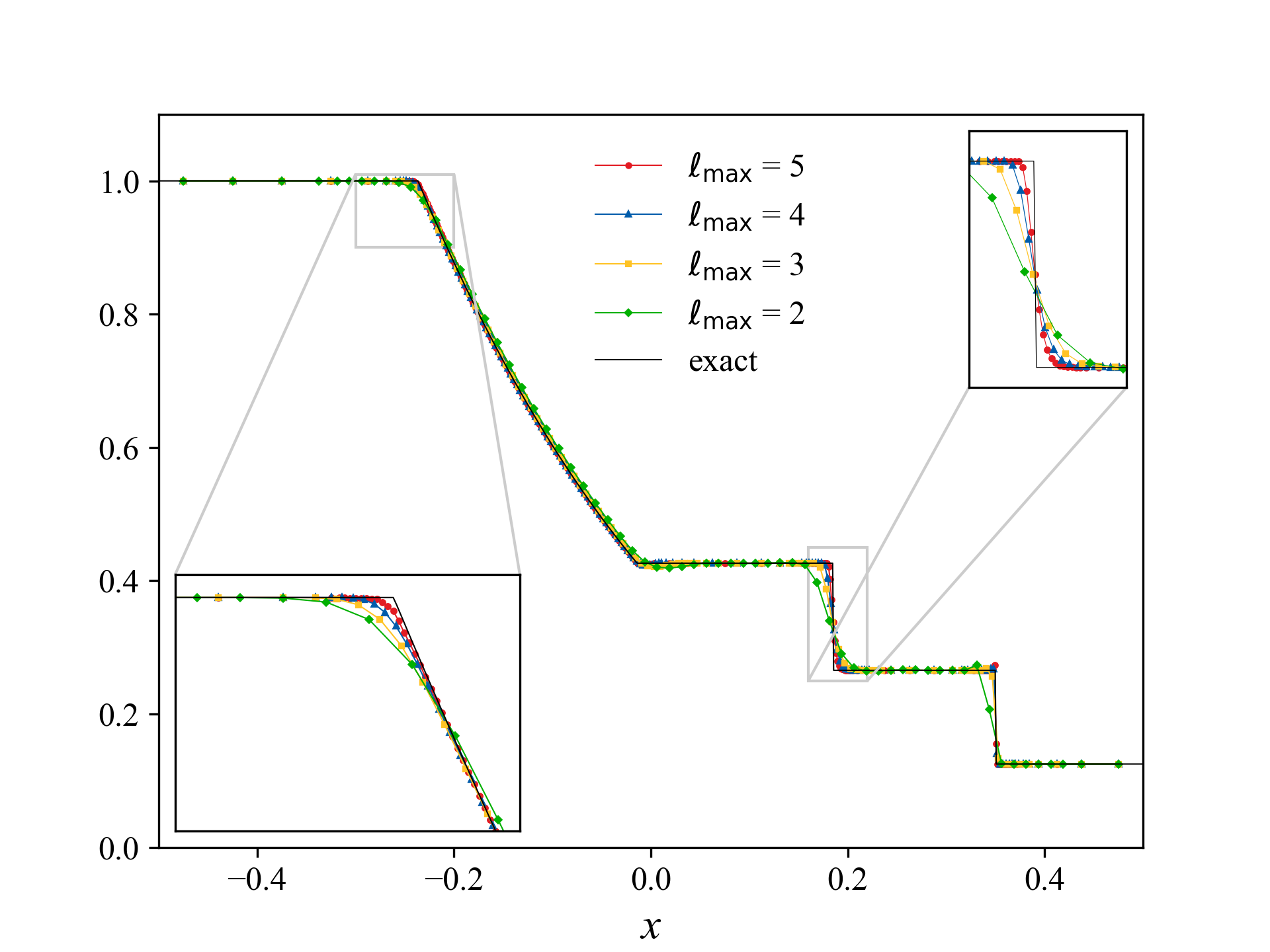}
\caption{Density profile of the Sod shock problem employing the CNI scheme with $\ell_{\max }=2 \sim 5$.}\label{sodLevel}
\end{figure}

The Sod shock tube problem~\cite{sod1978survey} is solved by the present 2D algorithm, and the resulting profiles are interpolated along the centerline of the computational domain $[-0.5,0.5] \times[-0.1,0.1]$, utilizing a root Cartesian mesh number of $20 \times 4$. To validate the algorithm's robustness, the problem is also assessed on a rotated domain to ensure the independence of outcomes from coordinate rotations. Initial values of $(\rho, u, v, p)$ are set to $(1,0,0,1)$ on the left half-side and $(0.125,0,0,0.1)$ on the right half-side. The CNI CESE scheme is employed for this analysis. The investigation focuses on a case with a maximum refinement level of $l_{\max }=4$, as depicted in Fig.~\ref{sodAllVar}. Notably, while the mesh level remains unaltered in regions without disturbances, automatic refinement occurs near shocks, discontinuities, and rarefaction waves. The numerical results exhibit close agreement with the exact solution, mirroring outcomes obtained using a uniform mesh with $\Delta x=\frac{1}{320}$, which is excluded from the plots for clarity. Figure~\ref{sodLevel} illustrates the results at varying maximum refinement levels, highlighting that increasing the level enhances the proximity of simulated results to the exact solution. This test underscores the algorithm's accuracy in capturing essential wave structures like shocks and contact discontinuities.

\subsection{Regular shock reflection}
\begin{figure}[!htbp]
\centering

\begin{subfigure}{\textwidth}
  \centering
  \includegraphics[width=14.5cm, trim=6cm 0cm 6cm 0.1cm, clip]{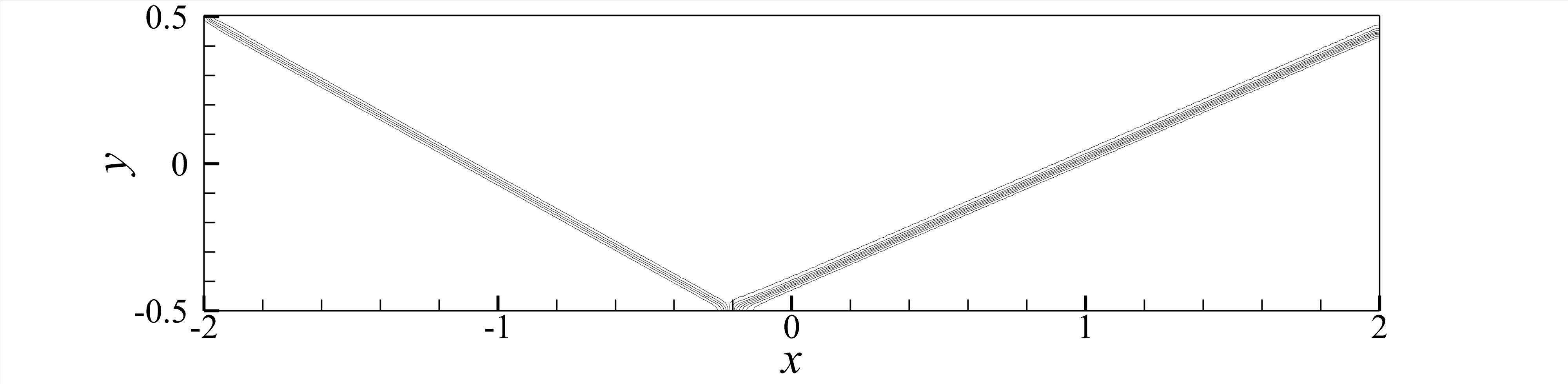}
  \caption{Cartesian root meshes, with 15 equally spaced density contours ranging from 1.4 to 3.8.}
\end{subfigure}

\begin{subfigure}{\textwidth}
  \centering
  \includegraphics[width=14.5cm, trim=6cm 0cm 6cm 0.1cm, clip]{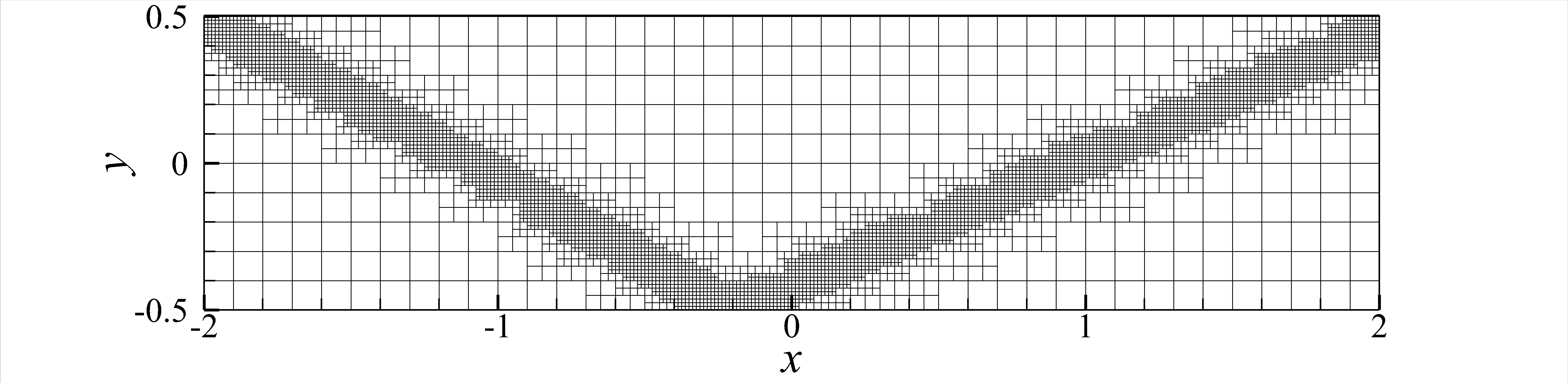}
  \caption{Cartesian root meshes and the adapted meshes.}
\end{subfigure}

\begin{subfigure}{\textwidth}
  \centering
  \includegraphics[width=14.5cm, trim=6cm 0cm 6cm 0.1cm, clip]{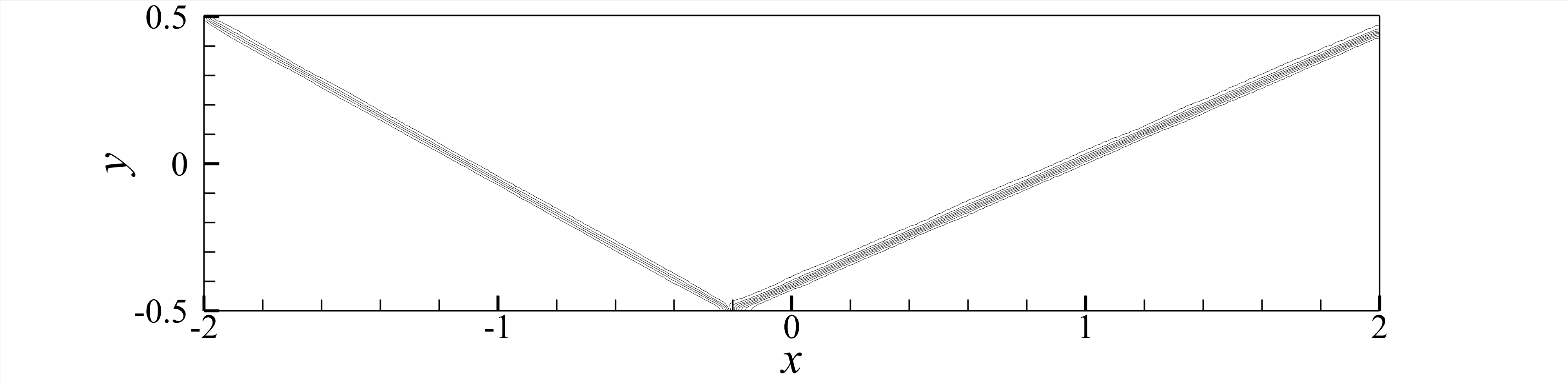}
  \caption{Perturbed root meshes, with 15 equally spaced density contours ranging from 1.4 to 3.8.}
\end{subfigure}

\begin{subfigure}{\textwidth}
  \centering
  \includegraphics[width=14.5cm, trim=6cm 0cm 6cm 0.1cm, clip]{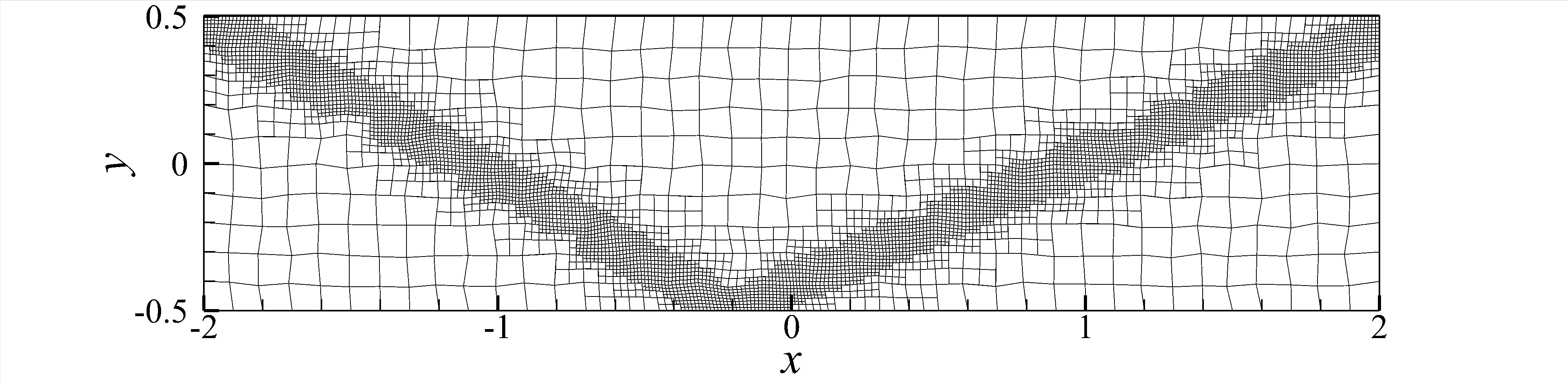}
  \caption{Perturbed root meshes and the adapted meshes.}
\end{subfigure}

\caption{Regular reflection problem solved with the CNI scheme on (a, b) Cartesian or (c, d) perturbed root meshes, of number $40 \times 10$ with $\ell_{\max }=3$.}
\label{regRef}
\end{figure}

\begin{figure}[H]
\centering
\includegraphics[width=11cm, trim=2cm 1cm 2cm 1cm, clip]{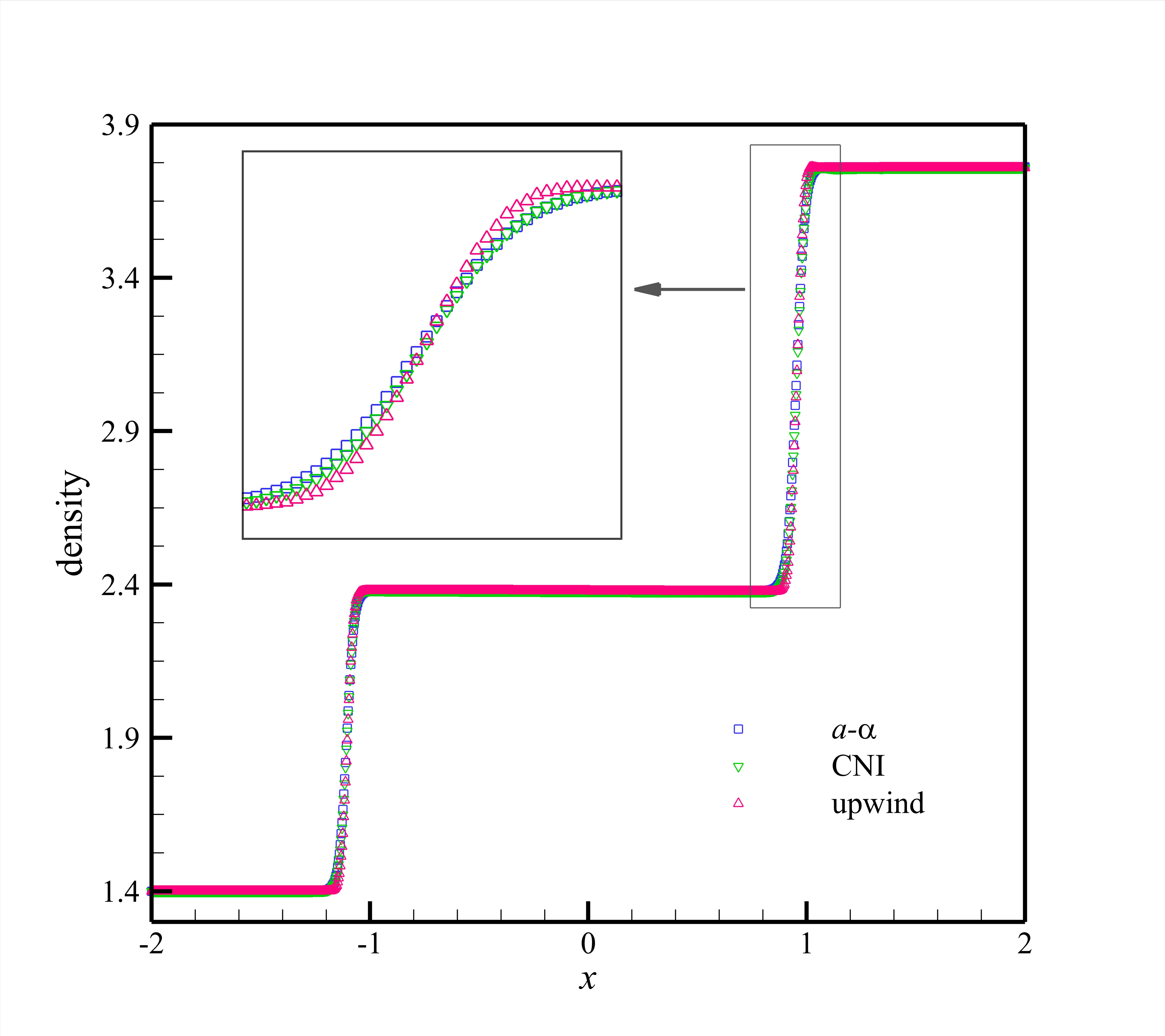}
\caption{Densities interpolated along $y=0$ for the three different CESE schemes. Uniform root meshes of $40 \times 10$ with $\ell_{\max }=3$.}
\label{regRefSchemes}
\end{figure}

The regular shock reflection problem serves as a typical illustration of steady flow phenomena~\cite{sidilkover2018towards,zhang2007new}. The physical domain spans $[-0.5,0.5] \times[-2,2]$, with a root Cartesian mesh number of $40 \times 10$ and a maximum refinement level of 3. In order the examine whether the results are sensitive to the mesh adaptation originated from different root meshes (uniform or skewed meshes), a test with perturbed meshes is also conducted with a perturbation amplitude of $10 \%$ (Fig.~\ref{regRef}d). Inlet boundary conditions are prescribed along the left and top boundaries, with $(\rho, u, v, p)_{\text {left }}$ as $(1.4,2.9,0,1)$ and $(\rho, u, v, p)_{\text {top }}$ as $(2.38,2.6193$, $0.50632,2.13948)$. The lower boundary is reflective, while the right boundary acts as a supersonic outlet. The simulation proceeds until a state of steady flow is achieved. In Fig.~\ref{regRef}, density contours computed using the CNI scheme and adapted meshes for both Cartesian and perturbed scenarios reveal precise shock detection through mesh refinement. Despite mesh skewing in the perturbed case, no discernible difference is evident in the density contour. Figure~\ref{regRefSchemes} compares results from different schemes by examining the density distribution along the centerline. Results from all schemes closely align, with the upwind CESE scheme displaying a slightly sharper density profile near shocks owing to its lower numerical dissipation~\cite{shen2018positivity}. These tests also showcase that although the $a$-$\alpha$ scheme is Courant number-sensitive, it still yields accurate results. Although excessive numerical dissipation arising from small Courant numbers in the $a$-$\alpha$ scheme can notably impact outcomes near shocks and discontinuities where local gradient plays a crucial role. In AMR, cells with small Courant numbers exist in locally smooth areas and appear less affected by Courant numbers when appropriately refined.

\FloatBarrier

\subsection{2D Riemann problem}
This problem serves to assess the algorithm's ability to dynamically adapt meshes during unsteady computation, commonly referred to as ``on flight'' adaptation. The domain is defined as $[-1,1] \times[-1,1]$, with initial root meshes of $34 \times 34$. Zero-gradient boundary conditions are enforced at all boundaries. Initially, the gas parameters within each quadrant are uniform: $(\rho, u, v, p)_{\mathrm{LD}}=(0.138,1.206,1.206,0.029), (\rho, u, v, p)_{\mathrm{RD}}=(0.5323,0,1.206,0.3)$, $(\rho, u, v, p)_{\mathrm{RU}}=(1.5,0,0,1.5)$, and $(\rho, u, v, p)_{\mathrm{LU}}=(0.0 .5223,1.206,0,0.3)$. The simulation is conducted until reaching a time of $t=1.1$.

\begin{figure}[H]
\centering
\begin{subfigure}{0.47\textwidth}
  \centering
  \includegraphics[width=0.9\linewidth, trim=0cm 0cm 0cm 1cm, clip]{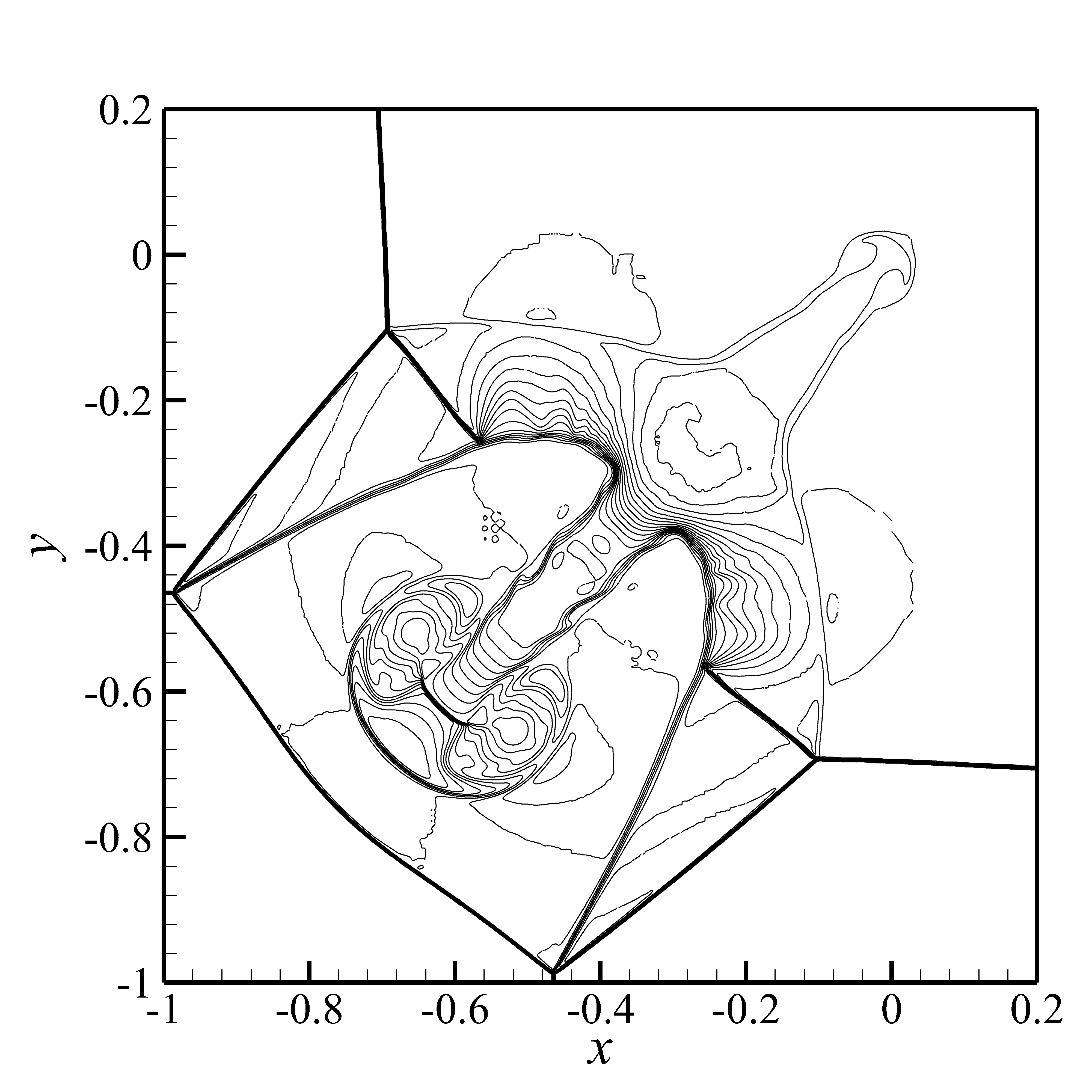}
  \caption{$a$-$\alpha$ CESE, AMR}
\end{subfigure}
\begin{subfigure}{0.47\textwidth}
  \includegraphics[width=0.9\linewidth, trim=0cm 0cm 0cm 1cm, clip]{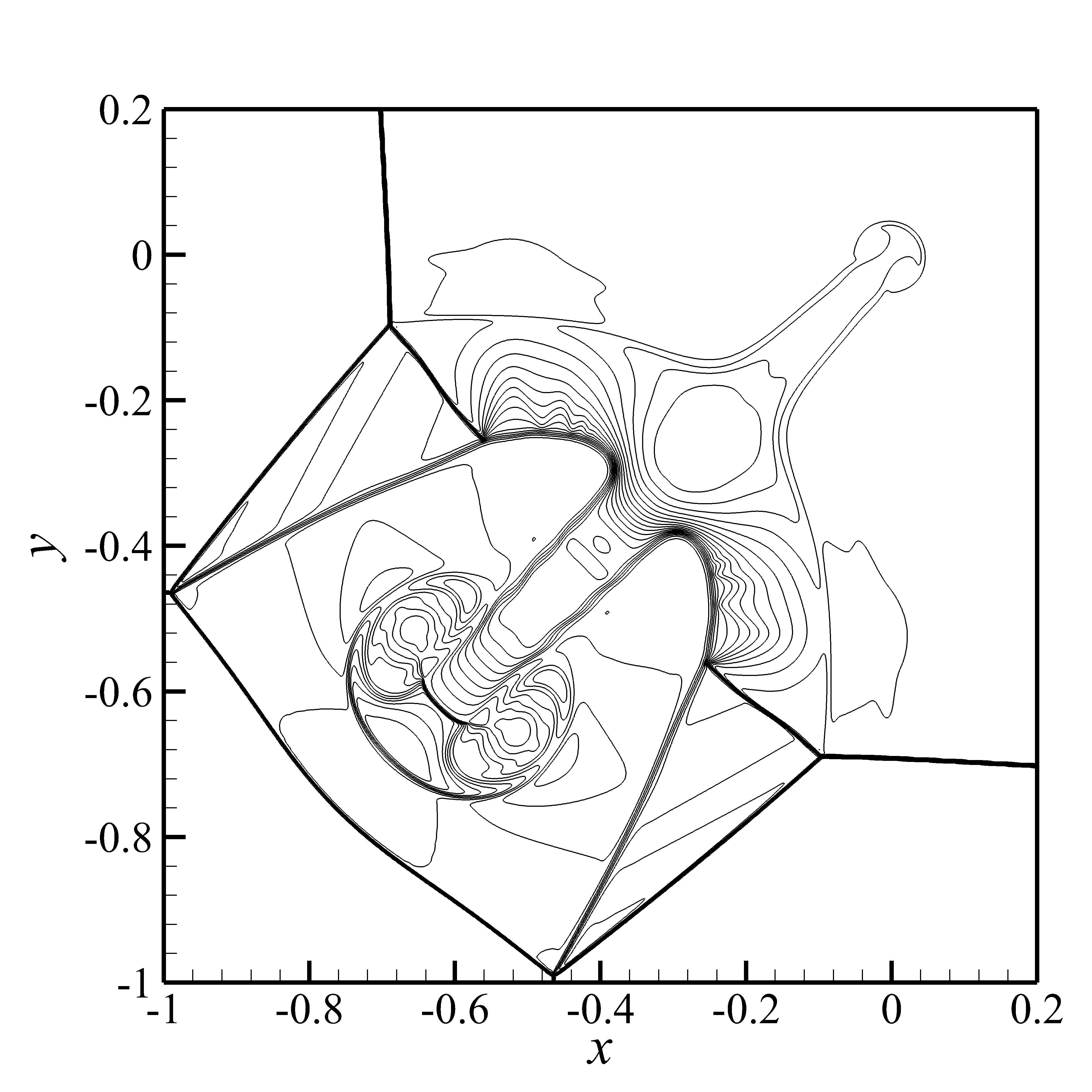}
  \caption{$a$-$\alpha$ CESE, uniform}
\end{subfigure}
\begin{subfigure}{0.47\textwidth}
  \centering
  \includegraphics[width=0.9\linewidth, trim=0cm 0cm 0cm 1cm, clip]{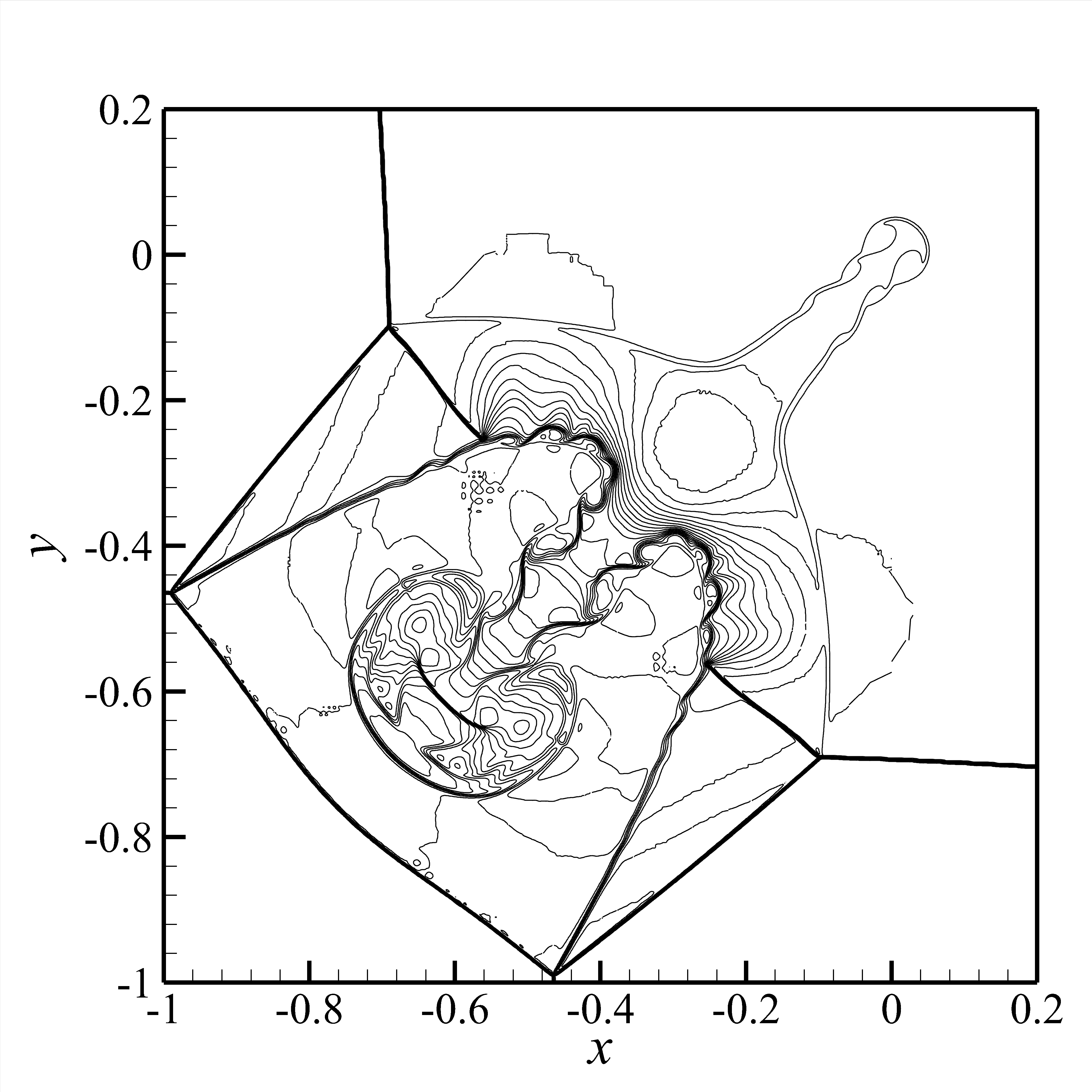}
  \caption{CNI CESE, AMR}
\end{subfigure}
\begin{subfigure}{0.47\textwidth}
  \centering
  \includegraphics[width=0.9\linewidth, trim=0cm 0cm 0cm 1cm, clip]{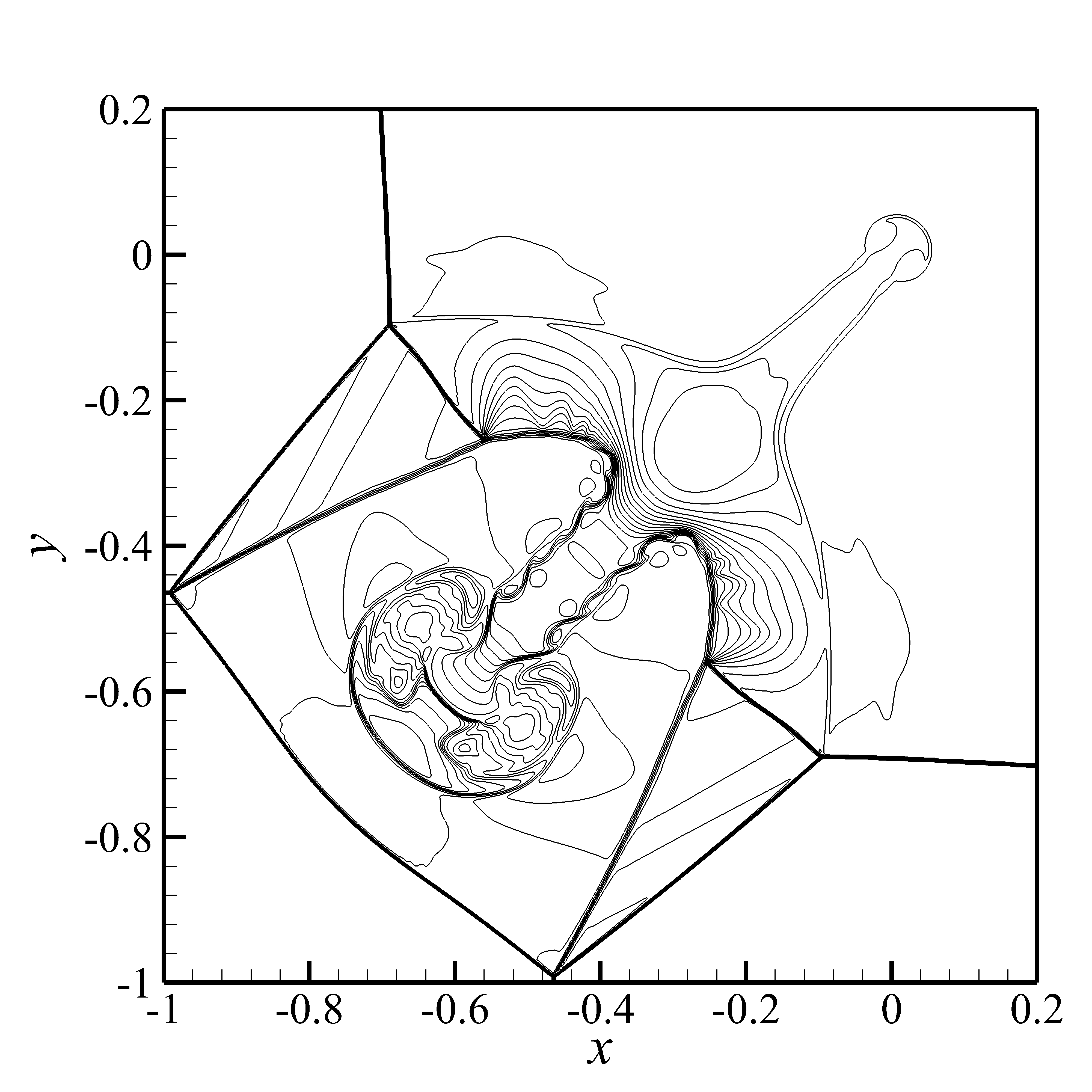}
  \caption{CNI CESE, uniform}
\end{subfigure}
\begin{subfigure}{0.47\textwidth}
  \centering
  \includegraphics[width=0.9\linewidth, trim=0cm 0cm 0cm 1cm, clip]{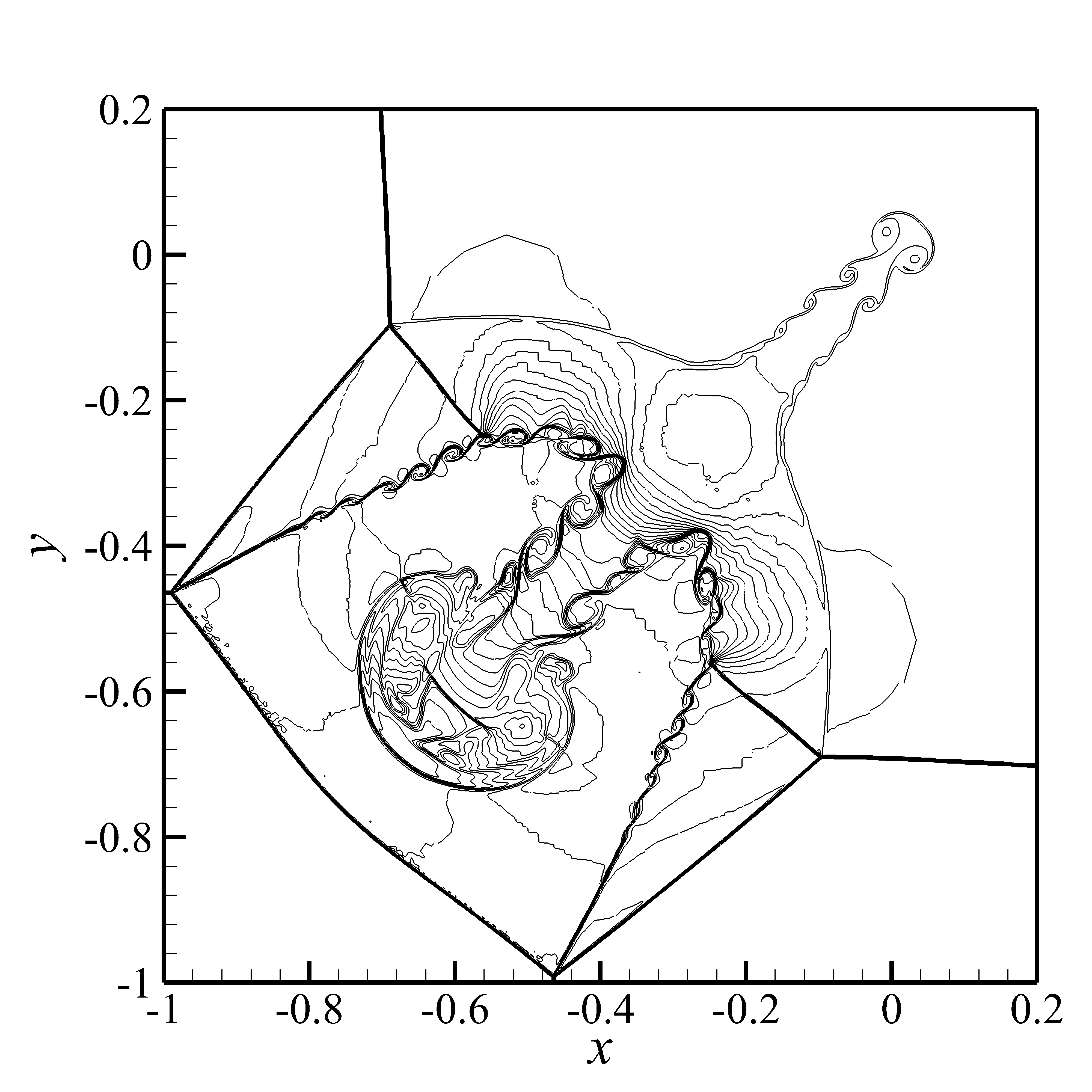}
  \caption{HLLC CESE, AMR}
\end{subfigure}
\begin{subfigure}{0.47\textwidth}
  \centering
  \includegraphics[width=0.9\linewidth, trim=0cm 0cm 0cm 1cm, clip]{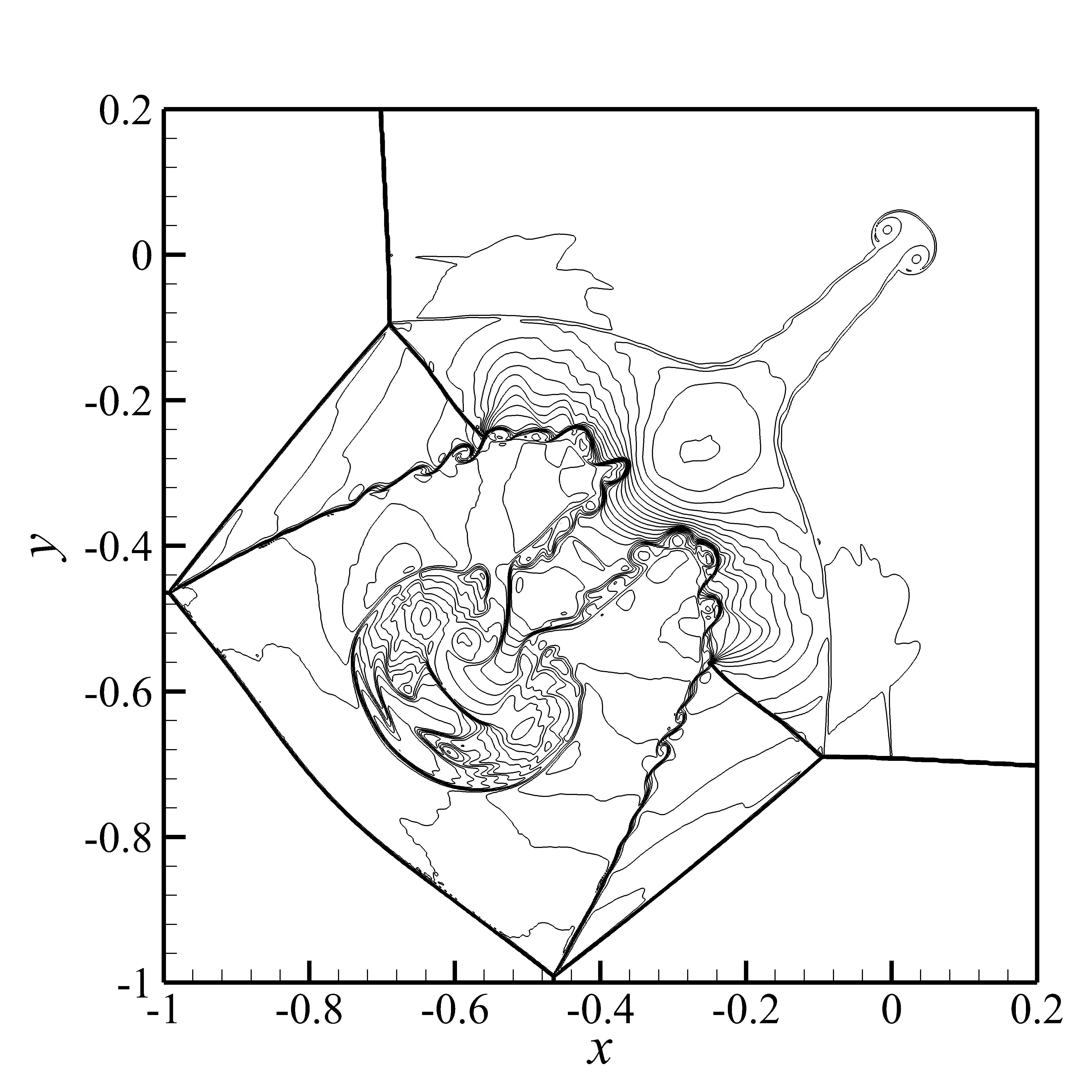}
  \caption{HLLC CESE, uniform}
\end{subfigure}
\caption{2D Riemann problems computed by different CESE schemes with root meshes $34 \times 34$ and $l_{\max}=5$ or uniform meshes $1088 \times 1088$. Thirty equally spaced density contours from 0.1 to 1.8 are shown.}
\label{2DRimann}
\end{figure}

\begin{figure}[H]
  \centering
  \includegraphics[width=0.45\linewidth, trim=0cm 0cm 0cm 1cm, clip]{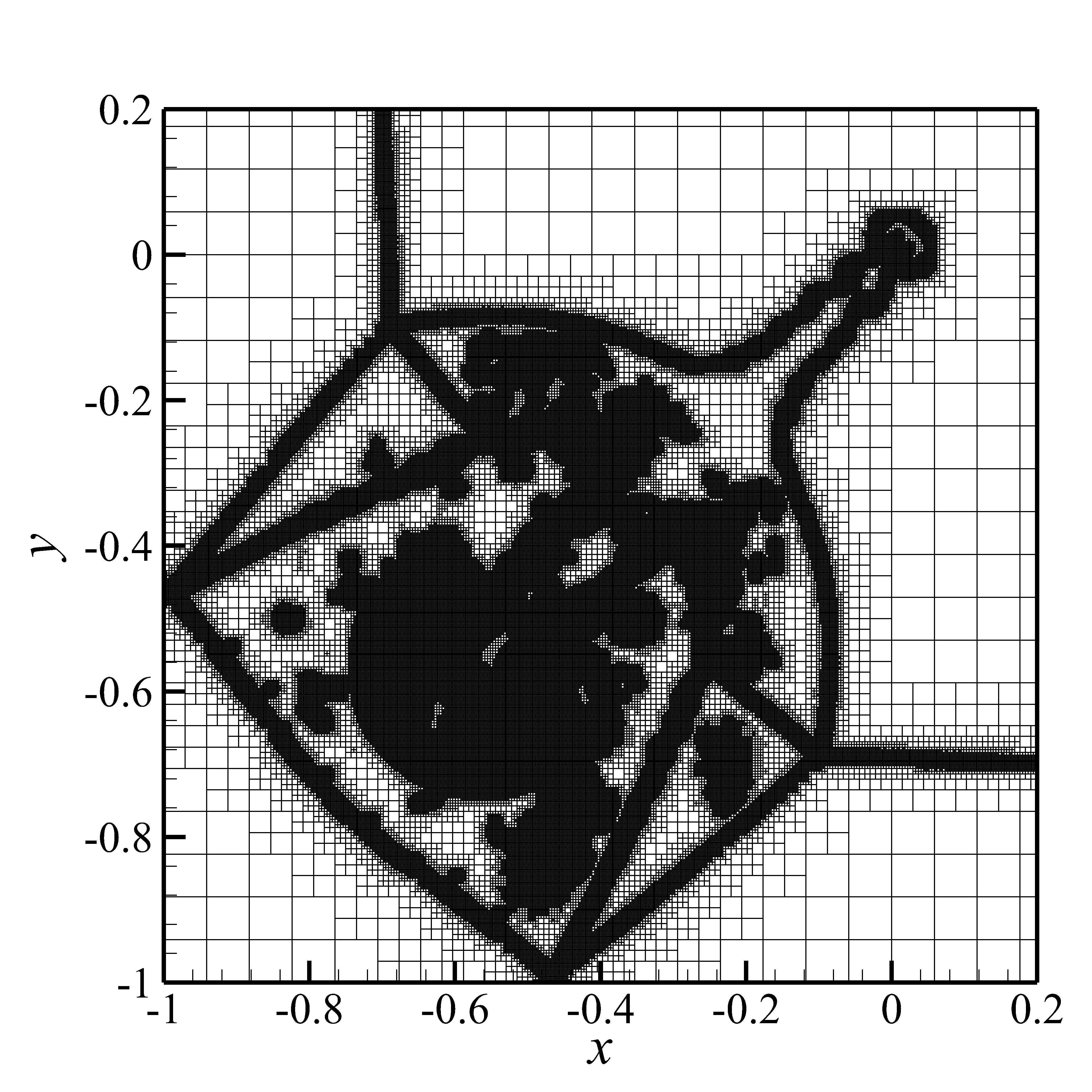}
  \caption{Adapted meshes for HLLC CESE.}
\label{2DRimannMesh}
\end{figure}

The obtained results, as depicted in Fig.~\ref{2DRimann}, showcase the outcomes of three schemes, along with the corresponding meshes when employing the upwind CESE scheme (Fig.~\ref{2DRimannMesh}). The results achieved through AMR closely capture essential flow structures compared to simulations utilizing uniform meshes, and are similar to simulations with similar resolutions as referenced in Shen et al.~\cite{shen2016characteristic}, indicating the favorable effectiveness of the current algorithm. Furthermore, the CNI and upwind CESE schemes demonstrate superior resolution, as they exhibit more noticeable instabilities. To maintain clarity, subsequent investigations will focus specifically on the upwind CESE scheme.

\subsection{Shock wave over wedge}
\begin{figure}[H]
\centering
\includegraphics[width=11.5cm,trim=1cm 10cm 1cm 57cm, clip]{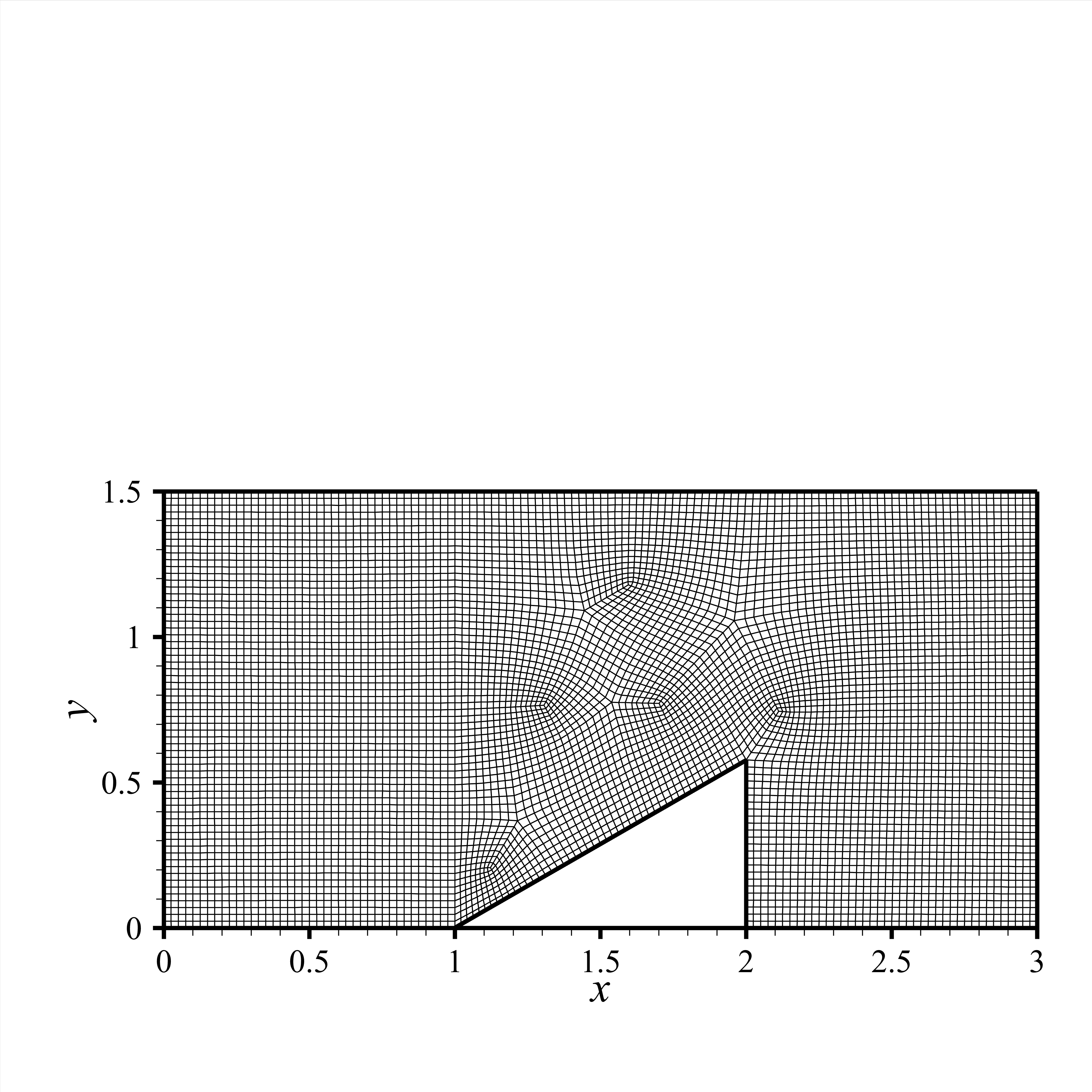}
\caption{Unstructured quadrilateral root meshes employed for shock passing a finite wedge.}\label{wedgeMesh}
\end{figure}

\begin{figure}[H]
\centering
\begin{subfigure}{0.47\textwidth}
  \centering
  \includegraphics[width=0.9\linewidth,trim=0cm 10cm 0cm 0.5cm, clip]{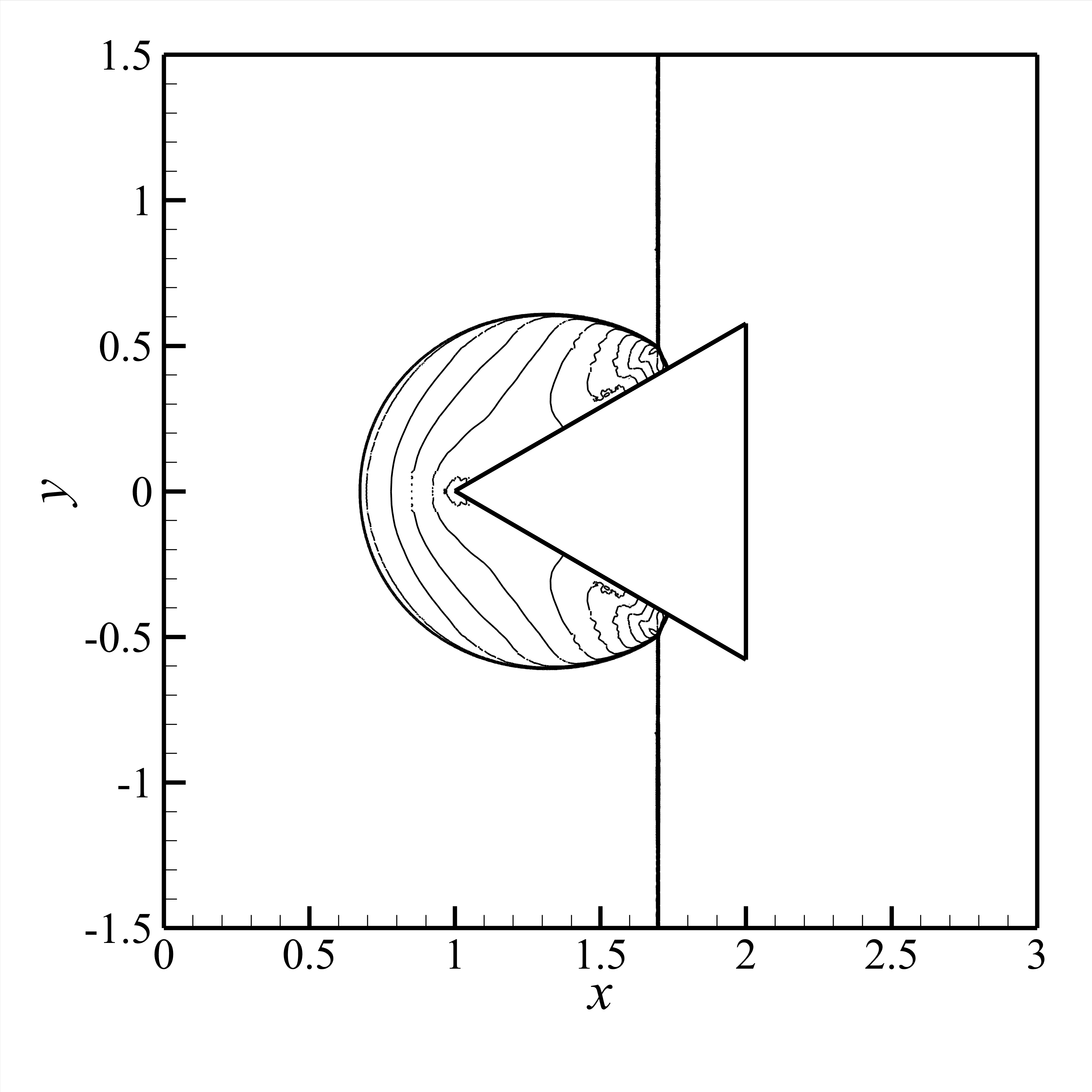}
  \caption{$t=0.521$, 30 equally spaced density contours from 1.39 to 2.88.}
\end{subfigure}
\begin{subfigure}{0.47\textwidth}
  \centering
  \includegraphics[width=0.9\linewidth,trim=0cm 10cm 0cm 0.5cm, clip]{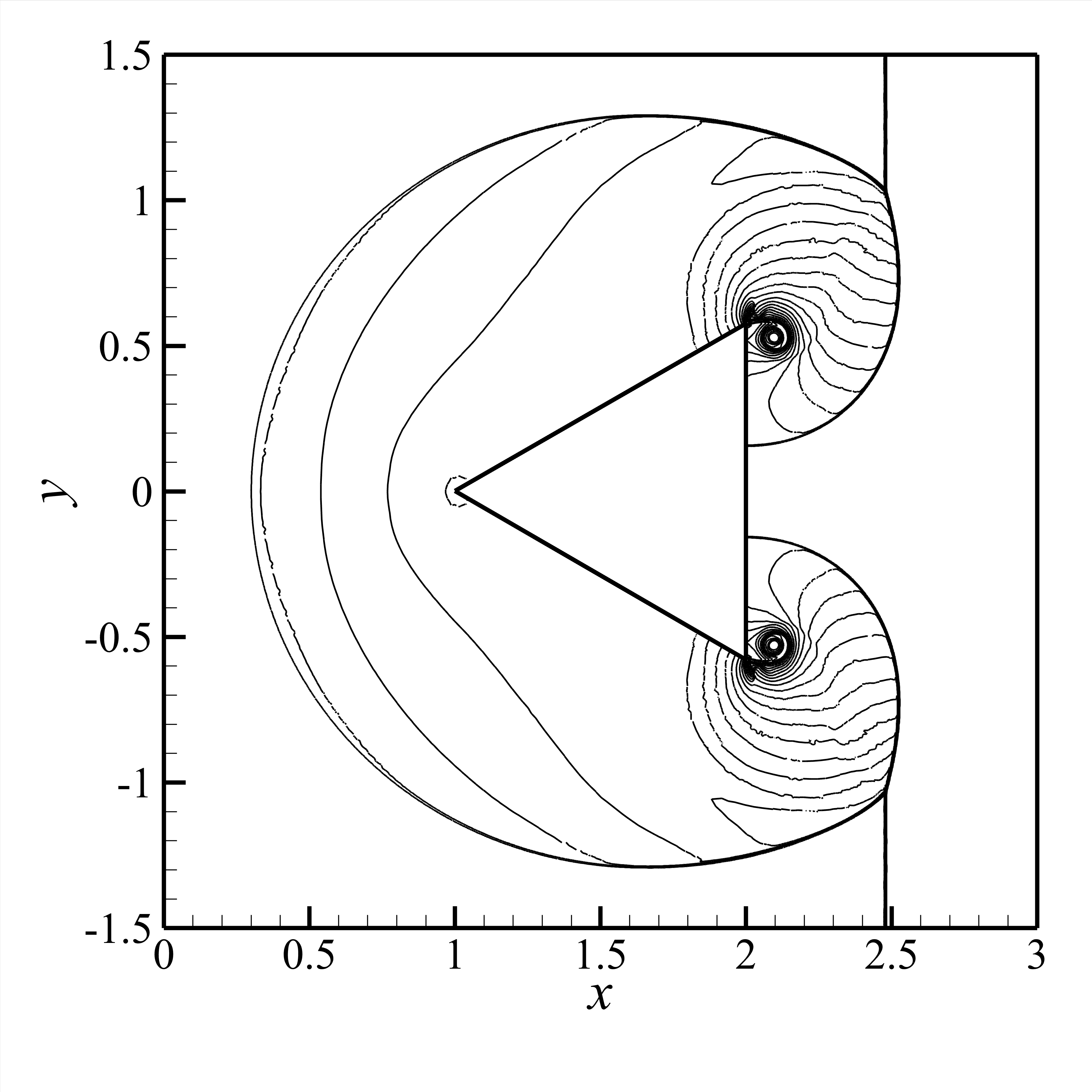}
  \caption{$t=1.104$, 30 equally spaced density contours from 0.6 to 2.9.}
\end{subfigure}
\begin{subfigure}{0.47\textwidth}
  \centering
  \includegraphics[width=0.9\linewidth,trim=0cm 10cm 0cm 0.5cm, clip]{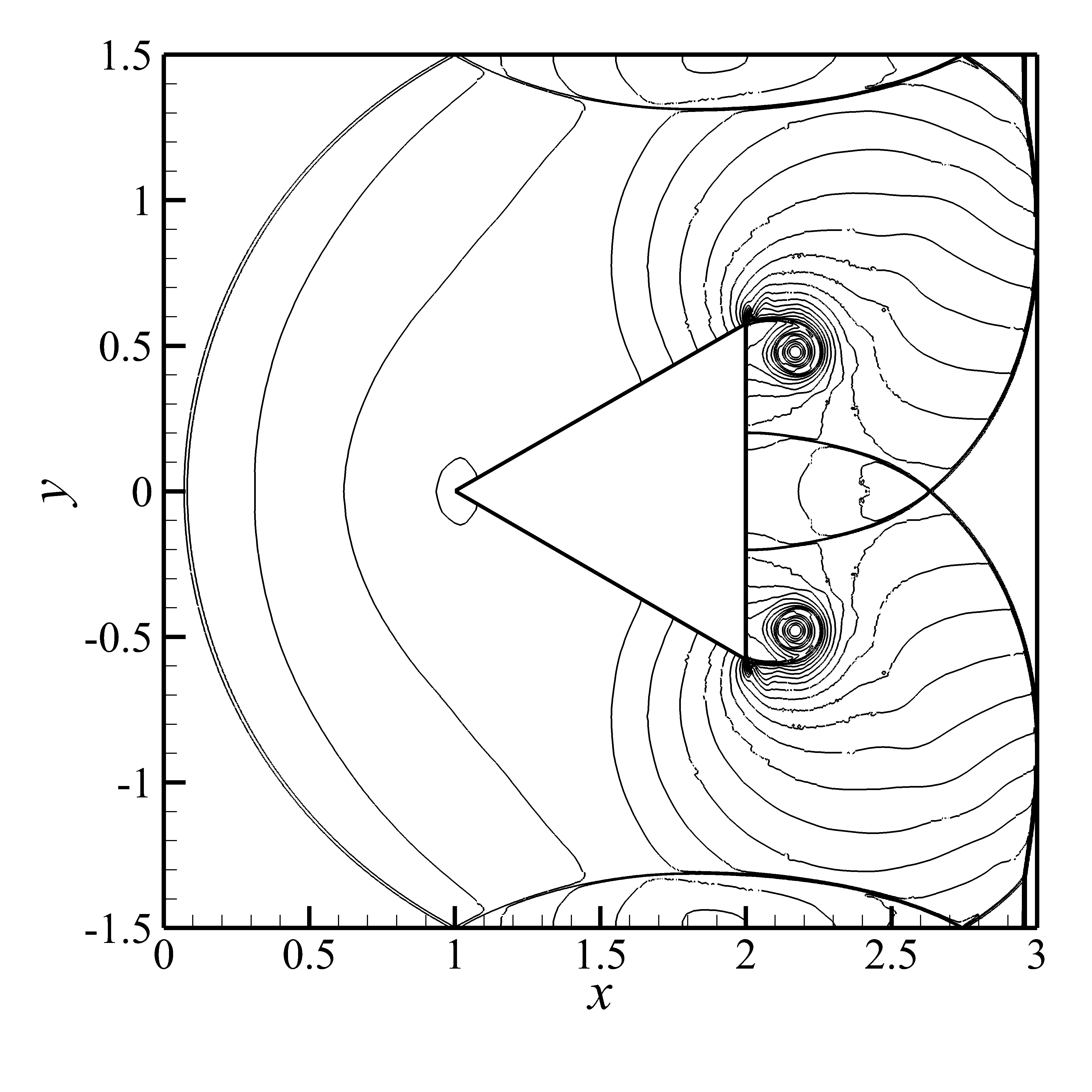}
  \caption{$t=1.46$, 30 equally spaced density contours from 0.5 to 2.9.}
\end{subfigure}
\caption{Density contours at different time steps as a shock wave passes an equilateral triangle, with $l_{\max}=3$ }
\label{wedgeRes}
\end{figure}

This problem addresses the computational challenges associated with unstructured root meshes within a complex computational domain, commonly known as Schardin's problem~\cite{schardin1957high}. A planar shock wave with a Mach number of 1.34 interacts with an equilateral triangle, undergoing diffraction. Due to symmetry, the analysis focuses solely on the upper half of the computational domain. The computational domain, spanning $[0,3] \times[0,1.5]$, is discretized using unstructured quadrilateral meshes as illustrated in Fig.~\ref{wedgeMesh}. Initially positioned at the wedge apex, the right-propagating shock wave encounters a quiescent gas region characterized by a density and pressure of 1.4 and 1, respectively. The obtained results, as depicted in Fig.~\ref{wedgeRes}, exhibit favorable agreement with those presented in a prior study by Zhang et al.~\cite{zhang2012maximum}.

\subsection{Double Mach reflection}
\begin{figure}[H]
\centering
\includegraphics[width=10cm,trim=0cm 7cm 0cm 48.5cm, clip]{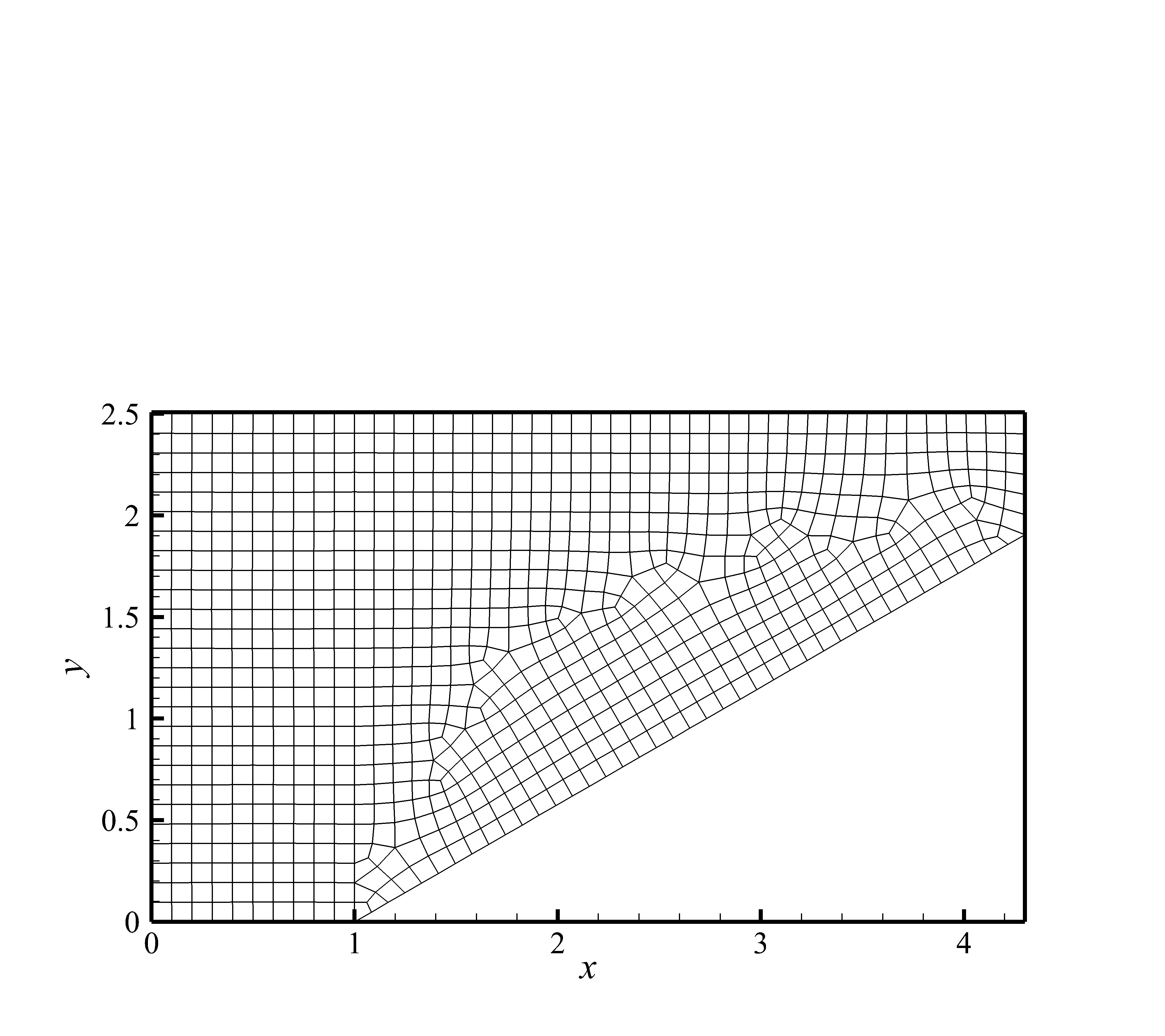}
\caption{Root mesh of the double Mach reflection problem with an average mesh size of $\frac{1}{10}$.}\label{doubleMachMesh}
\end{figure}

This problem focuses on the phenomenon of a Mach 10 shock interacting with a 30-degree wedge, resulting in double Mach reflections. This scenario serves as a prominent benchmark for evaluating high-resolution numerical schemes and has been extensively investigated by rotating the shock direction to enable the use of a simplified rectangular computational domain. In this work, we directly simulate this problem using a coarse unstructured root mesh, as visualized in Fig.~\ref{doubleMachMesh}, with an average mesh size approximately equal to $\frac{1}{10}$. The computational setup incorporates inlet boundary conditions on the left side, outflow conditions on the right side, and slip wall boundary conditions elsewhere. Initially, the right-propagating shock is situated at $x=1$.

\begin{figure}[H]
\centering
\begin{subfigure}{0.47\textwidth}
  \centering
  \includegraphics[width=0.9\linewidth,trim=5cm 5cm 5cm 5cm, clip]{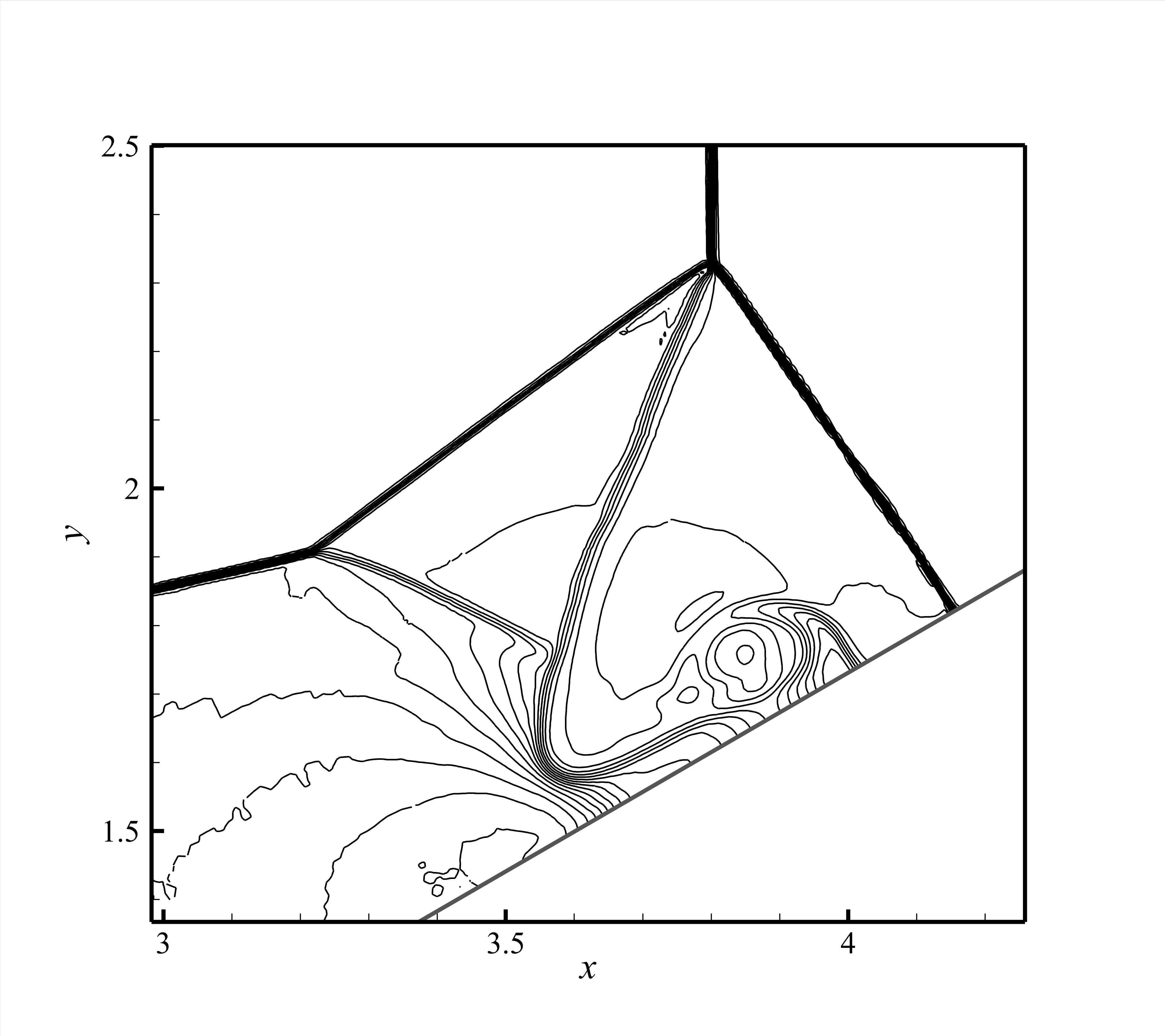}
  \caption{$\ell_{\text {max }}=4$}
\end{subfigure}
\begin{subfigure}{0.47\textwidth}
  \centering
  \includegraphics[width=0.9\linewidth,trim=5cm 5cm 5cm 5cm, clip]{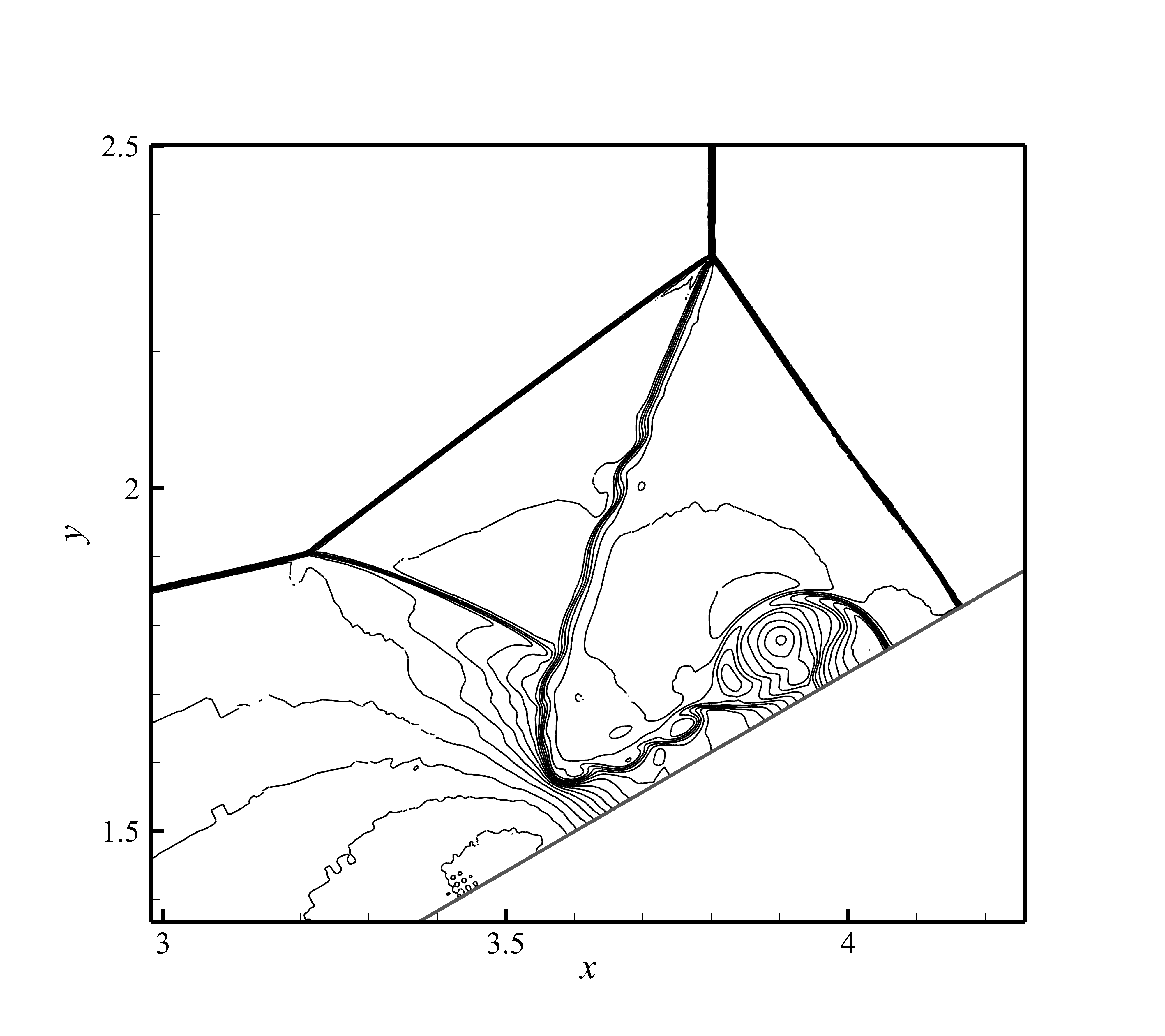}
  \caption{$\ell_{\text {max }}=5$}
\end{subfigure}
\begin{subfigure}{0.47\textwidth}
  \centering
  \includegraphics[width=0.9\linewidth,trim=5cm 5cm 5cm 5cm, clip]{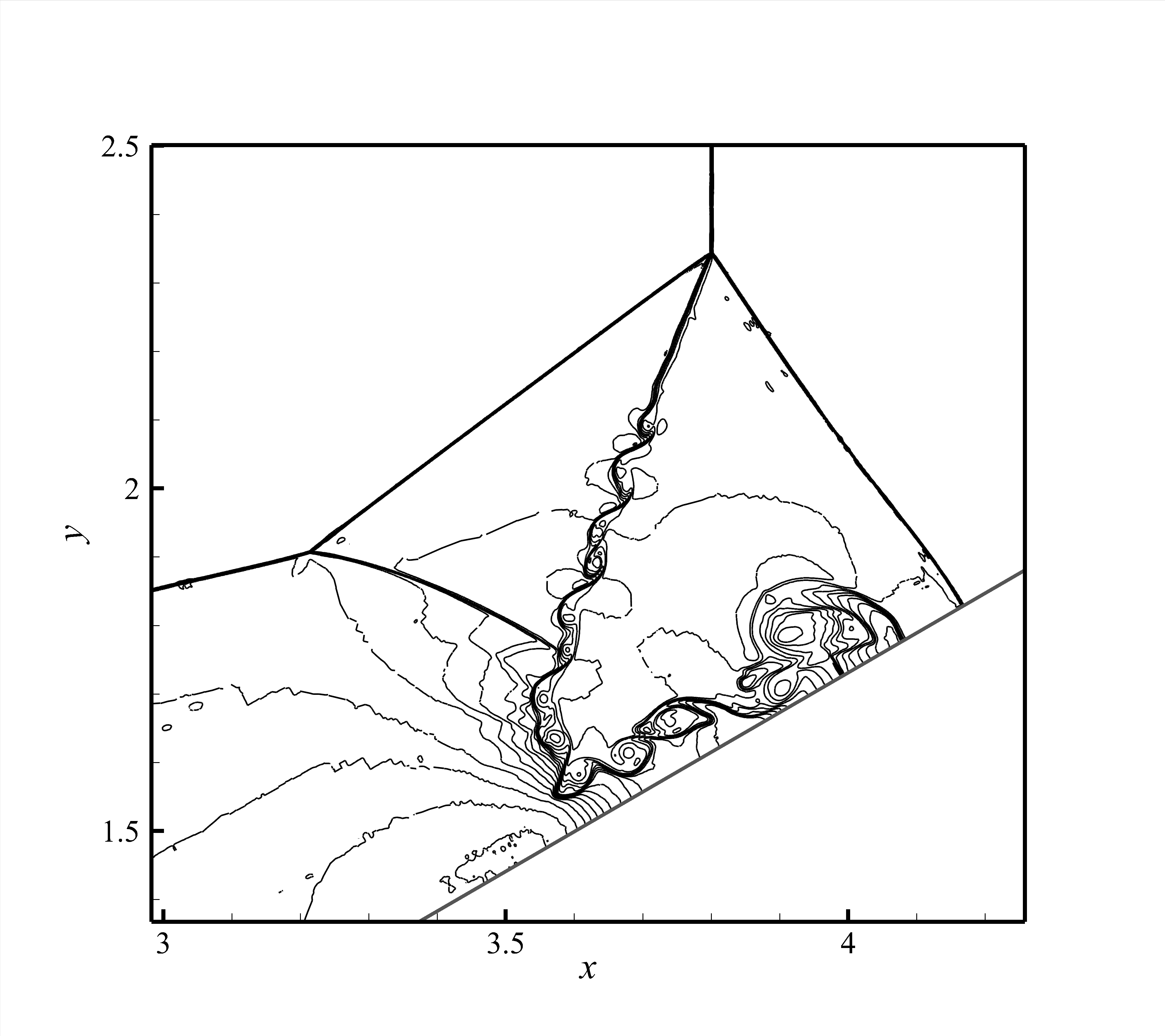}
  \caption{$\ell_{\text {max }}=6$}
\end{subfigure}
\begin{subfigure}{0.47\textwidth}
  \centering
  \includegraphics[width=0.9\linewidth,trim=5cm 5cm 5cm 5cm, clip]{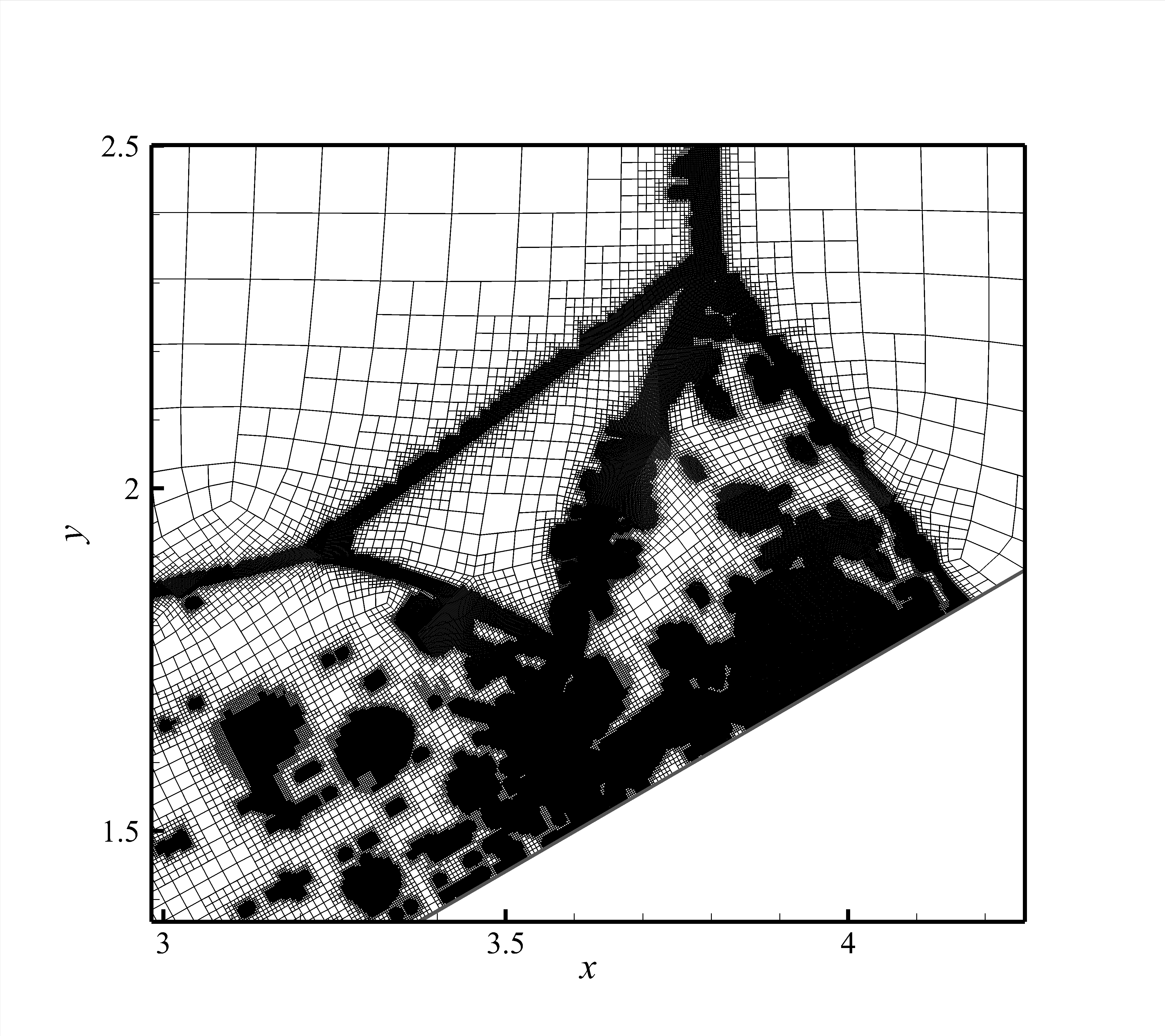}
  \caption{adapted meshes at $\ell_{\text {max }}=6$}
\end{subfigure}
\caption{Zoomed views near the Mach stem for $\ell_{\max }=4 \sim 6$. Density contours are plotted with 40 equally spaced intervals from 0.5 to 2.9.}
\label{doubleMachL}
\end{figure}

The problem is solved employing varying levels of mesh refinement, denoted by $\ell_{\max }=4 \sim 6$, corresponding to effective mesh size of $\frac{1}{160}, \frac{1}{320}$, and $\frac{1}{640}$, respectively, to assess the algorithm's computational efficiency. Density contours are illustrated in Fig.~\ref{doubleMachL}, revealing the emergence of Kelvin-Helmholtz instability near the slip line, accentuated with increasing refinement levels, and the shock becomes more distinct accordingly.

\FloatBarrier

\FloatBarrier

\subsection{Supersonic flow around a circular cylinder}
\begin{figure}[H]
\centering
\includegraphics[width=12cm,trim=6cm 5cm 5cm 2.5cm, clip]{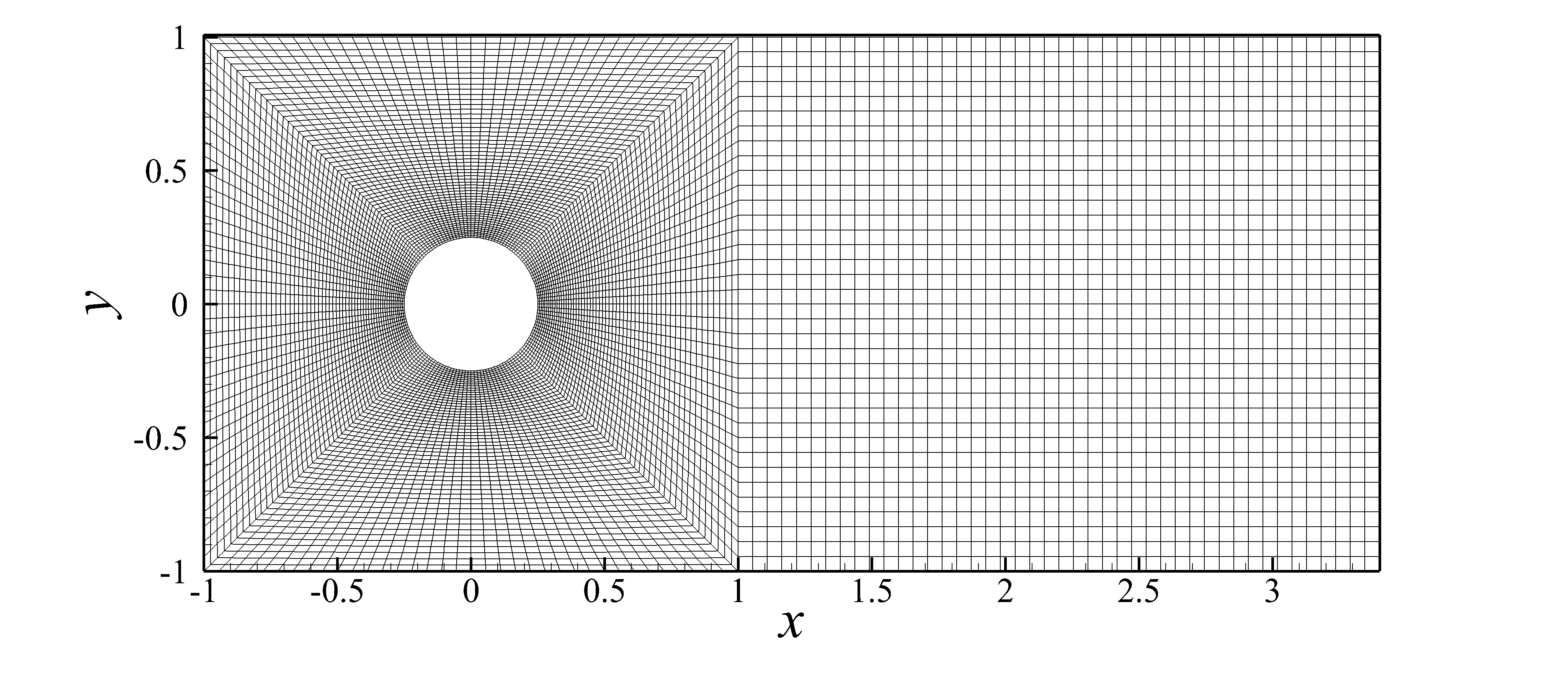}
\caption{Root meshes of the supersonic flow around a circular cylinder problem.}\label{superCylinMesh}
\end{figure}

In this section, we conduct a simulation of supersonic flow over a cylinder confined by a duct, following a setup similar to that described in Guermond et al.~\cite{guermond2018second}. The computational domain, depicted in Fig.~\ref{superCylinMesh}, spans $[-1,3.4] \times[-1,1]$ and is populated with root quadrilateral meshes. A circular cylinder with a radius of 0.25 is positioned at the origin $(0,0)$. The mesh topology conforms to the circular cylinder, extending outward, with uniform rectangular meshes employed in the wake of the cylinder. This scenario comprises intricate steady flow patterns alongside unsteady oscillating vortex structures. A maximum refinement level of $\ell_{\max }=3$ is implemented. To prevent excessive refinement near the cylinder, a constraint is imposed ensuring that the minimum mesh area remains above $4 \times 10^{-5}$; beyond this threshold, further refinement is restricted. This simulation aims to showcase the algorithm's proficiency in accommodating obstacles and capturing dynamic flow features adeptly.

The initial parameters encompass $\rho=1.4, p=1,(u, v)=(3,0)$. A Mach 3 inlet boundary condition is enforced at the left boundary, while supersonic outlets are prescribed at the right boundary of the computational domain. Additionally, the top and bottom boundaries, and the surface of the cylinder are designated as a slip wall.

\begin{figure}[t]
\centering
\includegraphics[width=16cm,trim=2cm 0.5cm 2cm 0.5cm, clip]{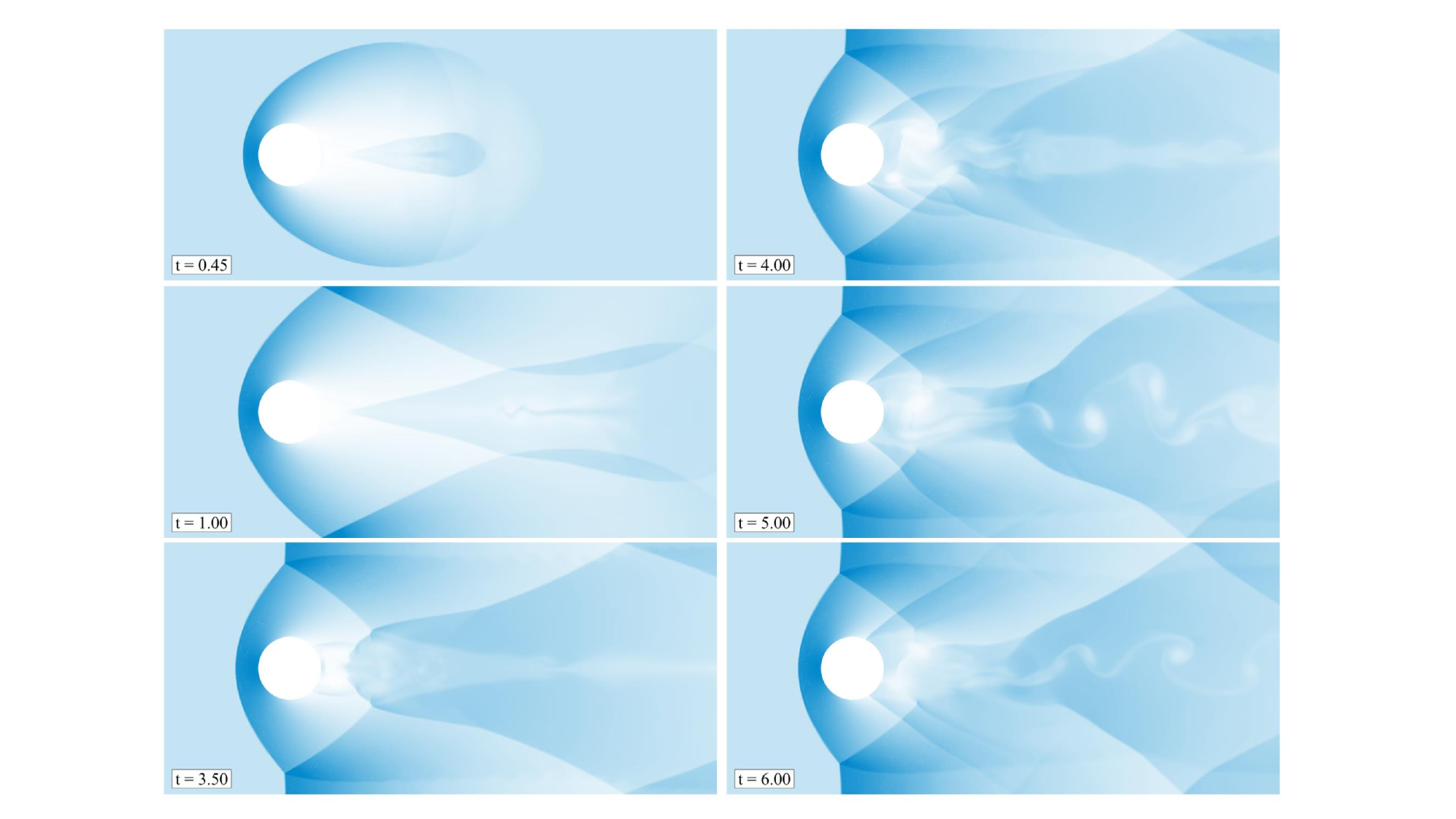}
\caption{Supersonic flow around a cylinder in the duct at different time steps, with $\ell_{\max }=3$.}\label{superCylinRho}
\end{figure}

In Fig.~\ref{superCylinRho}, the density contour plots at various time instances reveal the evolution of flow features. A bow shock emerges from the interaction of the supersonic flow with the cylinder, reflecting off the boundaries to form Mach stems. These reflected shocks propagate back into the channel, interacting with shears and generate a sequence of vortices. The intricate interplay between these shocks and vortex structures is distinctly observable. Analogous phenomena have been documented in prior studies such as Guermond et al.~\cite{guermond2018second} and Maier \& Kronbichler~\cite{maier2021efficient}, underscoring the efficacy of the current AMR algorithm for staggered schemes in ``on flight'' adapting to complex geometries seamlessly.

\section{Performance of the present AMR algorithm}
\label{secperf}

Here, we access the performance of the present AMR algorithms using the above examples of 2D Riemann problem and double Mach reflection problem, by comparing with the computations using the uniform meshes or with different refinement levels.

For the 2D Riemann problem, Tab.~\ref{tab2drie} compares the computational costs by using adapted meshes or uniform meshes with different schemes. The computational cost is counted as the computational time normalized by the time cost by the case using uniform meshes and the $a$-$\alpha$ scheme. It can be observed that for tests with uniform meshes, the CNI scheme is slightly slower than the $a$-$\alpha$ scheme. This is mainly caused by the interpolation step (Eq.~\eqref{defineqm}) in the CNI scheme. The computational cost by using HLLC CESE is significant higher than the other two central schemes, due to the reconstruction procedures (Eqs.~\eqref{wbap1}\&\eqref{wbap2}) and solving the additional Riemann problems. The advantages of obtaining superior resolutions are at the cost of the increase of computational time. When using a AMR approach, the computational costs by these three schemes ranges 8.6$\sim$13.2\% of their uniform counterparts. Nevertheless, it should be noted that the structured HLLC CESE has been successfully extended to multi-component flows~\cite{shen2017maximum,guan2018numerical,fan2019numerical}, while the interface could be very diffusive if computed by a central CESE scheme~\cite{qamar2012space}. The selection of appropriate scheme is suggested depend on the specific physics models and the balance of resolution requirement and the computational load.

\begin{table}[H]
\centering
\begin{tabular}{lcccc}
\hline  & uniform ($1088^2$) & AMR ($34^2$,$\ell_{\max }=5$) \\
\hline $a$-$\alpha$ CESE & 1.000 & 0.132 \\
CNI CESE & 1.220 & 0.140 \\
HLLC CESE & 2.627 & 0.226 \\
\hline
\end{tabular}
\caption{Normalized computational costs of uniform and AMR computation for different schemes in the 2D Riemann problem.}\label{tableCom}
\label{tab2drie}
\end{table}

\begin{figure}[t]
\centering
\includegraphics[width=12cm,trim=2cm 1cm 2cm 1cm, clip]{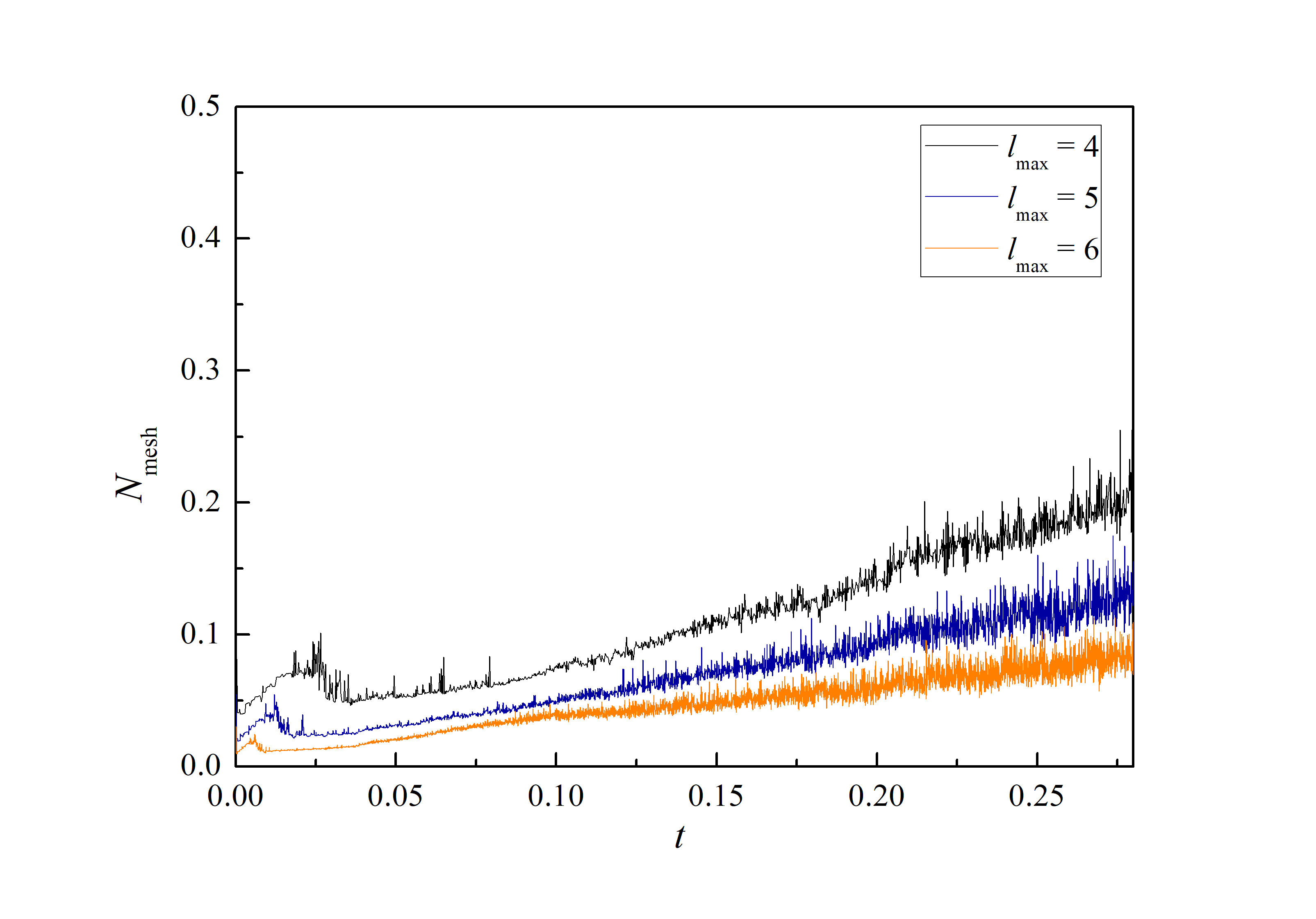}
\caption{Mesh numbers normalized by the number of equivalent uniform meshes in the double Mach reflection problem.}\label{doubleMachNorMesh}
\end{figure}

Next, we examine the computational performance for different refinement levels employed in addressing the double Mach reflection problem. 
The evolution of real-time mesh numbers for different maximum refinement levels, normalized by the equivalent numbers for uniform meshes, is elucidated in Fig.~\ref{doubleMachNorMesh}. The equivalent mesh number is defined as the total mesh number needed to be used for uniform meshes with resolution equals to the finest resolution resolved by the highest refinement level using the AMR simulation. As the Mach stem progresses forward during the simulation, the mesh number increases due to similarities in wave structures, except for the size of the Mach stem and shears, which amplify. Near the end of computation, the normalized mesh number for $\ell_{\max }=4$ remains within a $20 \%$ margin and around $8 \%$ for $\ell_{\max }=6$, underscoring the efficacy of the present AMR algorithms. As the maximum refinement level is increased, there is a corresponding decrease in the normalized mesh number, a phenomenon driven by the closer clustering of meshes around critical regions of interest.

\begin{figure}[t]
\centering
\includegraphics[width=13cm,trim=2cm 10.5cm 2cm 2.3cm, clip]{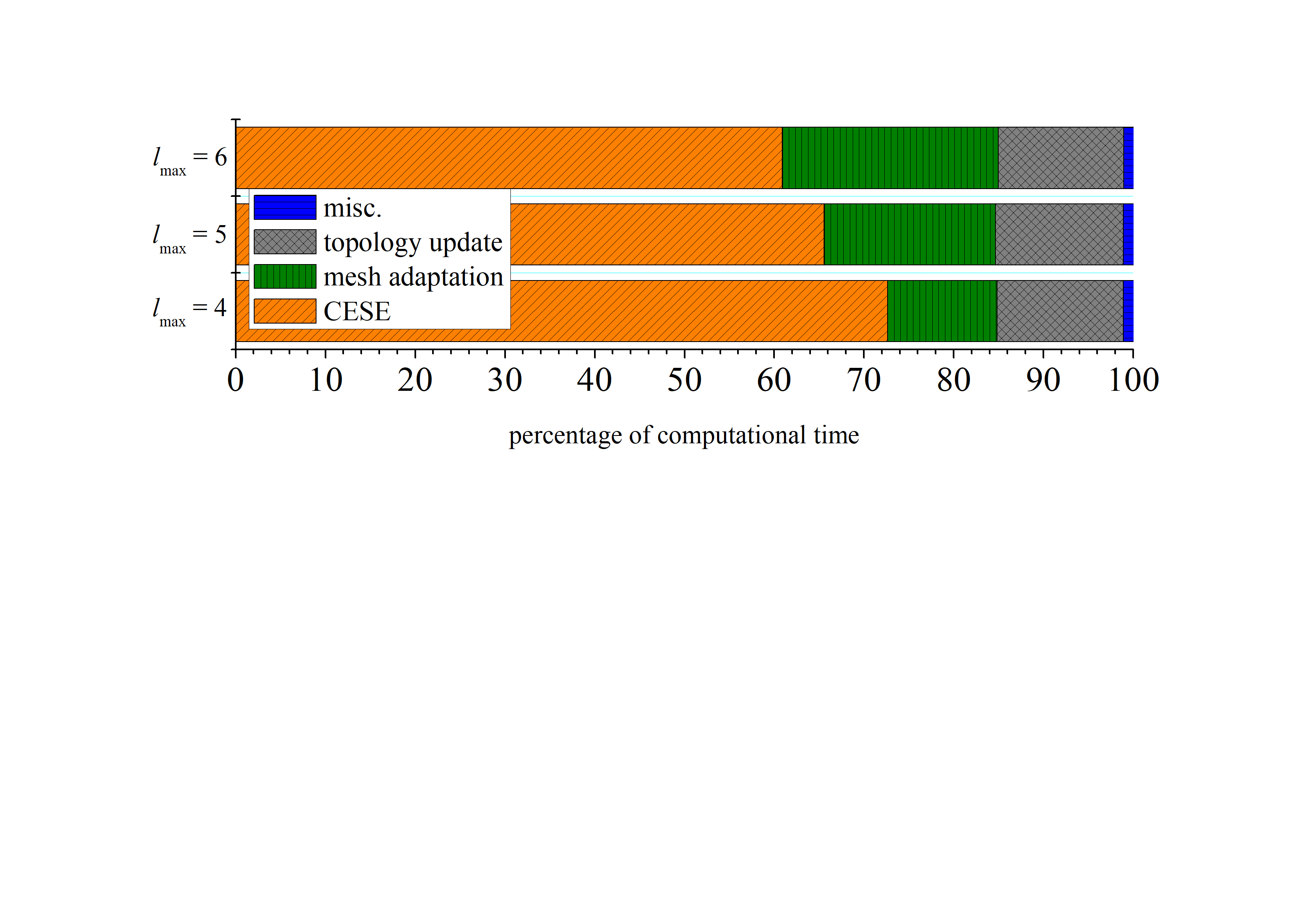}
\caption{Percentage of computational time cost by each part in the double Mach reflection problem.}\label{doubleMachTime}
\end{figure}

 The computational cost analysis, detailed in Fig.~\ref{doubleMachTime}, delineates the allocation of resources across four distinct components: the flow solver (CESE), AMR (mesh adaptation), redefine the conservations elements and solution elements for cell centers/vertices that have been affected (topology update), and miscellaneous parts (misc.) including input/output, apply boundary conditions, etc.

Despite the complexity and computational demands of adaptation processes, the cost associated with mesh refinement remains within acceptable limits even at elevated refinement levels such as $\ell_{\text {max }}=6$, which entail a total of seven layers of meshes including the root layer. Remarkably, the flow solver component stands out as the primary consumer of computational resources. Furthermore, when increasing the maximum refinement level by one, the computational resource required increases by approximately 5.6 times, contrasting with the 8-fold rise seen in fixed meshes (2 times in both spatial dimension and 2 times in temporal dimension, i.e., $2^{3}$). It should be reminded that, the mesh adaptation algorithm is executed in each time-step in the present study. Meanwhile, some studies have shown that the adaptation procedure can be executed every several time-steps~\cite{keppens2003adaptive,henshaw2008parallel}, which may further save the computational cost. Since the present study focuses on the novel strategy for mesh adaptation for staggered schemes, these enhancements could be further examined in the future studies.

\section{Conclusions}
\label{secconc}
The present study establishes the central and upwind CESE schemes for split quadrilateral meshes and introduces a tailored AMR strategy for staggered numerical schemes. The novel adaptation algorithm, using a cell-based approach and incorporating a sophisticated cell-tree-vertex data structure optimized for staggered schemes, streamlines the process of rapid topology construction. Various numerical tests demonstrate the capability of these advancements to accurately capture intricate flow phenomena, including shocks and discontinuities, within complex computational domains encompassing both steady-state and unsteady scenarios. The core concept of the proposed AMR algorithm shows promise for extension towards high-order compact staggered schemes. Future research efforts could explore enhancements in this area as well as considerations such as load-balancing to further improve the algorithm's effectiveness and applicability.

\section*{Acknowledgements}
This work was supported the opening project of the State Key Laboratory of Explosion Science and Technology (Beijing Institute of Technology, KFJJ23-20M).

\bibliographystyle{unsrt}
\bibliography{manuBib}

\begin{thebibliography}{10}

\bibitem{fan2022numerical}
E~Fan, Jiaao Hao, Ben Guan, Chih-Yung Wen, and Lisong Shi.
\newblock Numerical investigation on reacting shock-bubble interaction at a low
  mach limit.
\newblock {\em Combustion and Flame}, 241:112085, 2022.

\bibitem{cant2022unstructured}
RS~Cant, U~Ahmed, Jian Fang, N~Chakarborty, G~Nivarti, Charles Moulinec, and
  DR~Emerson.
\newblock An unstructured adaptive mesh refinement approach for computational
  fluid dynamics of reacting flows.
\newblock {\em Journal of Computational Physics}, 468:111480, 2022.

\bibitem{papoutsakis2018efficient}
Andreas Papoutsakis, Sergei~S Sazhin, Steven Begg, Ionut Danaila, and Francky
  Luddens.
\newblock An efficient adaptive mesh refinement (amr) algorithm for the
  discontinuous galerkin method: Applications for the computation of
  compressible two-phase flows.
\newblock {\em Journal of Computational Physics}, 363:399--427, 2018.

\bibitem{sun1999conservative}
M~Sun and K~Takayama.
\newblock Conservative smoothing on an adaptive quadrilateral grid.
\newblock {\em Journal of Computational Physics}, 150(1):143--180, 1999.

\bibitem{wang2022prediction}
He~Wang, Zhigang Zhai, and Xisheng Luo.
\newblock Prediction of triple point trajectory on two-dimensional unsteady
  shock reflection over single surfaces.
\newblock {\em Journal of Fluid Mechanics}, 947:A42, 2022.

\bibitem{schmidmayer2019adaptive}
Kevin Schmidmayer, Fabien Petitpas, and Eric Daniel.
\newblock Adaptive mesh refinement algorithm based on dual trees for cells and
  faces for multiphase compressible flows.
\newblock {\em Journal of Computational Physics}, 388:252--278, 2019.

\bibitem{schmidmayer2020ecogen}
Kevin Schmidmayer, Fabien Petitpas, S{\'e}bastien Le~Martelot, and {\'E}ric
  Daniel.
\newblock Ecogen: An open-source tool for multiphase, compressible,
  multiphysics flows.
\newblock {\em Computer Physics Communications}, 251:107093, 2020.

\bibitem{deiterding2003parallel}
Ralf Deiterding.
\newblock {\em Parallel adaptive simulation of multi-dimensional detonation
  structures}.
\newblock Dissertation. de, 2003.

\bibitem{gallier2017detonation}
S~Gallier, F~Le~Palud, F~Pintgen, R~M{\'e}vel, and JE~Shepherd.
\newblock Detonation wave diffraction in h2--o2--ar mixtures.
\newblock {\em Proceedings of the Combustion Institute}, 36(2):2781--2789,
  2017.

\bibitem{o2005introducing}
Brian~W O’shea, Greg Bryan, James Bordner, Michael~L Norman, Tom Abel, Robert
  Harkness, and Alexei Kritsuk.
\newblock Introducing enzo, an amr cosmology application.
\newblock In {\em Adaptive Mesh Refinement-Theory and Applications: Proceedings
  of the Chicago Workshop on Adaptive Mesh Refinement Methods, Sept. 3--5,
  2003}, pages 341--349. Springer, 2005.

\bibitem{khokhlov1999interaction}
Alexei~M Khokhlov, Elaine~S Oran, Almadena~Yu Chtchelkanova, and J~Craig
  Wheeler.
\newblock Interaction of a shock with a sinusoidally perturbed flame.
\newblock {\em Combustion and flame}, 117(1-2):99--116, 1999.

\bibitem{deiterding2009parallel}
Ralf Deiterding.
\newblock A parallel adaptive method for simulating shock-induced combustion
  with detailed chemical kinetics in complex domains.
\newblock {\em Computers \& Structures}, 87(11-12):769--783, 2009.

\bibitem{liang2014effects}
Jianhan Liang, Xiaodong Cai, Zhiyong Lin, and Ralf Deiterding.
\newblock Effects of a hot jet on detonation initiation and propagation in
  supersonic combustible mixtures.
\newblock {\em Acta Astronautica}, 105(1):265--277, 2014.

\bibitem{cai2016adaptive}
Xiaodong Cai, Jianhan Liang, Ralf Deiterding, Yonggang Che, and Zhiyong Lin.
\newblock Adaptive mesh refinement based simulations of three-dimensional
  detonation combustion in supersonic combustible mixtures with a detailed
  reaction model.
\newblock {\em international journal of hydrogen energy}, 41(4):3222--3239,
  2016.

\bibitem{peng2023three}
Han Peng and Ralf Deiterding.
\newblock A three-dimensional solver for simulating detonation on curvilinear
  adaptive meshes.
\newblock {\em Computer Physics Communications}, 288:108752, 2023.

\bibitem{henry2023pelec}
Marc~T Henry~de Frahan, Jon~S Rood, Marc~S Day, Hariswaran Sitaraman, Shashank
  Yellapantula, Bruce~A Perry, Ray~W Grout, Ann Almgren, Weiqun Zhang, John~B
  Bell, et~al.
\newblock Pelec: An adaptive mesh refinement solver for compressible reacting
  flows.
\newblock {\em The International Journal of High Performance Computing
  Applications}, 37(2):115--131, 2023.

\bibitem{katz2020preparing}
Max~P Katz, Ann Almgren, Maria~Barrios Sazo, Kiran Eiden, Kevin Gott, Alice
  Harpole, Jean~M Sexton, Don~E Willcox, Weiqun Zhang, and Michael Zingale.
\newblock Preparing nuclear astrophysics for exascale.
\newblock In {\em SC20: International Conference for High Performance
  Computing, Networking, Storage and Analysis}, pages 1--12. IEEE, 2020.

\bibitem{zhang2019amrex}
Weiqun Zhang, Ann Almgren, Vince Beckner, John Bell, Johannes Blaschke,
  Cy~Chan, Marcus Day, Brian Friesen, Kevin Gott, Daniel Graves, et~al.
\newblock Amrex: a framework for block-structured adaptive mesh refinement.
\newblock {\em The Journal of Open Source Software}, 4(37):1370, 2019.

\bibitem{zhang2021amrex}
Weiqun Zhang, Andrew Myers, Kevin Gott, Ann Almgren, and John Bell.
\newblock Amrex: Block-structured adaptive mesh refinement for multiphysics
  applications.
\newblock {\em The International Journal of High Performance Computing
  Applications}, 35(6):508--526, 2021.

\bibitem{macneice2000paramesh}
Peter MacNeice, Kevin~M Olson, Clark Mobarry, Rosalinda De~Fainchtein, and
  Charles Packer.
\newblock Paramesh: A parallel adaptive mesh refinement community toolkit.
\newblock {\em Computer physics communications}, 126(3):330--354, 2000.

\bibitem{stone2020athena++}
James~M Stone, Kengo Tomida, Christopher~J White, and Kyle~G Felker.
\newblock The athena++ adaptive mesh refinement framework: Design and
  magnetohydrodynamic solvers.
\newblock {\em The Astrophysical Journal Supplement Series}, 249(1):4, 2020.

\bibitem{khokhlov1998fully}
Alexei~M Khokhlov.
\newblock Fully threaded tree algorithms for adaptive refinement fluid dynamics
  simulations.
\newblock {\em Journal of Computational Physics}, 143(2):519--543, 1998.

\bibitem{bastian2021dune}
Peter Bastian, Markus Blatt, Andreas Dedner, Nils-Arne Dreier, Christian
  Engwer, Ren{\'e} Fritze, Carsten Gr{\"a}ser, Christoph Gr{\"u}ninger, Dominic
  Kempf, Robert Kl{\"o}fkorn, et~al.
\newblock The dune framework: Basic concepts and recent developments.
\newblock {\em Computers \& Mathematics with Applications}, 81:75--112, 2021.

\bibitem{lawlor2006parfum}
Orion~S Lawlor, Sayantan Chakravorty, Terry~L Wilmarth, Nilesh Choudhury, Isaac
  Dooley, Gengbin Zheng, and Laxmikant~V Kale.
\newblock Parfum: a parallel framework for unstructured meshes for scalable
  dynamic physics applications.
\newblock {\em Engineering with Computers}, 22:215--235, 2006.

\bibitem{chang1995method}
Sin-Chung Chang.
\newblock The method of space-time conservation element and solution
  element—a new approach for solving the navier-stokes and euler equations.
\newblock {\em Journal of computational Physics}, 119(2):295--324, 1995.

\bibitem{wang2010improved}
Gang Wang, Deliang Zhang, Kaixin Liu, and Jingtao Wang.
\newblock An improved ce/se scheme for numerical simulation of gaseous and
  two-phase detonations.
\newblock {\em Computers \& fluids}, 39(1):168--177, 2010.

\bibitem{shen2015robust}
Hua Shen, Chih-Yung Wen, Kaixin Liu, and Deliang Zhang.
\newblock Robust high-order space--time conservative schemes for solving
  conservation laws on hybrid meshes.
\newblock {\em Journal of Computational Physics}, 281:375--402, 2015.

\bibitem{shi2023numerical}
Lisong Shi, E~Fan, Hua Shen, Chih-Yung Wen, Shuai Shang, and Hongbo Hu.
\newblock Numerical study of the effects of injection conditions on rotating
  detonation engine propulsive performance.
\newblock {\em Aerospace}, 10(10):879, 2023.

\bibitem{chang2000application}
Sin-Chung Chang, Xiao-Yen Wang, and Wai-Ming To.
\newblock Application of the space--time conservation element and solution
  element method to one-dimensional convection--diffusion problems.
\newblock {\em Journal of Computational Physics}, 165(1):189--215, 2000.

\bibitem{chang2002courant}
S~Chang and X~Wang.
\newblock Courant number insensitive ce/se euler scheme.
\newblock In {\em 38th AIAA/ASME/SAE/ASEE Joint Propulsion Conference \&
  Exhibit}, page 3890, 2002.

\bibitem{shen2015characteristic}
Hua Shen, Chih-Yung Wen, and De-Liang Zhang.
\newblock A characteristic space--time conservation element and solution
  element method for conservation laws.
\newblock {\em Journal of Computational Physics}, 288:101--118, 2015.

\bibitem{shen2016characteristic}
Hua Shen and Chih-Yung Wen.
\newblock A characteristic space--time conservation element and solution
  element method for conservation laws ii. multidimensional extension.
\newblock {\em Journal of Computational Physics}, 305:775--792, 2016.

\bibitem{jiang2020space}
Yazhong Jiang, Chih-Yung Wen, and Deliang Zhang.
\newblock Space--time conservation element and solution element method and its
  applications.
\newblock {\em AIAA Journal}, 58(12):5408--5430, 2020.

\bibitem{fedorov2004evolution}
Alexander Fedorov and Anatoli Tumin.
\newblock Evolution of disturbances in entropy layer on blunted plate in
  supersonic flow.
\newblock {\em AIAA journal}, 42(1):89--94, 2004.

\bibitem{shen2017maximum}
Hua Shen, Chih-Yung Wen, Matteo Parsani, and Chi-Wang Shu.
\newblock Maximum-principle-satisfying space-time conservation element and
  solution element scheme applied to compressible multifluids.
\newblock {\em Journal of Computational Physics}, 330:668--692, 2017.

\bibitem{guan2018numerical}
Ben Guan, Yao Liu, Chih-Yung Wen, and Hua Shen.
\newblock Numerical study on liquid droplet internal flow under shock impact.
\newblock {\em AIAA journal}, 56(9):3382--3387, 2018.

\bibitem{fan2019numerical}
E~Fan, Ben Guan, Chih-Yung Wen, and Hua Shen.
\newblock Numerical study on the jet formation of simple-geometry heavy gas
  inhomogeneities.
\newblock {\em Physics of Fluids}, 31(2), 2019.

\bibitem{zhang2020effects}
ZJ~Zhang, CY~Wen, YF~Liu, DL~Zhang, and ZL~Jiang.
\newblock Effects of different particle size distributions on aluminum
  particle--air detonation.
\newblock {\em AIAA Journal}, 58(7):3115--3128, 2020.

\bibitem{yang2018upwind}
Yun Yang, Xue-Shang Feng, and Chao-Wei Jiang.
\newblock An upwind cese scheme for 2d and 3d mhd numerical simulation in
  general curvilinear coordinates.
\newblock {\em Journal of Computational Physics}, 371:850--869, 2018.

\bibitem{wang19993}
Xiao-Yen Wang and Sin-Chung Chang.
\newblock A 3-d non-splitting structured/unstructured euler solver based on the
  space-time conservation element and solution element method.
\newblock In {\em 14th Computational Fluid Dynamics Conference}, page 3278,
  1999.

\bibitem{zhang2002space}
Zeng-Chan Zhang, ST~John Yu, and Sin-Chung Chang.
\newblock A space-time conservation element and solution element method for
  solving the two-and three-dimensional unsteady euler equations using
  quadrilateral and hexahedral meshes.
\newblock {\em Journal of Computational Physics}, 175(1):168--199, 2002.

\bibitem{wen2018extension}
Chih-Yung Wen, H~Saldivar Massimi, and Hua Shen.
\newblock Extension of ce/se method to non-equilibrium dissociating flows.
\newblock {\em Journal of Computational Physics}, 356:240--260, 2018.

\bibitem{shen2018positivity}
Hua Shen and Matteo Parsani.
\newblock Positivity-preserving ce/se schemes for solving the compressible
  euler and navier--stokes equations on hybrid unstructured meshes.
\newblock {\em Computer Physics Communications}, 232:165--176, 2018.

\bibitem{jiang2012solving}
Chaowei Jiang, Shuxin Cui, and Xueshang Feng.
\newblock Solving the euler and navier--stokes equations by the amr--cese
  method.
\newblock {\em Computers \& Fluids}, 54:105--117, 2012.

\bibitem{liu2022numerical}
Zhipeng Liu, Chaowei Jiang, Xueshang Feng, Pingbing Zuo, and Yi~Wang.
\newblock Numerical simulation of solar magnetic flux emergence using the
  amr--cese--mhd code.
\newblock {\em The Astrophysical Journal Supplement Series}, 264(1):13, 2022.

\bibitem{fu2013simulation}
Zheng Fu, Kai-Xin Liu, and Ning Luo.
\newblock Simulation of shock-induced instability using an essentially
  conservative adaptive ce/se method.
\newblock {\em Chinese Physics B}, 23(2):020202, 2013.

\bibitem{van1979towards}
Bram Van~Leer.
\newblock Towards the ultimate conservative difference scheme. v. a
  second-order sequel to godunov's method.
\newblock {\em Journal of computational Physics}, 32(1):101--136, 1979.

\bibitem{shu1988efficient}
Chi-Wang Shu and Stanley Osher.
\newblock Efficient implementation of essentially non-oscillatory
  shock-capturing schemes.
\newblock {\em Journal of computational physics}, 77(2):439--471, 1988.

\bibitem{gottlieb2001strong}
Sigal Gottlieb, Chi-Wang Shu, and Eitan Tadmor.
\newblock Strong stability-preserving high-order time discretization methods.
\newblock {\em SIAM review}, 43(1):89--112, 2001.

\bibitem{guo2004extension}
Yanhu Guo, Andrew~T Hsu, Jie Wu, Zhigang Yang, and Ayo Oyediran.
\newblock Extension of ce/se method to 2d viscous flows.
\newblock {\em Computers \& fluids}, 33(10):1349--1361, 2004.

\bibitem{shi2017assessment}
Lisong Shi, Hua Shen, Peng Zhang, Deliang Zhang, and Chih-Yung Wen.
\newblock Assessment of vibrational non-equilibrium effect on detonation cell
  size.
\newblock {\em Combustion Science and Technology}, 189(5):841--853, 2017.

\bibitem{feng2007novel}
Xueshang Feng, Yufen Zhou, and ST~Wu.
\newblock A novel numerical implementation for solar wind modeling by the
  modified conservation element/solution element method.
\newblock {\em The Astrophysical Journal}, 655(2):1110, 2007.

\bibitem{wen2023space}
Chih-Yung Wen, Yazhong Jiang, and Lisong Shi.
\newblock {\em Space--Time Conservation Element and Solution Element Method:
  Advances and Applications in Engineering Sciences}.
\newblock Springer Nature, 2023.

\bibitem{chang2003multi}
S~Chang and X~Wang.
\newblock Multi-dimensional courant number insensitive euler solvers for
  applications involving highly nonuniform meshes.
\newblock In {\em 39th AIAA/ASME/SAE/ASEE Joint Propulsion Conference and
  Exhibit}, page 5285, 2003.

\bibitem{li2011multi}
Wanai Li, Yu-Xin Ren, Guodong Lei, and Hong Luo.
\newblock The multi-dimensional limiters for solving hyperbolic conservation
  laws on unstructured grids.
\newblock {\em Journal of Computational Physics}, 230(21):7775--7795, 2011.

\bibitem{toro2013riemann}
Eleuterio~F Toro.
\newblock {\em Riemann solvers and numerical methods for fluid dynamics: a
  practical introduction}.
\newblock Springer Science \& Business Media, 2013.

\bibitem{toro1994restoration}
Eleuterio~F Toro, Michael Spruce, and William Speares.
\newblock Restoration of the contact surface in the hll-riemann solver.
\newblock {\em Shock waves}, 4:25--34, 1994.

\bibitem{levy1993use}
David~W Levy, Kenneth~G Powell, and Bram van Leer.
\newblock Use of a rotated riemann solver for the two-dimensional euler
  equations.
\newblock {\em Journal of Computational Physics}, 106(2):201--214, 1993.

\bibitem{nishikawa2008very}
Hiroaki Nishikawa and Keiichi Kitamura.
\newblock Very simple, carbuncle-free, boundary-layer-resolving, rotated-hybrid
  riemann solvers.
\newblock {\em Journal of Computational Physics}, 227(4):2560--2581, 2008.

\bibitem{ren2003robust}
Yu-Xin Ren.
\newblock A robust shock-capturing scheme based on rotated riemann solvers.
\newblock {\em Computers \& Fluids}, 32(10):1379--1403, 2003.

\bibitem{colella1984piecewise}
Phillip Colella and Paul~R Woodward.
\newblock The piecewise parabolic method (ppm) for gas-dynamical simulations.
\newblock {\em Journal of computational physics}, 54(1):174--201, 1984.

\bibitem{geuzaine2009gmsh}
Christophe Geuzaine and Jean-Fran{\c{c}}ois Remacle.
\newblock Gmsh: A 3-d finite element mesh generator with built-in pre-and
  post-processing facilities.
\newblock {\em International journal for numerical methods in engineering},
  79(11):1309--1331, 2009.

\bibitem{sod1978survey}
Gary~A Sod.
\newblock A survey of several finite difference methods for systems of
  nonlinear hyperbolic conservation laws.
\newblock {\em Journal of computational physics}, 27(1):1--31, 1978.

\bibitem{sidilkover2018towards}
David Sidilkover.
\newblock Towards unification of the vorticity confinement and shock capturing
  (tvd and eno/weno) methods.
\newblock {\em Journal of Computational Physics}, 358:235--255, 2018.

\bibitem{zhang2007new}
Shuhai Zhang and Chi-Wang Shu.
\newblock A new smoothness indicator for the weno schemes and its effect on the
  convergence to steady state solutions.
\newblock {\em Journal of Scientific Computing}, 31:273--305, 2007.

\bibitem{schardin1957high}
H~Schardin.
\newblock High frequency cinematography in the shock tube.
\newblock {\em The Journal of Photographic Science}, 5(2):17--19, 1957.

\bibitem{zhang2012maximum}
Xiangxiong Zhang, Yinhua Xia, and Chi-Wang Shu.
\newblock Maximum-principle-satisfying and positivity-preserving high order
  discontinuous galerkin schemes for conservation laws on triangular meshes.
\newblock {\em Journal of Scientific Computing}, 50(1):29--62, 2012.

\bibitem{guermond2018second}
Jean-Luc Guermond, Murtazo Nazarov, Bojan Popov, and Ignacio Tomas.
\newblock Second-order invariant domain preserving approximation of the euler
  equations using convex limiting.
\newblock {\em SIAM Journal on Scientific Computing}, 40(5):A3211--A3239, 2018.

\bibitem{maier2021efficient}
Matthias Maier and Martin Kronbichler.
\newblock Efficient parallel 3d computation of the compressible euler equations
  with an invariant-domain preserving second-order finite-element scheme.
\newblock {\em ACM Transactions on Parallel Computing}, 8(3):1--30, 2021.

\bibitem{qamar2012space}
Shamsul Qamar, Munshoor Ahmed, and Ishtiaq Ali.
\newblock The space-time ce/se method for solving reduced two-fluid flow model.
\newblock {\em Communications in Computational Physics}, 12(4):1070--1095,
  2012.

\bibitem{keppens2003adaptive}
Rony Keppens, M~Nool, G~T{\'o}th, and JP~Goedbloed.
\newblock Adaptive mesh refinement for conservative systems: multi-dimensional
  efficiency evaluation.
\newblock {\em Computer Physics Communications}, 153(3):317--339, 2003.

\bibitem{henshaw2008parallel}
William~D Henshaw and Donald~W Schwendeman.
\newblock Parallel computation of three-dimensional flows using overlapping
  grids with adaptive mesh refinement.
\newblock {\em Journal of Computational Physics}, 227(16):7469--7502, 2008.

\end{thebibliography}

\end{document}